\documentclass{article}
\usepackage{graphicx,epsfig}
\usepackage{graphicx,subfigure}
\usepackage{arxiv}
\usepackage{amsmath}
\usepackage{amssymb}
\usepackage{titlesec}
\usepackage[utf8]{inputenc} 
\usepackage[T1]{fontenc}    
\usepackage{url}            
\usepackage{booktabs}       
\usepackage{amsfonts,bm}       
\usepackage{microtype}      
\usepackage[mathscr]{euscript}
\usepackage[dvipsnames]{xcolor}
\pdfoutput=1
\usepackage{hyperref}
\hypersetup{
    colorlinks=true,
    linkcolor=black,
    citecolor=black,
    filecolor=black,
    urlcolor=black,
}
\makeatother
\usepackage[numbers, compress]{natbib}
\title{Impact of a floating flexible plate on the stability of double-layered falling flow}
\author{
  Md. Mouzakkir Hossain\\
  Department of Mathematics\\
  SRM Institute of Science and Technology\\
  Kattankulathur-603203, India\\
  \texttt{mouzakkir123@gmail.com} \\
\And
  Sukhendu Ghosh*\\
  Department of Mathematics\\
  Indian Institute of Technology Jodhpur\\
  Rajasthan-342037, India\\
  \texttt{sukhendu.math@gmail.com} \\
     \And
  Harekrushna Behera*\\
  Department of Mathematics\\
  SRM Institute of Science and Technology\\
  Kattankulathur-603203, India\\
  \texttt{hkb.math@gmail.com} \\
}

\begin{document}
\maketitle

\begin{abstract}
The hydrodynamic stability behaviour of a two-layer falling film is explored with a floating flexible plate on the top surface. The stress balance at the surface is modeled using a modified membrane equation. There is an insoluble surfactant at the liquid-liquid interface of the flow system. The linear instability of perturbation waves is captured by numerically solving the generalized Orr-Sommerfeld eigenvalue problem using Chebyshev spectral collocation technique. Four different types of linear stability modes are identified: namely surface mode (SM), interface mode (IM), interface surfactant mode (ISM), and shear mode (SHM). The floating flexible plate has an inferior impact on all the existing modes in the longwave zone. However, a significant influence is noticed for the finite wave numbers. The surface and interface wave instabilities can be suppressed in the smaller wavenumber zone by imposing higher structural rigidity and uniform thickness of the flexible plate.
The stabilizing nature of the surface mode becomes more powerful when the top layer viscosity is dominant.
A new interfacial instability emerges when the top layer is less viscous than the lower one.
At moderate Reynolds numbers, the behaviour of interface mode is different in two different zones $mr<1$ and $mr>1$, where $m$ and $r$ denote the viscosity and density ratio, respectively. Further, the unstable shear modes induced by the top and bottom layers are detected under the low inclination angles with a strong inertia force. The occurrence of the shear modes demands the bottom layer's viscosity and density to be very high than that of the upper layer. The influence of characteristic parameters of the flexible plate on the lower layer shear mode is not very sensitive. Finally, the competition between the different modes for dominance of stability boundaries is also discussed.

\end{abstract}


\section{Introduction}\label{Introduction}
The hydrodynamic instability of falling flow models receives considerable attention due to its numerous applications in coating technology \cite{weinstein2004coating},  chemical industry \cite{drazin2002introduction, truzzolillo2017hydrodynamic}, and in many other industrial and natural setups \cite{wang1984thin, sivapuratharasu2016inertial}. The surface wave has crucial importance in improving the heat and mass transfer rates in the process equipment.
The surface of a liquid film exhibits different dynamical structures known as free surface instabilities. Several researchers have been conducted to expose the physical mechanism of such instabilities and provide different explicit/implicit control strategies to stimulate or subdue single/double-layered free surface flow instabilities relying on the applications in recent decades. There are several real-world circumstances in which it is necessary to control the surface wave instability of fluid flows by using either surface active or surface inactive agents, depending on the applications \cite{li1997effect, blyth2004effect, gao2007effect, anjalaiah2013thin, samanta2014effect}. For example, in coating technology, the stable film flows of solitary or superposed layers play a vital role in ensuring the product quality of manufactured photographic films \cite{weinstein2004coating}.

A large number of studies providing various control strategies for surface wave instability of single/multi-layer, laminar, and Newtonian/non-Newtonian fluid flows over an inclined bed are available in the literature. In the single-layer falling film over an inclined wall, the instability emerges owing to the inertial influence, whereas the instability (interfacial) in double or multi-layer liquid films is due to kinetic reasons \cite{kao1965role, kao1968role}. One of the strategies for controlling the instability is by incorporating an insoluble surfactant as a surface active agent on the fluid surface. \citet{blyth2004effect} used the insoluble surfactant at the single-layer fluid surface and showed the stability impact of surfactant on the Yih mode by solving the corresponding Orr--Sommerfeld boundary value problem.
They found that the surfactant mode (Marangoni mode) associated with the surfactant arises from the local variation in surface tension induced by the surfactant agent.
Another idea for controlling surface wave energy is to impose an external shear at the free surface.
This process is physically applicable in the airway occlusion process \cite{otis1993role}. In view of this, \citet{samanta2014shear} examined the influence of external shear on the gravity-driven falling film down a rigid incline. It was explored that the flow-directed external force promotes longwave instability by reducing the critical Reynolds number. Later, \citet{bhat2019linear} imposed the external force at the top surface of the surfactant-laden single-layer fluid film down a rigid bed. Their aim was to observe how the insoluble surfactant at the fluid surface could be used to weaken external shear-induced wave energy. They identified the existence of another mode, named shear mode, other than the surface and surfactant modes. The high inertia force and small inclination angle mainly cause the emergence of shear mode. It was concluded that if the applied force acts in the reverse flow direction with a higher Marangoni force, a potent stabilization of the surface mode is possible. Also, they discussed the physical mechanism of surface wave instability using energy budget analysis based on the idea of \citet{kelly1989mechanism}. An extensive review of the works in the surface-active agents on the fluid surface has been provided by \citet{benjamin1964effects}, \citet{lin1970stabilizing}, and \citet{pascal2019stability}.

Concurrently, several researchers have shown an interest in studying the instability control mechanism in double-layer fluid flow systems due to its extensive use in various chemical and technical processes, such as liquid-liquid extraction, food production, liquid coating, etc. \cite{tuck1983continuous, suy1995stability}. Under certain flow configurations, interfacial and free surface instabilities may be generated and degrade the quality of the final product in coating processes and related applications.
Consequently, controlling these instabilities has enormous practical implications. In the case of two-layer, \citet{gao2007effect} considered the insoluble surfactant in both free surface and interface in the inertialess limit. They identified four modes but discussed only interface and surface modes. The impact of surface/interfacial surfactant was to reduce/enhance the inertialess instability in the longwave region. Moreover, the two-layer flows are widely accessible in the sense that they can be found in a wide range of natural circumstances and are applicable in various biological and technological systems such as in the petroleum and plastics industries \cite{espinosa1999bolus, malamataris2002solitary, criminale2003mathematical}. Interfacial instabilities can occur under specific operating requirements, causing unfavorable outcomes such as spoilage of the end product. The stability of the interfacial wave between the fluids is a crucial consideration in this respect, as a stable interface is required to achieve the necessary automated and border effects in the flow system.
\citet{kao1965stability} first proposed the concept of double-layered fluid flow by examining the impacts of depth ratio, viscosity ratio, and density ratio on the primary instability generated by the surface and interface modes.
Later, \citet{chen1993wave}  examined the linear stability of double-layered falling film at the low Reynolds number. He observed the key role of viscosity stratification at the interface mode and confirmed the occurrence of instability when the viscosity of the lower layer is higher than that of the upper layer. \citet{samanta2014effect} further extended the work of \citet{gao2007effect} by incorporating inertia force up to moderate values of Reynolds number. He studied the effect of insoluble surfactants on the primary instability induced by the free surface and interface mode. 
It was shown that the interfacial instability becomes weaker due to the appearance of interfacial surfactant when the top layer is less viscous compared to the bottom one. \citet{bhat2020linear} examined surfactant's influence on the two-layer falling film flows over a slippery incline.
They pointed out the double role of the viscosity ratio in the surface mode instability. There are studies with various flow configurations where researchers have used the external shear as a surface-active agent to regulate the free surface-wave instability \cite{ sani2020effect, samanta2021effect, samanta2021instability, pal2022linear, hossain2022linear}.

Apart from the above instability control strategy, the placement of a floating flexible membrane/plate at the top surface is another control technique to stabilize the most unstable mode. It draws remarkable significance in the scientific community for several uses \cite{hosoi2004peeling, trinh2014elastic, sani2021effect, selvan2021hydroelastic}. Physically, the elastic cover prevents the transition to turbulence in the laminar flow by reducing the flow velocity in the boundary layers below the plate. 
\citet{sani2021effect} applied an elastic membrane on the liquid surface to stabilize the primary instability of the film flows over an inclined porous bed. They observed that the floating elastic membrane follows the instability mechanism, similar to the Marangoni mechanism of the surfactant-contaminated single-layer falling liquid film.
 
Moreover, hydroelastic instability analysis in a gravity-driven fluid flow in the presence of a floating elastic plate/membrane can be widely used in the study of coating of elastic surfaces, collapsible tubes, and cardiovascular, and lung flow \cite{carpenter1990effect, luo1996numerical, davies1997numerical}. 

On the other hand, on a large scale, the floating elastic plate can serve as a practical model for the frozen ice or very large floating structure (VLFS) on the sea \cite{squire2007ocean, squire2011past}. It was initially used by \citet{kheysin1963moving, kheysin1964problem} to construct a wave-ice interaction model. Later, many researchers \cite{kheysin1963moving,sahoo2012mathematical, das2018dynamics, das2018flexural} modeled a thin ice-sheet as a floating flexible plate and explored various aspects of flexural gravity waves under a thin ice-sheet. To define the VLFS, they employed the Euler-Bernoulli plate or beam model \cite{meylan1997forced, das2018flexural}. The wave interaction with VLFS was investigated because of their vast use in the utilization of ocean space for numerous humanitarian activities such as airstrips, ice-crossing, and several other transportation links \cite{takizawa1985deflection, watanabe2004hydroelastic}.

Following the above ideas, \citet{selvan2021hydroelastic} considered the floating flexible plate at the surface of gravity-viscous liquid film over a rigid bed to regulate the hydrodynamic instability in both linear and nonlinear regimes. Their investigation confirmed that the flexible plate on a falling fluid surface could be utilized as an alternative for passive/active control when the compressive force is absent/present.
They used the numerical technique (Chebyshev collocation method) to solve the Orr-Sommerfeld eigenvalue problem and confirmed that the floating elastic plate suppresses the unstable Yih mode \cite{yih1963stability}, associated with the disturbance wave superimposed on the liquid surface. Recently, \citet{hossain2023instability} investigated the effect of infinitely extended floating flexible structure on both viscous and inviscid sheared-layered film flow instability over a horizontal slippery surface. Using energy budget analysis \cite{kelly1989mechanism, hossain2022linear, khan2021poiseuille}, they found that the plate's characteristic parameters (uniform rigidity and thickness) dampen the external shear-induced wave energy of the liquid surface, whereas compressive force encourages the flexural destabilization.  

The above discussion motivates us to examine the dynamics and stability behaviour of the two-layer gravity-driven flow
having a floating flexible plate on the top layer surface. In this manuscript, we have extended the work of \citet{selvan2021hydroelastic} relating the idea of \citet{bhat2020linear} by considering a floating elastic plate at the top surface of a two-layer gravity-driven falling film over a rigid incline having an insoluble surfactant at the interface between the upper and lower layer.
The Euler-Bernoulli plate theory \cite{ohmatsu2005overview} is used to describe the flexible structure. The modified Orr-Sommerfeld boundary value problem (OS-BVP) is derived using the normal mode approach and the linear perturbation technique. The numerical technique Chebyshev spectral collocation is used to solve it.
The primary goal of the present paper is to investigate how floating elastic plates affect different modes, including surface mode, interface surfactant mode, and interface mode, which have not yet been investigated.
The behaviour of these modes is discussed explicitly when the flow parameters thickness ratio, viscosity ratio, and density ratio fluctuate in the presence of a floating elastic plate. Also, we have investigated the characteristics of the shear mode. It emerges when the inertia force is strong enough and has a low angle of inclination, provided the lower layer's viscosity and density are very high compared to the upper layer. 

The followings are the manuscript's layout: Section~\ref{MF} describes the governing equations and the boundary conditions. Next,
Section~\ref{BVP} obtains the Orr-Sommerfeld eigenvalue problem using the normal mode analysis. Section~\ref{NR} details numerical results, and finally, a conclusion is made in Section~\ref{CON}. 

\section{Mathematical Formulation}  \label{MF}
 \begin{figure}[ht!]
\begin{center}
\includegraphics[width=12cm]{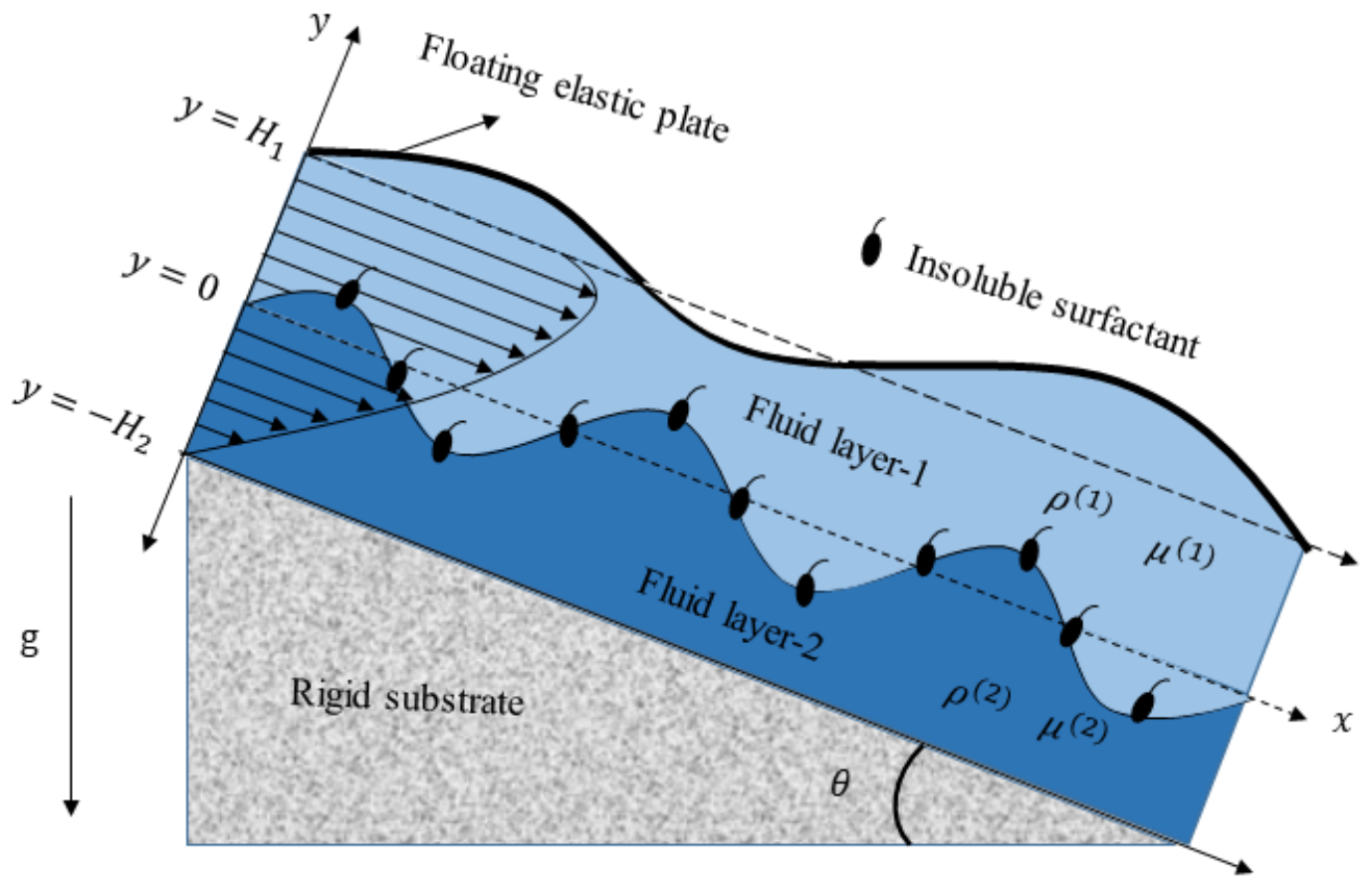}
\end{center}
	\caption{Schematic diagram for a two-layer thin film flow down a rigid substrate in the presence of a floating elastic plate. The insoluble surfactant is considered at the interface between the upper and lower layer. }\label{f1}
\end{figure}
A double-layered, incompressible, immiscible, Newtonian fluid flows over an inclined rigid bed of an angle $\theta$ (Fig.~\ref{f1}). Fluid-1 corresponds to the top layer of thickness $H_1$ with viscosity $\mu^{(1)}$ and density $\rho^{(1)}$ and the bottom layer of thickness $H_2$, is assumed to have the density $\rho^{(2)}$ and the viscosity $\mu^{(2)}$, is referred to as fluid-2.    
Further, the deformations of the top surface and interface are represented as $f^{(1)}$ and $f^{(2)}$, respectively. A thin flexible plate is placed on the top surface of the upper layer. A Cartesian coordinate system is adopted, where the origin is on the liquid-liquid interface. The $x$ axis is chosen along the streamwise, and the $y$ axis is taken in the cross-streamwise flow direction for both liquid layers.      
The dimensional forms of equations of motion for both fluid layers are as follows:

\begin{subequations}
\begin{align}
&u^{(j)}_x+v^{(j)}_y=0,
\\
&\rho^{(j)}\bigg[u^{(j)}_t+u^{(j)}~u^{(j)}_x+v^{(j)}~u^{(j)}_y\bigg]=-p^{(j)}_x+\mu^{(j)}~(u^{(j)}_{xx}+u^{(j)}_{yy})+\rho^{(j)}\,g\,\sin{\theta},
\\
&\rho^{(j)}\bigg[v^{(j)}_t+u^{(j)}~v^{(j)}_x+v^{(j)}~v^{(j)}_y\bigg]=-p^{(j)}_y+\mu^{(j)}~(v^{(j)}_{xx}+v^{(j)}_{yy})-\rho^{(j)}\,g\,\cos{\theta},
\end{align} \label{new1}
\end{subequations}
where fluid-1 and fluid-2 are marked by $j=1$ and $j=2$, respectively. The stramwise and cross-streamwise velocity of the $j$\textsuperscript{th} layer liquid are depicted by $u^{(j)}$ and $v^{(j)}$, correspondingly. $p^{(j)}$ represents the fluid pressure of the $j$\textsuperscript{th} layer and $g$ is the gravitational constant. The dimensional form of kinematic boundary conditions at the free surface and interface is given as
\begin{align}
v^{(j)}=f^{(j)}_t+u^{(j)}~f^{(j)}_x~~~~\mbox{at}~~~~y=f^{(j)}(x,t).\label{n1}
\end{align}\
It is assumed that the elastic plate's length is greater than the disturbance wavelength of the top layer. The exerted force by the elastic plate balances the pressure difference between the liquid and the surrounding air, which may be formulated by Euler- Bernoulli's beam equation \cite{magrab2012vibrations}.  

At the top surface, the dynamic conditions yield the following tangential and normal shear stress balance equations in dimensional form as   
\begin{align}
&\mu^{(1)}\bigg[-4u^{(1)}_xf^{(1)}_x+(u^{(1)}_y+v^{(1)}_x)~\bigg\{1-(f^{(1)}_x)^2\bigg\}\bigg]=0~~~~\text{at}~~~~ y = f^{(1)}(x,t),\label{ee4}\\
&p^{(1)} = \frac{2\mu^{(1)}}{\big[1+\big(f^{(1)}_x\big)^2\big]}
\biggl\{u^{(1)}_x~(f^{(1)}_x)^2-(u^{(1)}_y+v^{(1)}_x)~f^{(1)}_x+v^{(1)}_y\biggr\}+ EI\frac{\partial^2}{\partial x^2}\Biggl[\frac{f^{(1)}_{xx}}{\left\{1+(f^{(1)}_x)^2\right\}^{\frac{3}{2}}}\Biggr]\nonumber\\
&\hspace{2cm}-N\frac{f^{(1)}_{xx}}{\left\{1+(f^{(1)}_x)^2\right\}^{\frac{3}{2}}}-\rho_p\, d_p\,f^{(1)}_{tt}~~~\text{at}~~~~y=f^{(1)}(x,t). \label{ee6}
\end{align}
The structural rigidity of the elastic plate is determined by $EI$, where $E$ is the Young's modulus, $I=d_p^3/(12(1-\nu^2))$, $\nu$ is the Poisson's ratio, and $d_p$ is the uniform plate thickness. Further, $N$ stands for the in-axis compressive forces, and $\rho_p$ symbolizes the plate density \cite{das2018flexural, boral2021role}.
The balance of the shear stress and normal stress jump at the liquid-liquid interface yields the following boundary conditions \cite{kao1965stability, samanta2014effect, bhat2020linear, sani2020effect}:
\begin{align}
&\mu^{(2)}\,\bigg[-4u^{(2)}_xf^{(2)}_x+(u^{(2)}_y+v^{(2)}_x)\bigg\{1-(f^{(2)}_x)^2\bigg\}\bigg]-\sigma_x\sqrt{1+ (f^{(2)}_x)^{2} }=-4u^{(1)}_xf^{(2)}_x+(u^{(1)}_y+v^{(1)}_x)\nonumber\\
&\hspace{9.5cm}\bigg\{1-(f^{(2)}_x)^2\bigg\}~~~\text{at}~~~~y=f^{(2)}(x,t),\label{ee7}
\end{align}
\begin{align}
&p^{(2)}-\frac{2\mu^{(2)}}{\big[1+(f^{(2)}_x)^2\big]}\bigg[u^{(2)}_x\biggl\{1-(f^{(2)}_x)^2\biggr\}+(u^{(2)}_y+v^{(2)}_x)~f^{(2)}_x\bigg]+\sigma\frac{~f^{(2)}_{xx}}{\big[1+(f^{(2)}_x)^2\big]^{3/2}}=p^{(1)}\nonumber\\
&\hspace{3cm}-\frac{2\mu^{(1)}}{\big[1+(f^{(1)}_x)^2\big]}\bigg[u^{(1)}_x\bigg\{1-(f^{(2)}_x)^2\bigg\}+(u^{(1)}_y+v^{(1)}_x)~f^{(2)}_x\bigg]~~~\text{at}~~~~y=f^{(2)}(x,t).\label{ee8} 
\end{align}
At the top-bottom layer interface, the streamwise cross-streamwise velocity components of the upper and lower layer liquids are continuous, which yield 
\begin{align}
u^{(1)}=u^{(2)}~~~~\mbox{and}~~~~v^{(1)}=v^{(2)}~~~~~\mbox{at}~~~~~y=f^{(2)}(x,t).
\end{align}
Along the rigid wall below the double-layered fluid flow, there is no velocity slip and no penetration of fluid into the solid substrate, yielding in
\begin{align}
u^{(2)}=0~~~~\mbox{and}~~~~v^{(2)}=0~~~~~\mbox{at}~~~~~y=-H_2.\label{new9}
\end{align}
The deformation equation of the surfactant concentration $\Gamma(x,t)$ at the liquid-liquid interface \cite{frenkel2002stokes, blyth2004evolution} can be written as follows:  
\begin{align}
\Gamma_t+u^{(2)}~\Gamma_x+\Gamma\biggl(u^{(2)}_x+u^{(2)}_y\,f^{(2)}_x\biggr)=\mathcal{D}_s\Gamma_{xx}~~~~~\mbox{at}~~~~~y=f^{(2)}(x,t),\label{new2}
\end{align}
where $\mathcal{D}_s$ represents the surfactant diffusivity at the liquid-liquid interface. Now, the governing equations and the boundary conditions (Eqs.~\eqref{new1}-\eqref{new2}) are made into non-dimensional form \cite{anjalaiah2013thin, samanta2014effect, selvan2021hydroelastic}
by assuming $U_c$ (the average velocity of the flow system) and $H_1$ as the velocity and length scales, respectively, for both liquid layers, and $\rho^{(1)}U_c^2$ and $\rho^{(2)}U_c^2$ are the scales for the pressure $p^{(1)}$ and $p^{(2)}$, respectively.   

The dimensionless forms of equations for both liquid layers are written as:
\begin{align}
&u^{(j)}_x+v^{(j)}_y=0,\label{e10}
\\
&u^{(j)}_t+u^{(j)}~u^{(j)}_x+v^{(j)}~u^{(j)}_y=-p^{(j)}_x+\frac{1}{Re_j}~(u^{(j)}_{xx}+u^{(j)}_{yy})+G,
\\
&v^{(j)}_t+u^{(j)}~v^{(j)}_x+v^{(j)}~v^{(j)}_y=-p^{(j)}_y+\frac{1}{Re_j}~(v^{(j)}_{xx}+v^{(j)}_{yy})-G\cot\theta,
\\
&v^{(j)}=f^{(j)}_t+u^{(j)}~f^{(j)}_x~~~~\mbox{at}~~~~y=f^{(j)}(x,t),\\
&-4u^{(1)}_xf^{(1)}_x+(u^{(1)}_y+v^{(1)}_x)~\left\{1-(f^{(1)}_x)^2\right\}=0~~~~\text{at}~~~~ y = f^{(1)}(x,t),\\
&Re_1p^{(1)} = \frac{2}{\big[1+\big(f^{(1)}_x\big)^2\big]}
\bigg[u^{(1)}_x~(f^{(1)}_x)^2-(u^{(1)}_y+v^{(1)}_x)~f^{(1)}_x+v^{(1)}_y\bigg]+\alpha Re_1 \frac{\partial^2}{\partial x^2}\Biggl[\frac{f^{(1)}_{xx}}{\left\{1+(f^{(1)}_x)^2\right\}^{\frac{3}{2}}}\Biggr]\nonumber\\
&\hspace{2cm}-\beta Re_1\frac{f^{(1)}_{xx}}{\left\{1+(f^{(1)}_x)^2\right\}^{\frac{3}{2}}}-\gamma Re_1f^{(1)}_{tt}~~~\text{at}~~~~y=f^{(1)}(x,t),
\\
&m\,\biggr[-4u^{(2)}_xf^{(2)}_x+(u^{(2)}_y+v^{(2)}_x)\left\{1-(f^{(2)}_x)^2\right\}\biggl]+\frac{m Ma}{Ca}\Gamma_x\,\sqrt{1+ (f^{(2)}_x)^{2} } = -4u^{(1)}_xf^{(2)}_x+(u^{(1)}_y+v^{(1)}_x) \nonumber\\
&\hspace{9.5cm}\bigg\{1-(f^{(2)}_x)^2\bigg\}~~~\text{at}~~~~y=f^{(2)}(x,t),
\\
&Re_1p^{(1)}+\frac{2}{\big[1+(f^{(1)}_x)^2\big]}\bigg[u^{(1)}_x\bigg\{1-(f^{(2)}_x)^2\bigg\}+(u^{(1)}_y+v^{(1)}_x)~f^{(2)}_x\bigg]=mRe_2p^{(2)}+\frac{2m}{\big[1+(f^{(2)}_x)^2\big]}\bigg[u^{(2)}_x\bigg\{1-\nonumber\\
&\hspace{2.5cm}(f^{(2)}_x)^2\bigg\}+(u^{(2)}_y+v^{(2)}_x)~f^{(2)}_x\bigg]+\frac{m[1-Ma(\Gamma-1)]~f^{(2)}_{xx}}{Ca\big[1+(f^{(2)}_x)^2\big]^{3/2}}~~~\text{at}~~~~y=f^{(2)}(x,t),
\\
&u^{(1)}=u^{(2)}~~~~\mbox{and}~~~~v^{(1)}=v^{(2)}~~~~~\mbox{at}~~~~~y=f^{(2)}(x,t),
\\
&u^{(2)}=0~~~~\mbox{and}~~~~v^{(2)}=0~~~~~\mbox{at}~~~~~y=-\delta.\label{e19}
\end{align}
Note that $\alpha=EI/\rho^{(1)}U_c^2H^3_1$ and $\gamma=m_p/\rho^{(1)}H_1$ define the nondimensional form of structural rigidity and uniform thickness, respectively of the floating elastic plate. Whereas the compressive force of the upper layer liquid is denoted by $\beta=N/\rho^{(1)}U_c^2H_1$.  Further,
the ratio of inertia to the viscous force of the $j$\textsuperscript{th} layer fluid is marked by Reynolds number $Re_j=\rho^{(j)}U_cH_1/\mu^{(j)}$. The Reynolds numbers are related by $Re_2 = (r/m)Re_1$, where density ratio ($\rho^{(2)}/ \rho^{(1)}$), viscosity ratio ($\mu^{(2)}/\mu^{(1)}$), and depth ratio ($H_2/H_1$) are symbolized by $r$, $m$, and $\delta$, respectively.
Here $G=g\sin\theta H_1/U^2_c$ defines the Galileo number \cite{bhat2020linear, anjalaiah2013thin}. Furthermore, the Marangoni number $Ma=RT\Gamma/\sigma_0$ is associated with the interface surfactant and $Ca=U_c\mu^{(2)}/\sigma_0$ is the Capillary number of the lower layer fluid, where $\sigma_0$ refers the base interfacial tension.

The dimensionless form of the surfactant concentration $\Gamma(x,t)$ at the liquid-liquid interface is obtained as
\begin{align}
&\Gamma_t+u^{(2)}~\Gamma_x+\Gamma\biggl(u^{(2)}_x+u^{(2)}_y\zeta_x\biggr)=\frac{1}{Pe}\Gamma_{xx}~~~~~\mbox{at}~~~~~y=\zeta^{(2)}(x,t),\label{e20}
\end{align}
where $Pe = U_c H_1/\mathcal{D}_s$ is the P\'eclet number. The surfactant concentration $\Gamma(x,t)$ diffuses into the fluid at the interface, causing changes in the fluid interfacial tension $\sigma$. As a result, the relationship between surfactant concentration and interfacial tensions can be described as $ \sigma=\sigma_0-RT(\Gamma-\Gamma_0)$ with the universal gas constant $R$ and the absolute temperature $T$. The surface tension and surfactant concentration scales are the reference values $\sigma_0$ and $\Gamma_0$, respectively.
    
To explore the linear instability of the considered fluid flow model, the basic flow variables $(u^{(j)}, v^{(j)})$ $=$ $(U^{(j)}(y), 0)$ and $p^{(j)} = P^{(j)}(y)$ are substituted in the governing equations associated with the boundary conditions.
Assuming a unidirectional, locally parallel flow, known as base flow with fixed liquid layer thicknesses, the following analytical solutions \cite{samanta2014effect, bhat2020linear, sani2021effect} are identified in the nondimensional form:  
\begin{align}
&U^{(1)}(y)=\mathcal{K}\biggl[y-\frac{y^2}{2}+\frac{\delta(r\delta+2)}{2m}\biggr], \quad
U^{(2)}(y)=\frac{\mathcal{K}}{m}\left[y-\frac{ry^2}{2}+\frac{\delta(r\delta+2)}{2}\right],\label{ee13}\\
&P^{(1)}(y)=\frac{\mathcal{K}}{Re_1}\cot\theta~\big(1-y\big),~~~~\mbox{and}~~~~P^{(2)}(y)=\frac{\mathcal{K\cot\theta}}{rRe_1}~\big(1-ry\big).\label{ee14}
\end{align}
Notably, the inclusion of the elastic plate at the free surface and insoluble surfactant at the interface has no influence on the base profiles. This fact is similar to the case of single-layer fluid flow \cite{blyth2004effect, bhat2019linear, selvan2021hydroelastic}. To formulate the average velocity of the double-layered film flows over the impermeable bed, the dimensional base velocities are integrated over the film thicknesses. Next, the resulting equations are averaged with respect to the total film thickness, and the final equation is obtained as 
\begin{align}
    U_c=\frac{\rho^{(1)}g\sin\theta H^2_1}{\mathcal{K}\mu}, \quad\textrm{where}\quad \frac{1}{\mathcal{K}}=\frac{1}{\delta+1}\left[\frac{1}{3}+\frac{\delta^2}{2m}+\frac{\delta}{m}+\frac{\delta^3r}{3m}+\frac{\delta^2r}{2m}\right].
\end{align}

\section{\bf{Orr-Sommerfeld boundary value problem}}\label{BVP}

The OS-BVP for the double-layer flows over an impermeable bottom with the floating elastic plate on the top layer surface is obtained by assuming an infinitesimal perturbation to the base flow. Our primary goal is to interpret the linear stability analysis for infinitesimal perturbations of arbitrary wave numbers. A precise derivation of such an OS-BVP can be found in the works of  \citet{kao1965stability}, \citet{gao2007effect}, and \citet{samanta2014effect}. It can be obtained in the following manner for the current flow setup:
\allowdisplaybreaks
\begin{subequations}
\begin{align}	
&\bigg(\phi^{(j)}_{yyyy}-2k^2\phi^{(j)}_{yy}+k^4\phi^{(j)}\bigg)\,-\mathsf{i}kRe_j\bigg[(U^{(j)}-c)~(\phi^{(j)}_{yy}-k^2\phi^{(j)})-U^{(j)}_{yy}\phi^{(j)}\bigg]=0,\label{ee16a}\\
&\phi^{(j)}+(U^{(j)}-c)\eta^{(j)}=0~~~\mbox{at}~~y=1~~\mbox{for}~~j=1~~\mbox{and}~~y=0~~\mbox{for}~~j=2,\label{ee16b}\\
&\phi^{(2)}_{y}+U^{(1)}_{y}\eta^{(2)}+\bigg(U^{(2)}-c-\frac{\mathsf{i}k}{Pe}\bigg)\zeta=0~~~\mbox{at}~~~y=0,\label{ee16c}\\
&(\phi^{(1)}_{yy}+k^2\phi^{(1)})+U^{(1)}_{yy}\eta^{(1)}=0~~~~\mbox{at}~~~~y=1,\label{ee16d}\\
&\phi^{(1)}_{yyy}-3k^2\phi^{(1)}_{y}-\mathsf{i}kRe_1\bigg[(U^{(1)}-c)\phi^{(1)}_{y}+ck^2\gamma\,\phi^{(1)} \bigg]+\mathsf{i}k\bigg[-\mathcal{K}\cot\theta-k^4 Re_1\alpha \nonumber \\
&\hspace{2cm} - k^2 Re_1\beta-c k^2\,\gamma\, Re_1U^{(1)}\bigg]\eta^{(1)}=0 ~~~~\mbox{at}~~~~y=1,\label{ee16e}\\
&\phi^{(1)}_y-\phi^{(2)}_y+(m-1)U^{(2)}_{y}\eta^{(2)}=0~~~~\mbox{at}~~~~y=0,\label{ee16f}\\
&\phi^{(1)}-\phi^{(2)}=0~~~~\mbox{at}~~~~y=0,\label{ee16g}\\
&\big[mU^{(2)}_{yy}-U^{(1)}_{yy}\big]\eta^{(2)}+m(\phi^{(2)}_{yy}+k^2\phi^{(2)})-(\phi^{(1)}_{yy}+k^2\phi^{(1)})+\frac{\mathsf{i}kmMa}{Ca}\zeta=0~~~~\mbox{at}~~~~y=0,\label{ee16h}
\\
&\phi^{(1)}_{yyy}-3k^2\phi^{(1)}_y-\mathsf{i}kRe_1\bigg[(U^{(1)}-c)\phi^{(1)}_{y}-U^{(1)}_{y}\phi^{(1)}\bigg]-m\bigg[\phi^{(2)}_{yyy}-3k^2\phi^{(2)}_{y}\nonumber\\
&-\mathsf{i}kRe_2\bigg\{(U^{(2)}-c)\phi^{(2)}_{y}-U^{(2)}_{y}\phi^{(2)}\bigg\}\bigg]+\mathsf{i}k\bigg[\mathcal{K}(r-1)\cot\theta+\frac{mk^2}{Ca}\bigg]\eta^{(2)}=0~~~~\mbox{at}~~~~y=0,\label{ee16i}\\
&\phi^{(2)}_{y}=0\quad \mbox{and} \quad \phi^{(2)}=0~~~~\mbox{at}~~~~y=-\delta.\label{ee16j}
\end{align}
\end{subequations}
The stream function amplitudes, top surface/interface deformations, and interface surfactant concentration are defined by $\phi^{(j)}$, $\eta^{(1)}/\eta^{(2)}$, and $\zeta$, respectively. Here $k$ signs the wavenumber, and the complex wave speed $c=(c_r+ic_i)$ of an infinitesimal disturbance, the solution of which is supposed to be in the normal mode form $\biggl(\mathcal{A}(y)\exp{[\mathsf{i}k(x-ct)]}\biggr)$, where $\mathcal{A}(y)$ is the perturbation amplitude. 
 \begin{figure}[ht!]
\begin{center}
\subfigure[]{\includegraphics*[width=7.2cm]{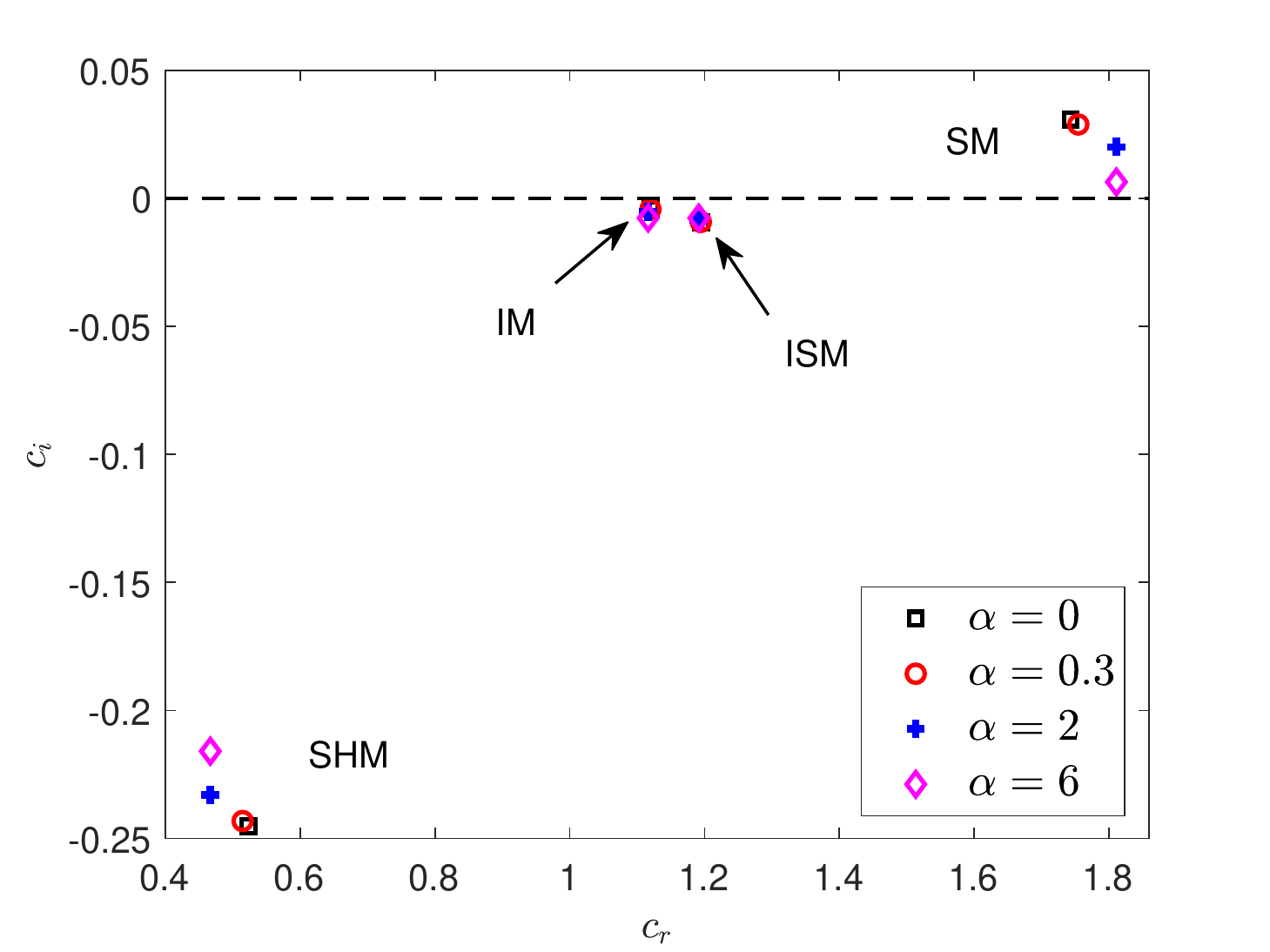}}
\subfigure[]{\includegraphics*[width=7.2cm]{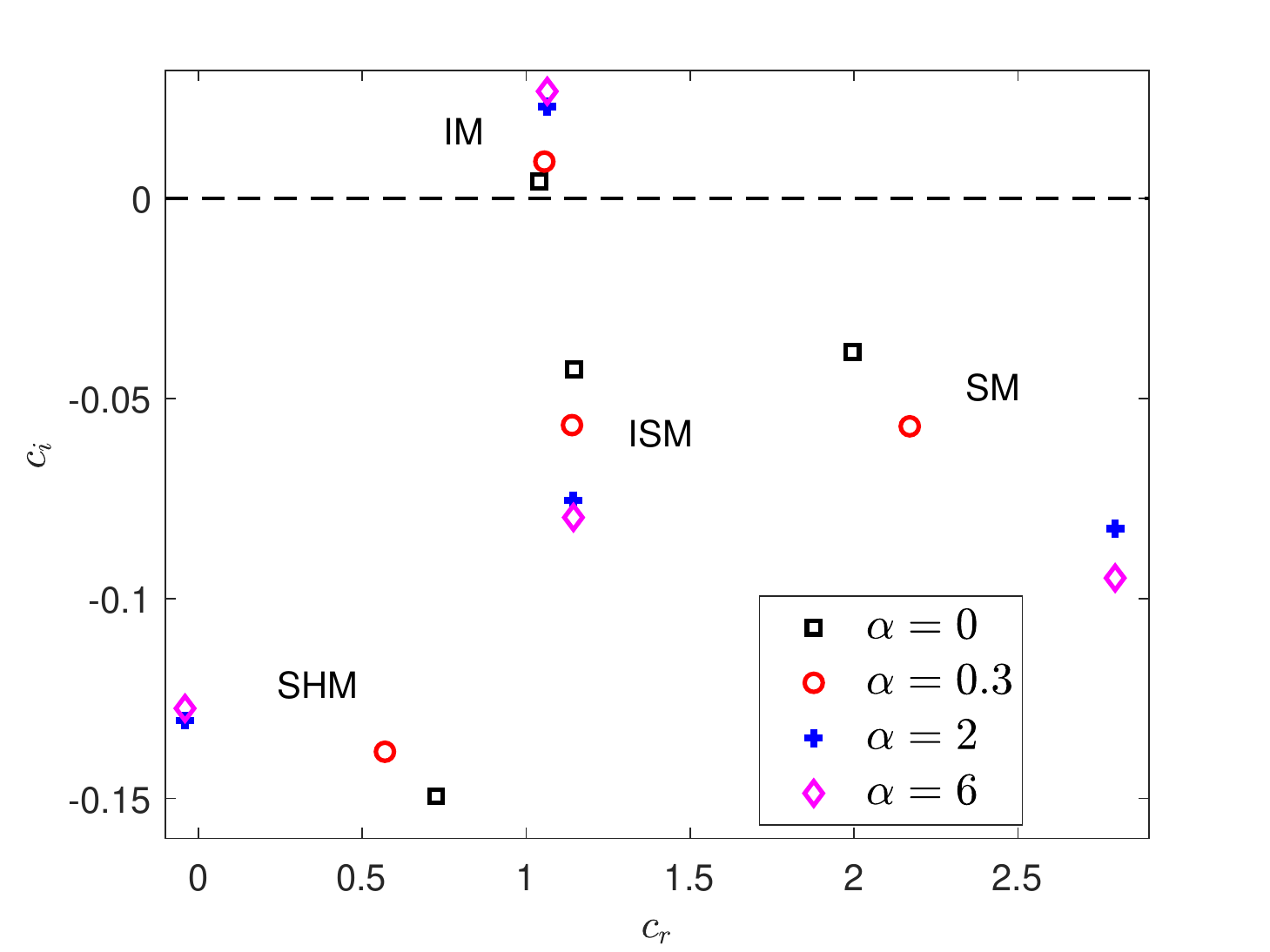}}
\subfigure[]{\includegraphics*[width=7.2cm]{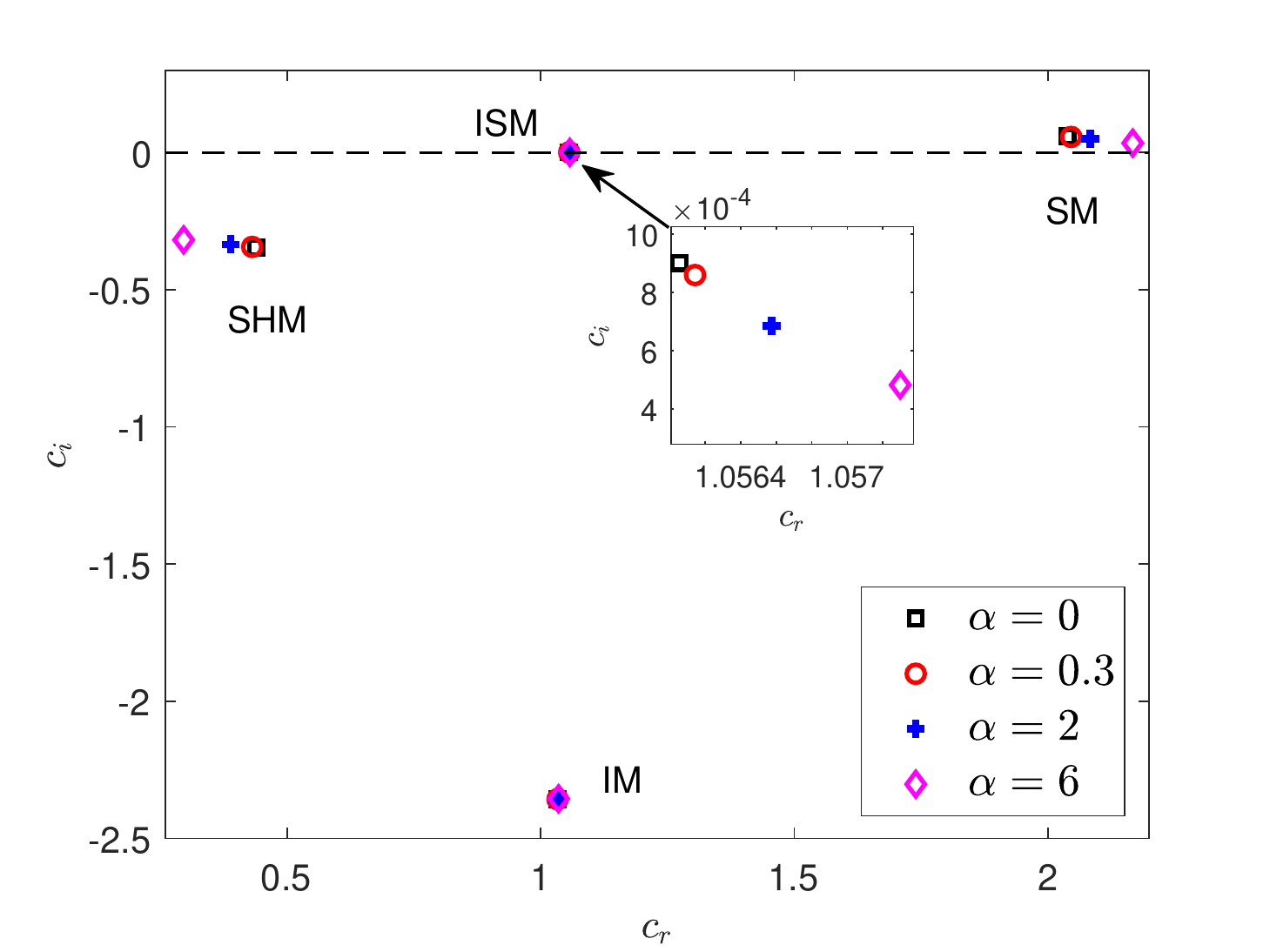}}
\subfigure[]{\includegraphics*[width=7.2cm]{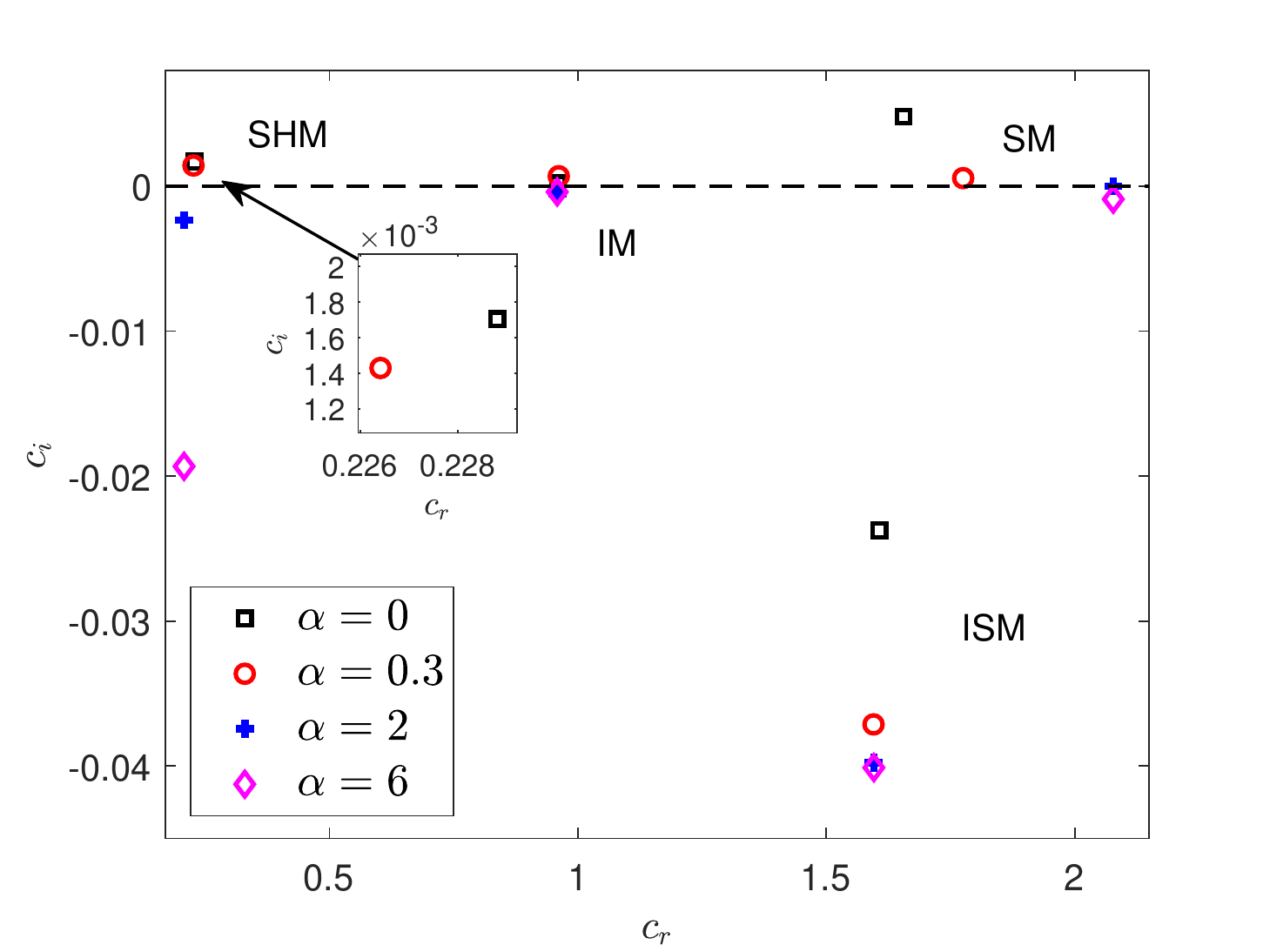}}
\end{center}
\caption{Distribution of eigenvalues showing the impact of structural rigidity ($\alpha$) on the most unstable (a) surface mode ($k=0.85$ and $m=0.6$), (b) interface mode ($k=0.95$ and $m=1.5$), (c) interface surfactant mode ($k=0.4$, $m=1.5$, $Ma=0.01$, and $Pe=800$), and (d) shear mode of the lower layer ($k=0.65$, $r=m=5$, $\theta=0.07~rad$, and $Re_1=8000$). The other constant flow parameters are $Re_1=20$, $\delta=1$, $r=1$, $Ca=1$, $Ma=0.1$, $\theta=0.2~ rad$, $\beta=\frac{1}{Re_1}$, $Pe=\infty$, and $\gamma=0.001$.}\label{f2}
\end{figure}
Here $\mathsf{i}\equiv \sqrt{-1}$ and the term $\omega=kc (=\omega_r+i\omega_i)$ defines the growth rate of the perturbation wave. It is important to emphasize that the fluid flow will be linearly stable if $\omega_i(=kc_i)$ is positive, unstable if negative, and neutrally stable if equal to zero.  

It is essential to highlight that when an insoluble surfactant is assumed in the place of a floating elastic plate on the top surface, and a slippery substrate is chosen as the bottom wall, the above system of Eqs.~\eqref{ee16a}-\eqref{ee16j} becomes identical with the OS-BVP of \citet{bhat2020linear} for the double-layer fluid flows over a slippery incline having insoluble surfactant at both the free surface and liquid-liquid interface. Further, the current OS-BVP also recovers the OS-BVP obtained by \citet{samanta2014effect} whenever the floating elastic plate is replaced by the insoluble surfactant at the top layer surface.

\section{\bf{Numerical results and discussion}}\label{NR}
 The numerical findings for a wide range of dimensionless flow parameters are explained in this section. The eigenmodes for different combinations of flow parameters are obtained by developing a Matlab 2022b subroutine that solves the aforementioned eigenvalue problem \eqref{ee16a}-\eqref{ee16j}. 
The current analysis identifies three different types of unstable eigenmodes for low to moderate Reynolds numbers, which are surface mode (SM), interface mode (IM), and interface surfactant mode (ISM). Apart from this, another unstable eigenmode called the shear mode (SHM), emerges when the inertia force is strong enough with a low inclination angle. The different types of modes are selected according to the phase speed $c_r$ using the criteria $c_r|_{SM}>c_r|_{ISM}>c_r|_{IM}>c_r|_{SHM}$.
The current work intends to investigate the characteristics of all such unstable modes in the presence of a floating elastic plate at the top surface and to explore the control mechanism of the plate on the unstable modes.

Fig.~\ref{f2} shows the effect of uniform rigidity parameter $\alpha$ of the elastic plate on the existing unstable SM, IM, ISM, and SHM. The momentum conservation inside the disturbed two-layer flow is mainly responsible for the existence of different kinds of unstable modes.
In the case of SM, the higher rigidity parameter $\alpha$ reduces the temporal growth rate as a result of the attenuation of the net force acting normally on the surface (see, Fig.~\ref{f2}(a)).
This results in the reduction of the amplitude of perturbed surface waves below the plate.
\begin{figure}[ht!]
	\begin{center}
\subfigure[]{\includegraphics[width=12cm,height=8cm]{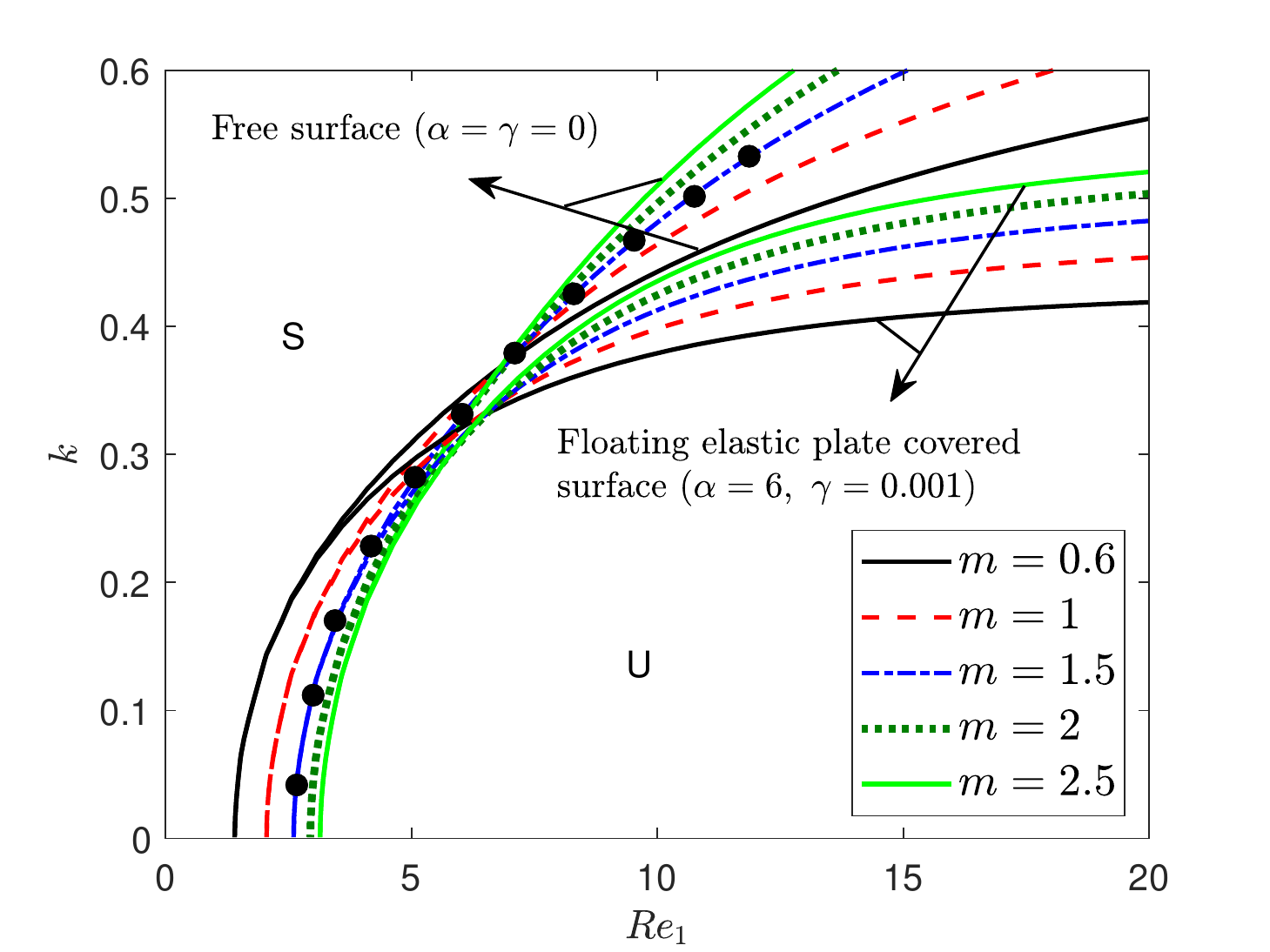}}
\end{center}
\caption{The marginal stability curves of surface mode in ($Re_1-k$) plane when the other fixed parameters are  $r=1$, $\delta=1$, $Ca=1$, $Ma=0$, $\theta=0.2~rad$, $\beta=\frac{1}{Re_1}$, and $Pe=\infty$. The solid circular points are the result of \citet{samanta2014effect} (see, Fig.~7(a) of their paper).   
	 }\label{f3}
\end{figure}
Further, it is displayed in Fig.~\ref{f2}(b) that the unstable interface mode exists when the viscosity of the bottom layer is greater than that of the top layer ($m=1.5$).
Here, the rigidity parameter $\alpha$ enhances the interfacial instability by raising the temporal growth rate of the most unstable IM. Next, Fig.~\ref{f2}(c) depicts the stabilizing nature of the most unstable ISM owing to a higher $\alpha$ value.
Finally, Fig.~\ref{f2}(d) confirms the occurrence of another unstable eigenmode, known as the shear mode (SHM), which emerges in the considered fluid model for a strong inertia force and low inclination, provided the viscosity and density of the lower layer are much greater than those of the upper layer.
The change in the most unstable SHM of the lower layer with $\alpha$ value is displayed. The floating plate reduces the lower layer shear mode instability by lowering the growth rate as well as phase speed. 
Fig.~\ref{f2} confirms that the floating plate characteristic parameter $\alpha$ (structural rigidity) has an influence on all the unstable modes emerging from the perturbed flow system. In the subsequent sections, we have elaborately discussed how the flexible plate properties trigger a significant change in the behaviour of all co-existing unstable modes.   

\subsection{\bf{Effect of the floating flexible plate on the surface mode}}
 In this section, the characteristic of the dominant surface mode is discussed briefly in various flow parameter regimes.
 The surface mode, known as Yih mode \cite{yih1963stability}, is generally responsible for the evolution or deformation of the free surface. Before going through the discussion of the SM characteristic behaviour, we have validated the surface mode result (see, Fig.~\ref{f3}) with the numerical result obtained by \citet{samanta2014effect}. The stability/instability boundary lines in $Re_1-k$ plane for different viscosity ratio $m$ are plotted for the cases of free surface and floating plate-covered surface when $r=1$, $\delta=1$, $Ca=1$, $Ma=0$, $\gamma=0$, $\beta=\frac{1}{Re_1}$, and $Pe=\infty$. The symbols $U$ and $S$ are basically used to define the unstable and stable zones, respectively. The result for $m=1.5$ matches the \citet{samanta2014effect} result very well in the limit $Ma=0$.
 \begin{figure}[ht!]
	\begin{center}
	 \subfigure[]{\includegraphics[width=7.2cm]{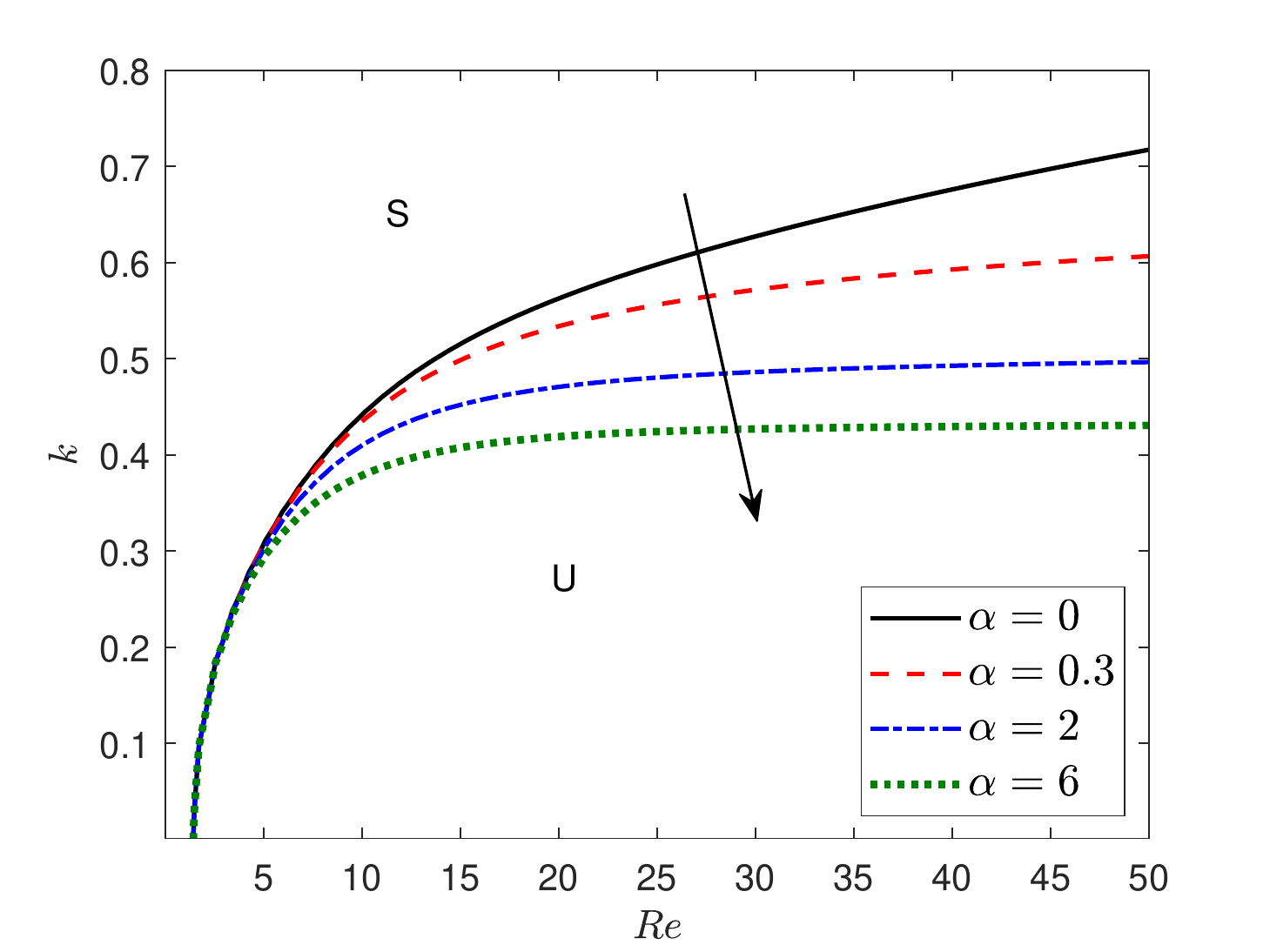}}
	 \subfigure[]{\includegraphics[width=7.2cm]{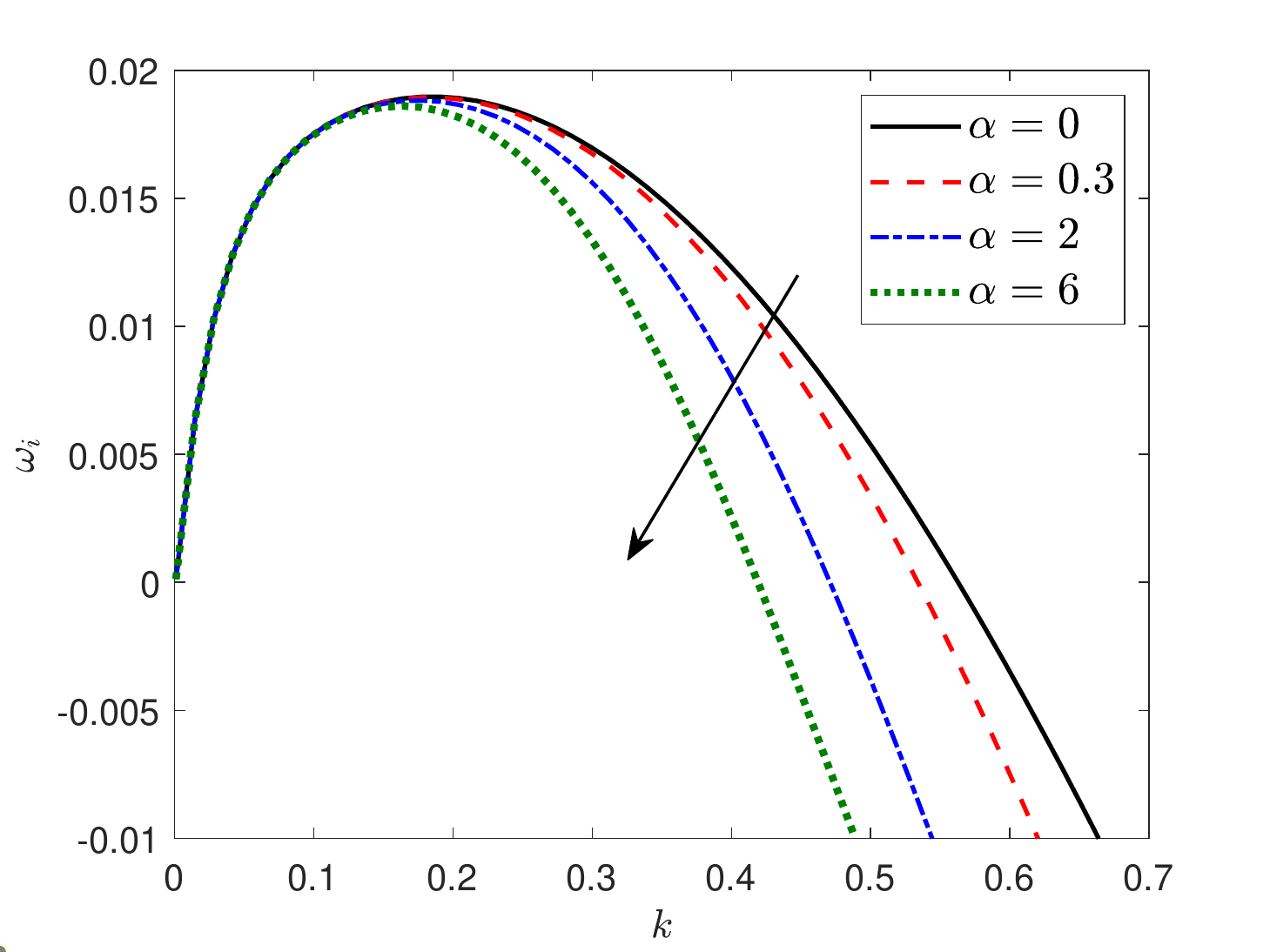}}
	  \subfigure[]{\includegraphics[width=7.2cm]{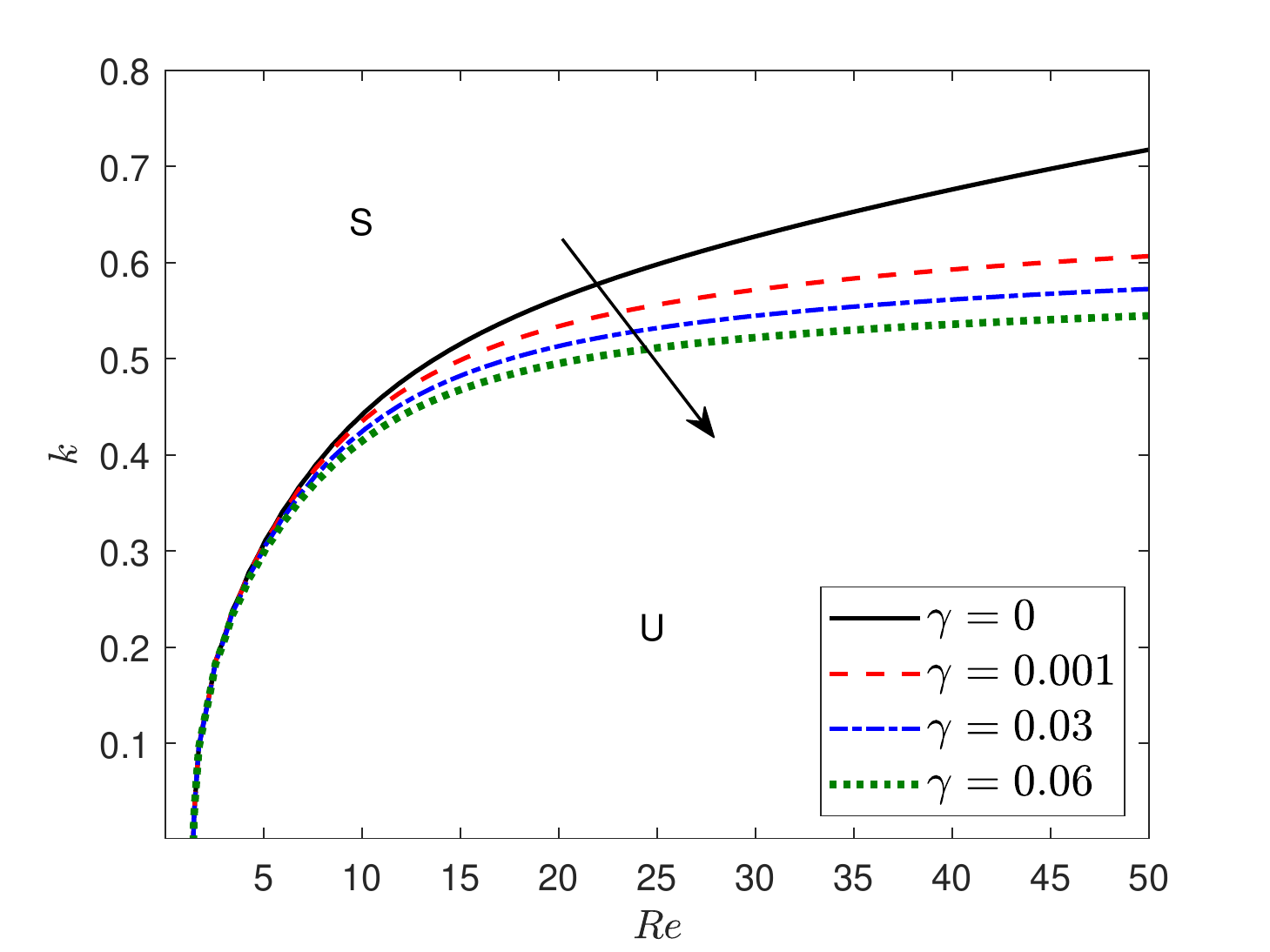}}
	 \subfigure[]{\includegraphics[width=7.2cm]{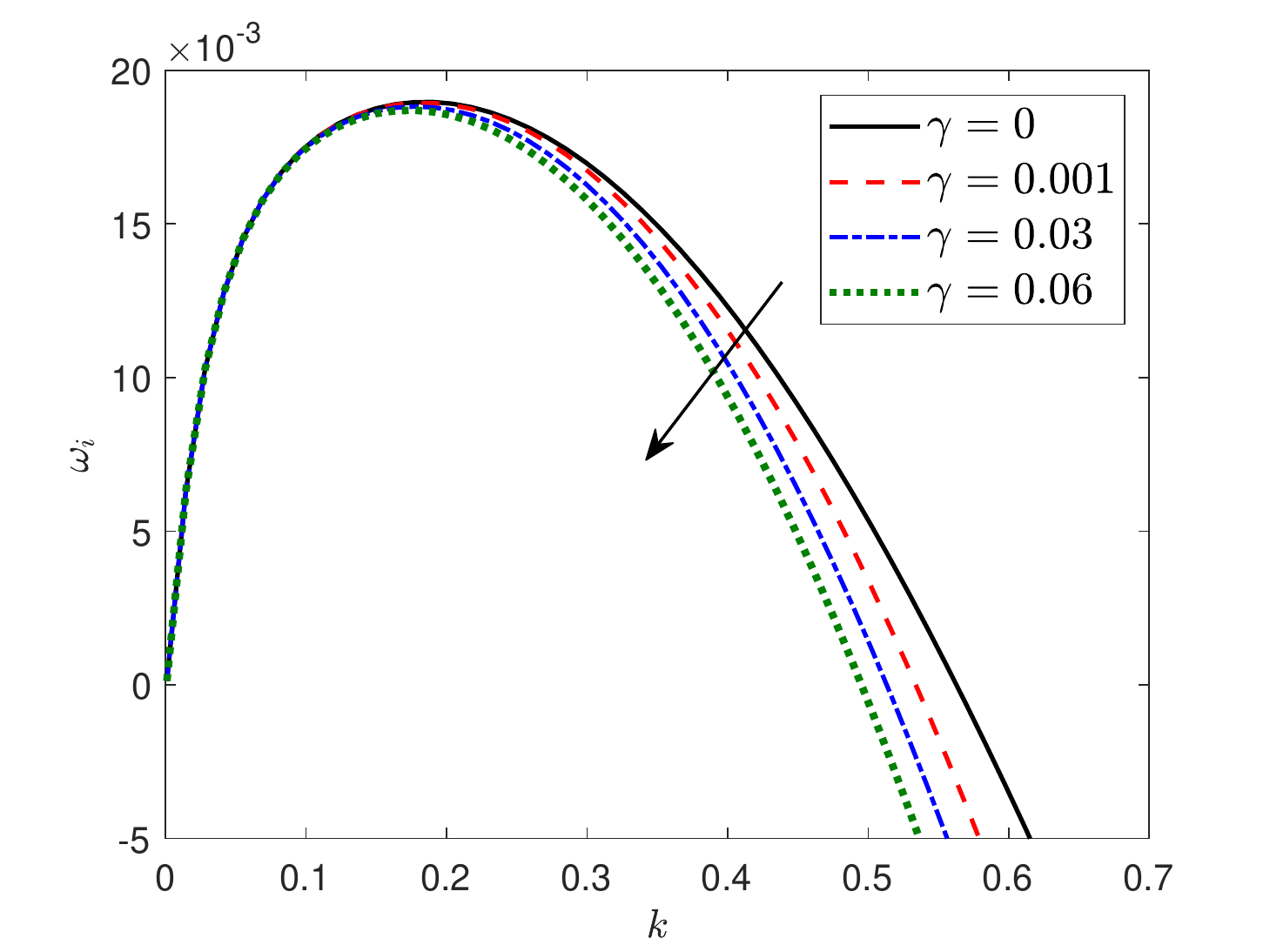}}
 \end{center}\vspace{-0.5cm}
	\caption{ ((a) and (c)) The variation of marginal stability curves ($\omega_i=0$) of the surface mode in ($Re_1-k$) plane for various values of structural rigidity  $\alpha$ (in (a)) and uniform thickness $\gamma$ (in (c)). The arrow is pointing out the direction of decreasing manner of an unstable regime with increasing $\alpha$ and $\gamma$. ((b) and (d)) The  corresponding temporal growth rate curves ($\omega_i$) as a function of wavenumber ($k$) of the surface mode  for various values of structural rigidity  $\alpha$ (in (b)) and uniform thickness $\gamma$ (in (d)) with $Re_1=20$. The arrow is pointing out the direction of decreasing manner of growth rate with increasing $\alpha$ and $\gamma$. The other fixed parameters are $r=1.1$, $m=0.6$, $\delta=1$, $Ca=1$, $Ma=0.1$, $\theta=0.2~rad$, $\beta=\frac{1}{Re_1}$, and $Pe=\infty$. The solid line marks the free surface flow (i.e., $\alpha=0$ and $\gamma=0$). The value of $\gamma=0.001$ when $\alpha\neq0$ and  $\alpha=0.3$ when $\gamma\neq0$. 
	}\label{f4}
\end{figure}

Indeed, the dual impact of $m$ on the surface mode instability is found. The higher values of $m$ (i.e., the lower layer has higher viscosity than the upper layer) supply the stabilizing effect near the criticality by reducing the unstable region but provide a destabilizing influence far away from the criticality by enhancing the range of unstable wavenumbers. As long as $m$ increases, the upper layer viscosity stress enhances to balance the lower layer viscous stress at the interface. Consequently, the upper layer flow-rate becomes slower than the lower layer and yields stabilizing behaviour of the surface mode.
 An important finding from Fig.~\ref{f3} is that the floating elastic plate significantly shrinks the unstable zone in the finite wavenumber region, whereas it has negligible influence in the longwave zone. 

\begin{figure}[ht!]
	\begin{center}
	  \subfigure[]{\includegraphics[width=7.2cm]{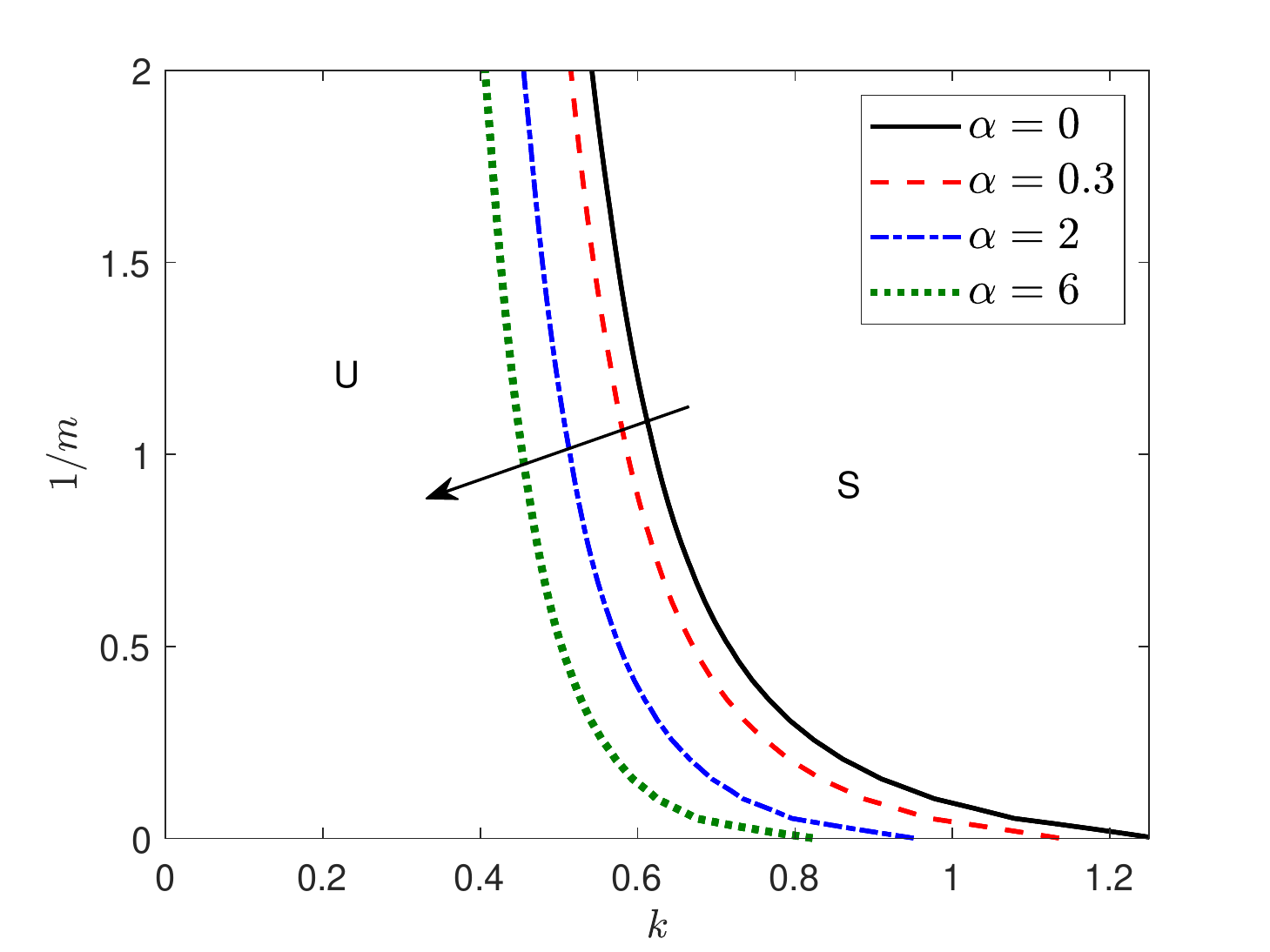}}
	   \subfigure[]{\includegraphics[width=7.2cm]{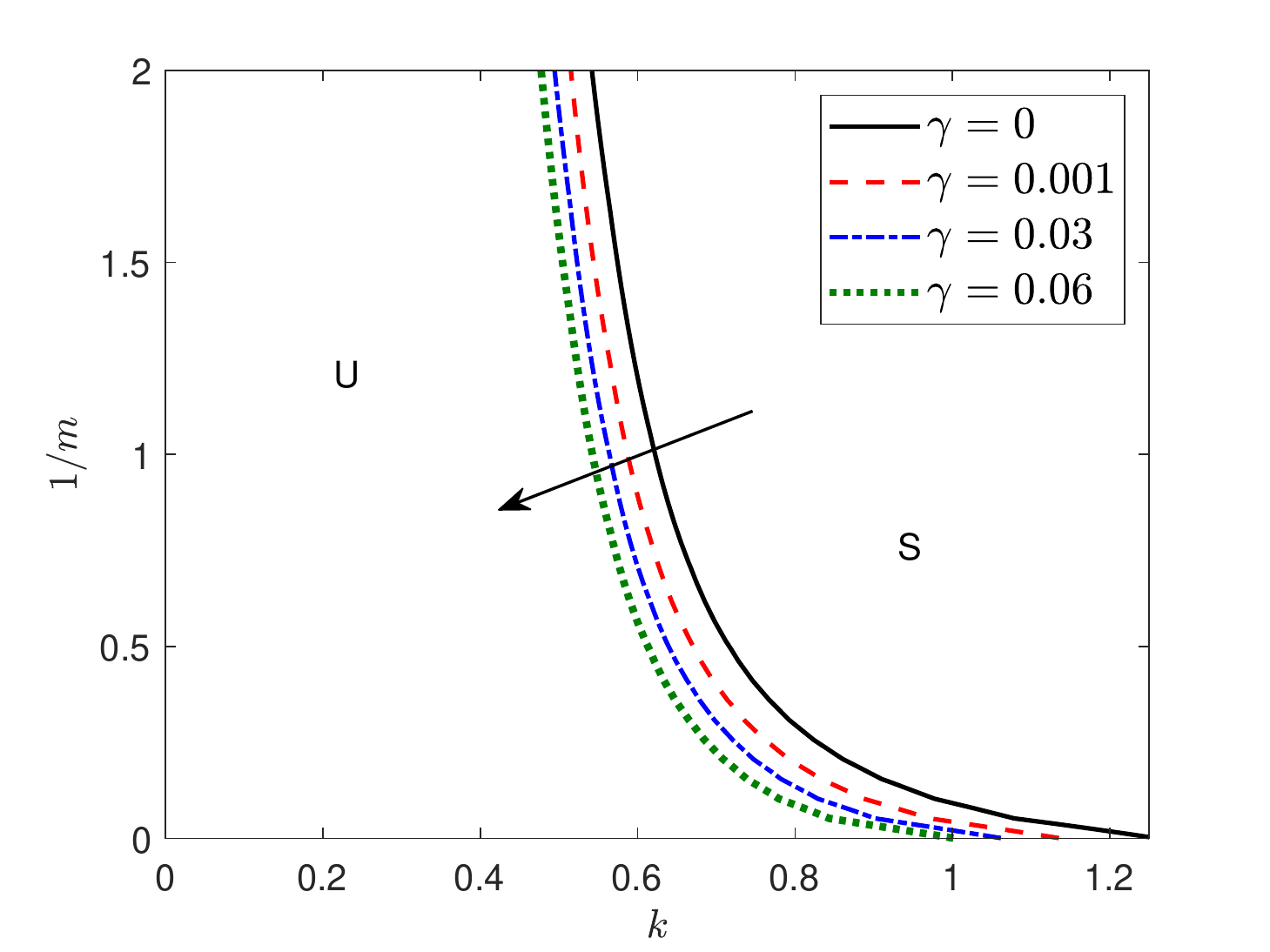}}
	  \subfigure[]{\includegraphics[width=7.2cm]{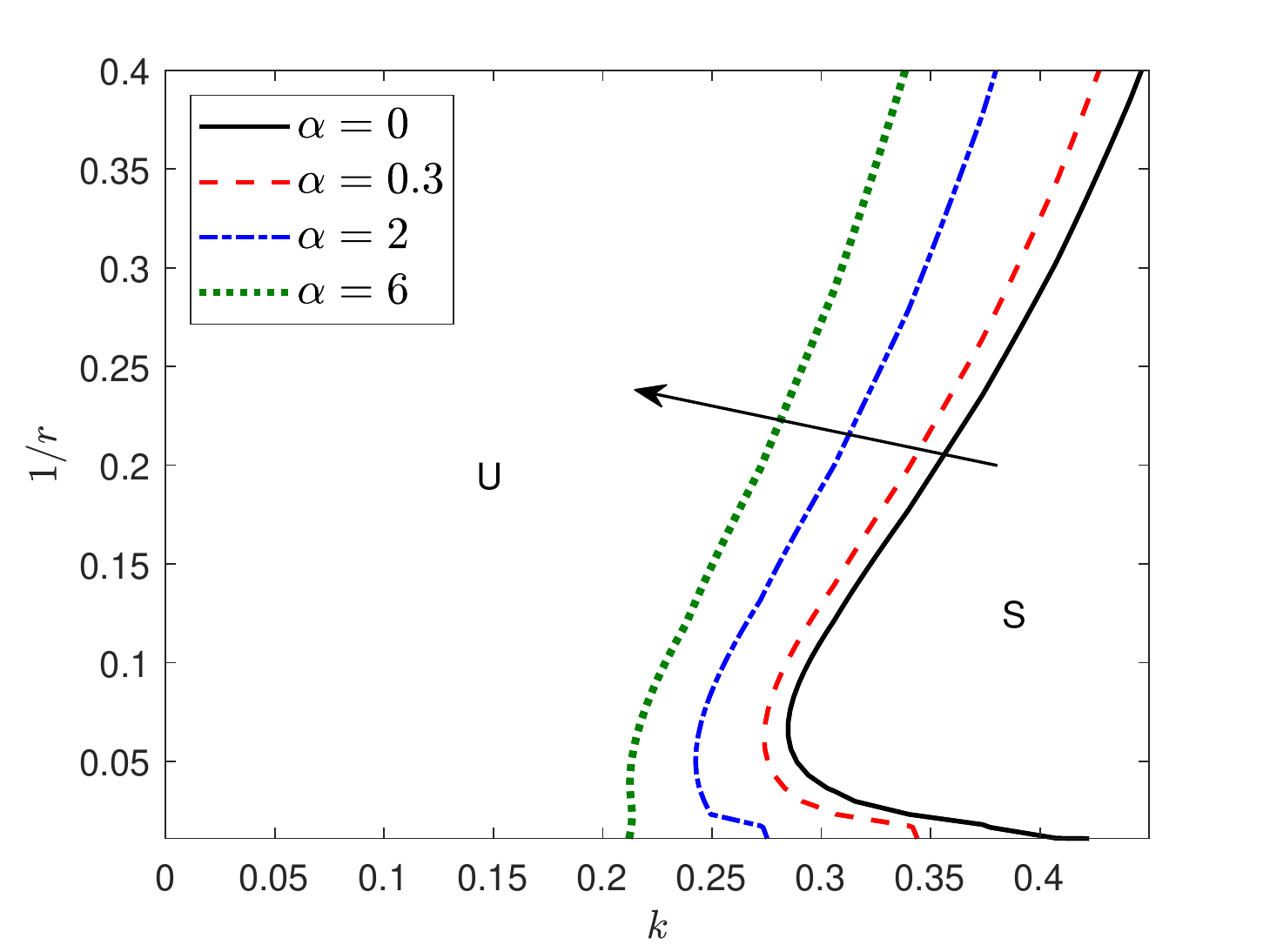}}
	 \subfigure[]{\includegraphics[width=7.2cm]{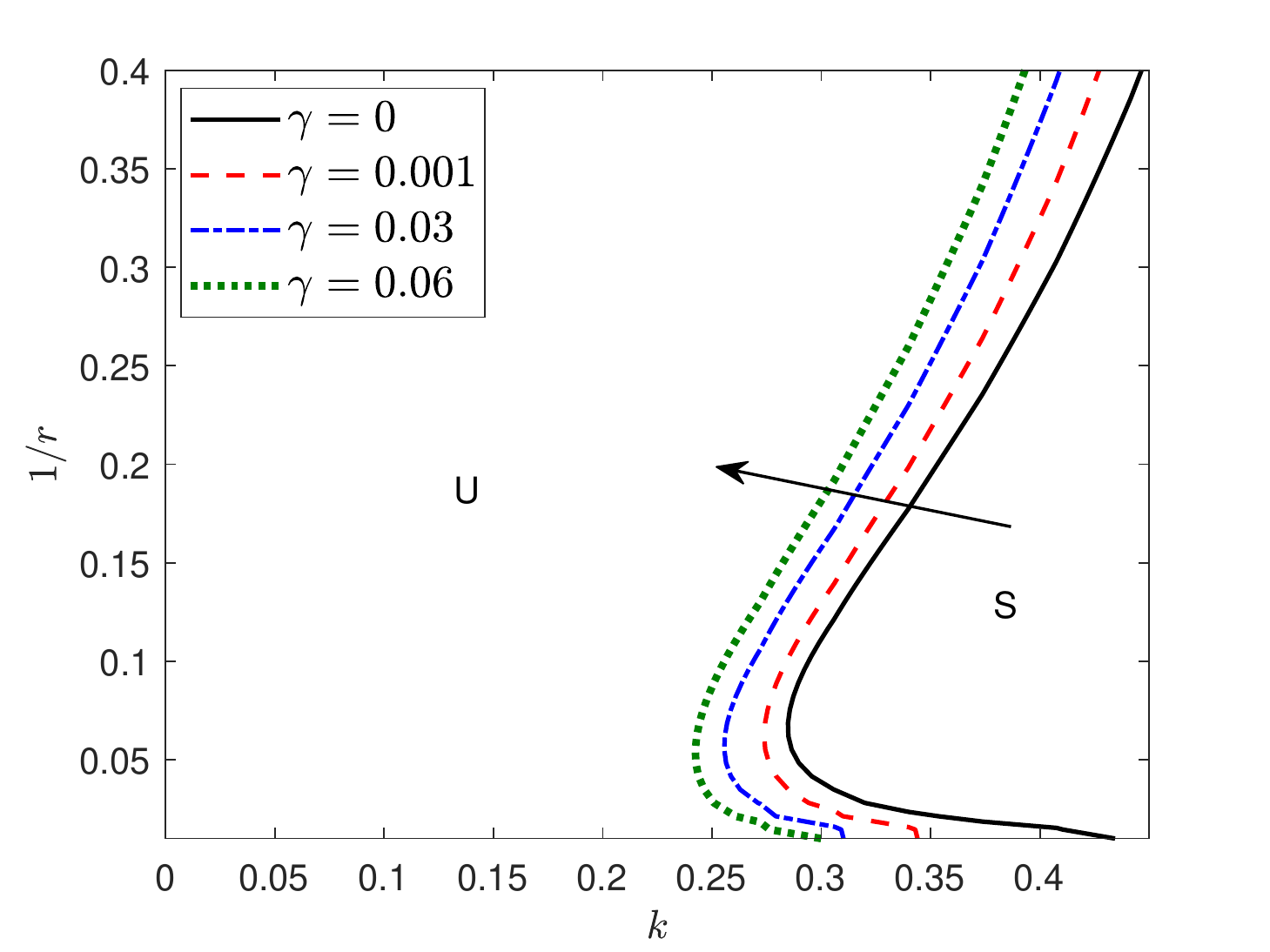}}
   \end{center}\vspace{-0.5cm}
	\caption{ ((a) and (b)) The variation of marginal stability curves of the surface mode in ($k-1/m$) plane for different values of structural rigidity  $\alpha$ (in (a)) and uniform thickness $\gamma$ (in (b)) with $r=1.1$. ((c) and (d)) The variation of marginal stability curves of the surface mode in ($k-1/r$) plane for various values of structural rigidity  $\alpha$ (in (c)) and uniform thickness $\gamma$ (in (d)) with $m=0.6$. The other fixed parameters are $Re_1=20$, $\delta=1$, $Ca=1$, $Ma=0.1$, $\theta=0.2~rad$, $\beta=\frac{1}{Re_1}$, and $Pe=\infty$. The arrow is pointing out the direction of decreasing unstable regime with increasing $\alpha$ and $\gamma$. The solid line marks the free surface flow (i.e., $\alpha=0$ and $\gamma=0$). The value of $\gamma=0.001$ when $\alpha\neq0$ and  $\alpha=0.3$ when $\gamma\neq0$.  
	 }\label{f5}
\end{figure}

The numerical test is conducted to examine the plate-covered surface wave dynamics in Fig.~\ref{f4}. Figs.~\ref{f4}(a) and (c) exhibit the change in the marginal line of the most unstable SM in the $Re_1-k$ domain for different structural rigidity parameters $\alpha$ and uniform thickness $\gamma$, respectively. Note that, in the longwave range ($k\rightarrow0$) there is no effect of floating plate characteristic parameters $\alpha$ and $\gamma$. The normal stress balance Eq.~\eqref{ee16e} at the surface contains the terms $k^5Re_1\alpha$ and $Re_1k^3\gamma(1+U^{(1)})$, which suggest a negligible effect of $\alpha$ and $\gamma$ in the longwave zone ($k\rightarrow0$). Although, at the finite wavenumber range, the impact of plate characteristic parameters is significant.  For finite wavenumber, the higher values of both parameters $\alpha$ and $\gamma$ reduce the surface wave instability by shrinking the unstable zone. To strengthen the above statement, the corresponding growth rate $\omega_i$ ($=\,kc_i$) results are illustrated in Figs.~\ref{f4}(b) and (d) by choosing $Re_1=20$.   
The stabilizing effect of $\alpha$ in the finite wavenumber range is more clear in the corresponding temporal growth rate results, as depicted in Fig.~\ref{f4}(b). The structural rigidity $\alpha$ reduces the maximum growth rate of surface mode. At stronger structural rigidity, the normal force acting downwards on the top surface dissipates the perturbation wave energy obtained from the liquid surface and makes the SM more stable.       
On the other hand, the higher plate thickness ($\gamma$) also attenuates the finite wavenumber unstable zone of surface mode (see, Fig.~\ref{f4}(c)) and yields a stabilizing effect on the surface waves. The corresponding growth rate results in Fig.~\ref{f4}(d) are fully consistent with Fig.~\ref{f4}(c). Here, the plate thickness parameter $\gamma$ reduces the maximum growth rate.    

The variation of neutral curves (see, Figs.~\ref{f5}(a) and (b)) of the SM in ($k-1/m$) plane is plotted  to decipher the impact of both rigidity $\alpha$ and plate thickness $\gamma$, respectively. 
It is observed that the unstable zone bandwidth gradually shrinks as the viscosity of the upper layer increases (i.e., increasing the value of $1/m$). 
Both the plate characteristic parameters $\alpha$ (see, Fig.~\ref{f5}(a)) and $\gamma$ (see, Fig.~\ref{f5}(b)) significantly attenuate the unstable zone triggered by the SM. This happens due to the balance of normal force and the viscous force in the vicinity of the top surface of the flow system. 
The neutral stability curves are displayed in ($k-1/r$) plane for various plate characteristic parameters $\alpha$, shown in Fig.~\ref{f5}(c) and $\gamma$, as in Fig.~\ref{f5}(d). It is found that the unstable bandwidth continuously amplifies with the increase of upper layer density (i.e., increasing value of $1/r$).
The comparatively lesser density of the lower layer strengthens the net driving force of the top layer liquid, which boosts surface wave instability. Further, the surface mode instability reduces as the plate rigidity parameter $\alpha$ increases, which is ensured by the  continuous depletion of the unstable region. Similar stabilizing nature of the SM is observed when uniform plate thickness parameter $\gamma$ increases.

Thus, three important scenarios can be concluded: (i) a floating elastic plate has the ability to stabilize the surface mode instability by reducing the perturbed surface wave energy, (ii) a potent destabilization of unstable surface mode is possible if compared to the top layer, bottom layer liquid is of greater viscosity, and (iii) higher dense upper layer fluid than the lower layer enhances the surface wave energy, which makes the surface mode more unstable.
\subsection{\bf{Effect of the floating flexible plate on the interface mode}}
In this subsection, the numerical investigation is repeated to explore the detailed interface wave dynamics of the elastic plate-contaminated fluid flow system over the rigid inclined plane. 
\begin{figure}[ht!]
	\begin{center}
	 \subfigure[$Ma=0$]{\includegraphics[width=5.4cm]{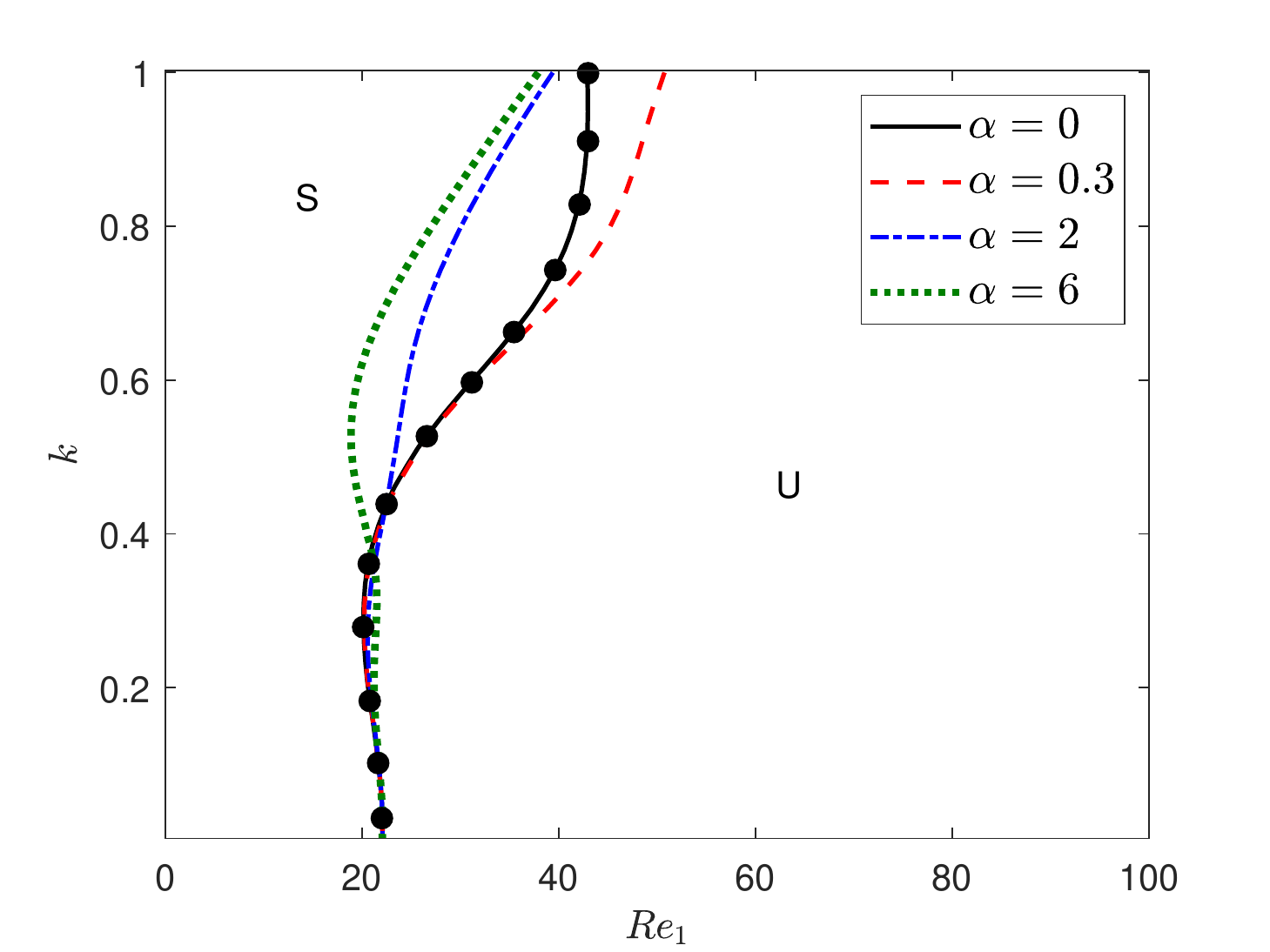}}
 	 \subfigure[$Ma=0.03$]{\includegraphics[width=5.4cm]{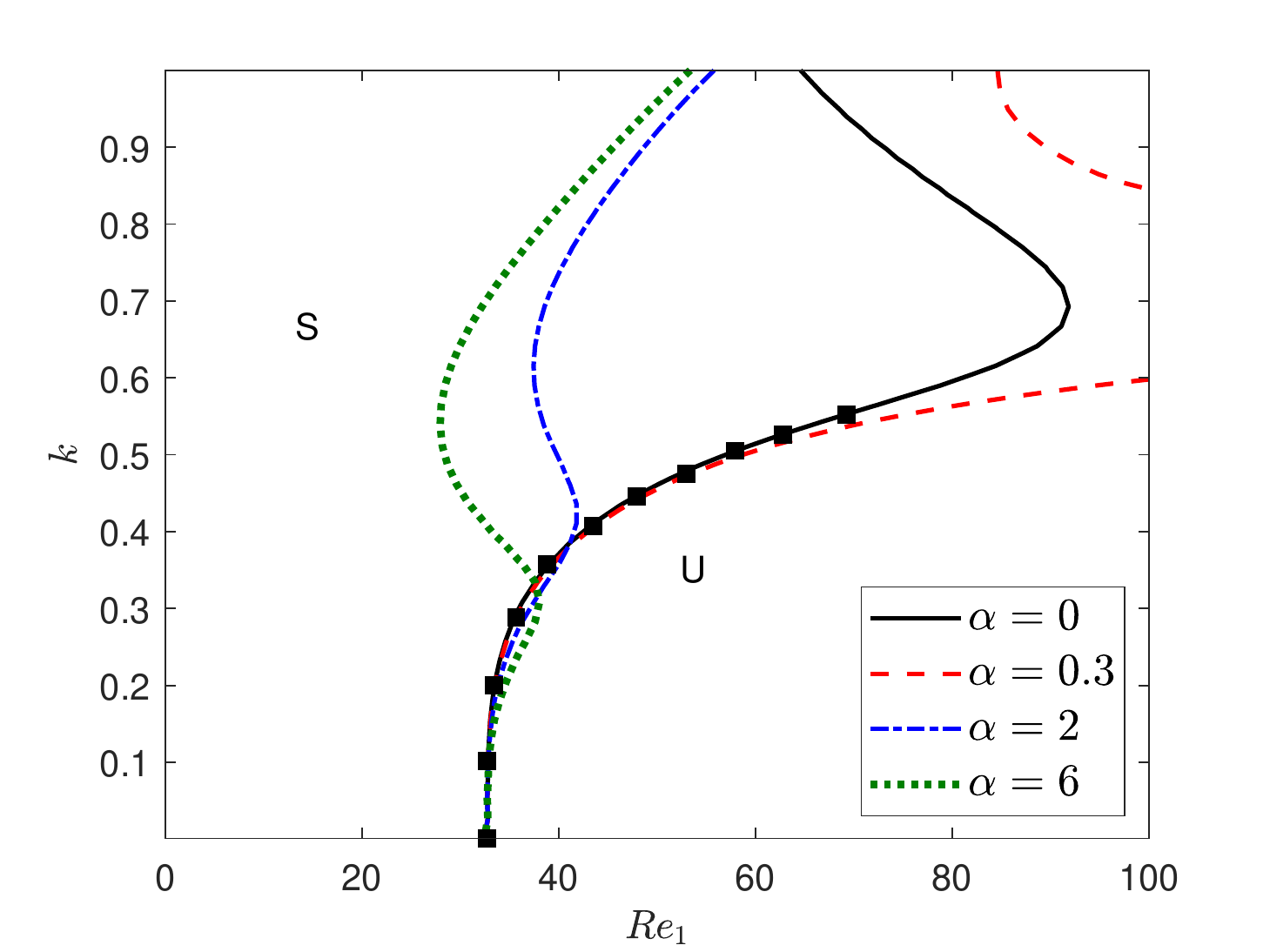}}
 	 \subfigure[$Ma=0.05$]{\includegraphics[width=5.4cm]{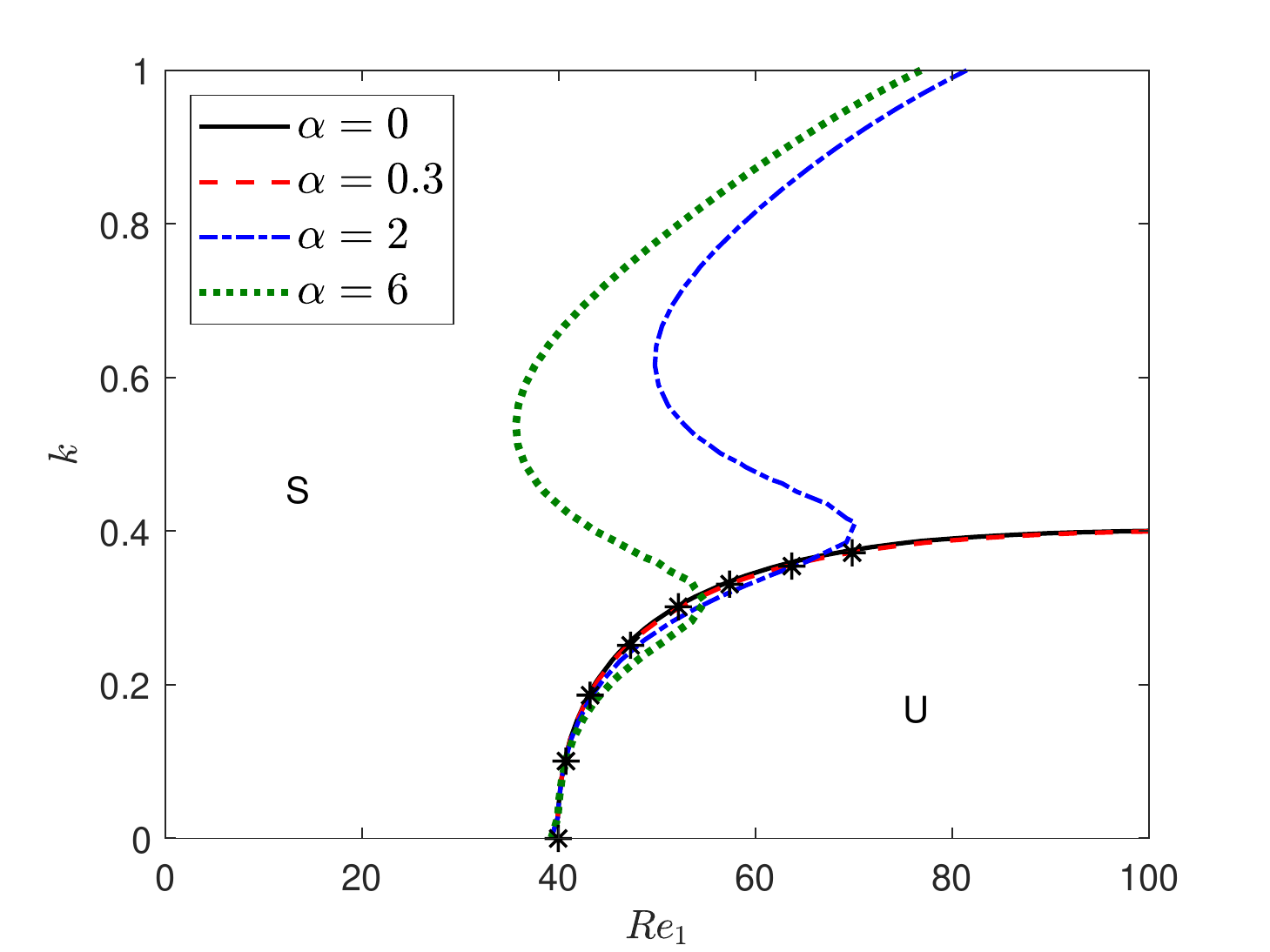}}
 	  \subfigure[$Ma=0$]{\includegraphics[width=5.4cm]{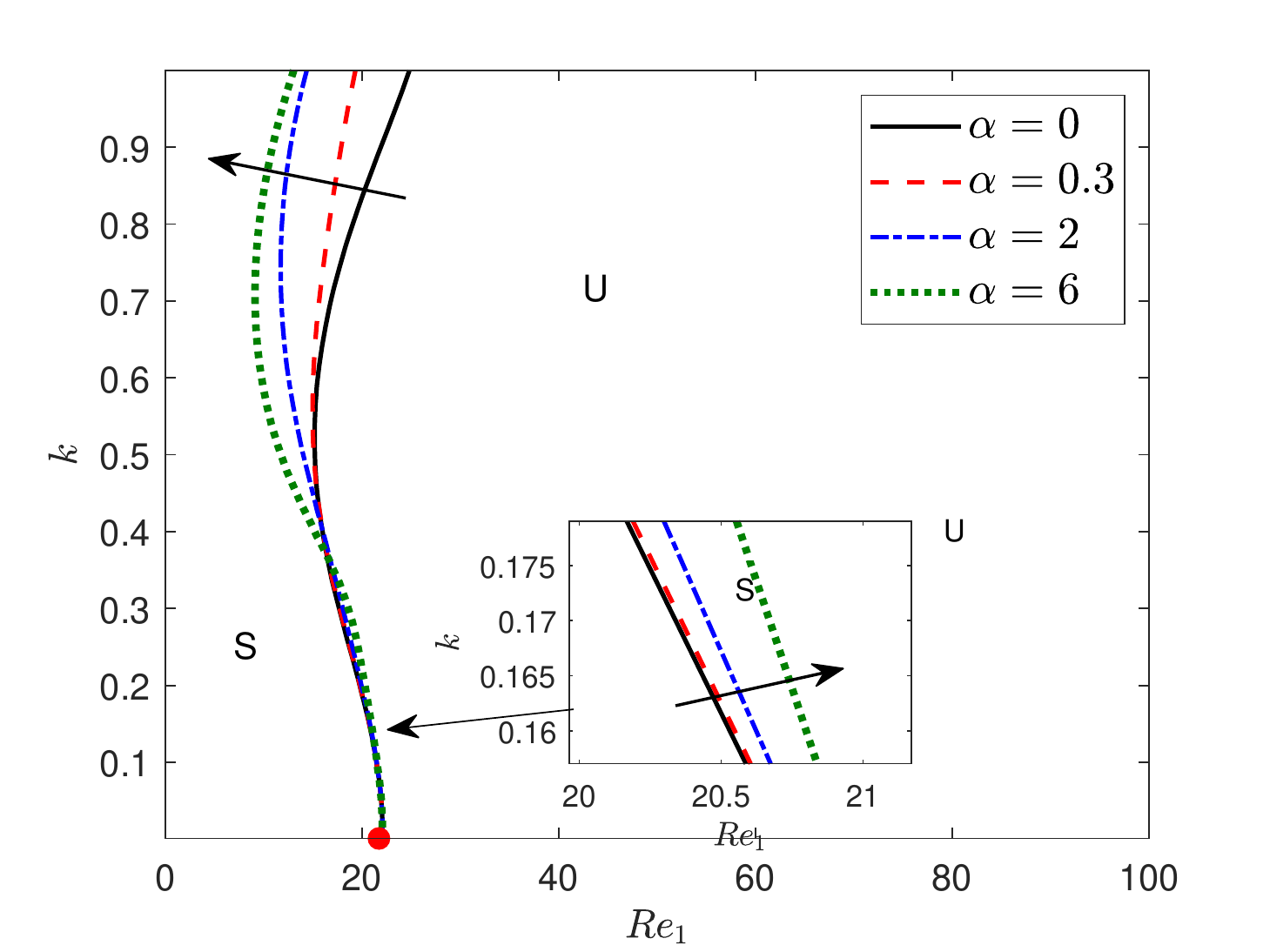}}
 	 \subfigure[$Ma=0.03$]{\includegraphics[width=5.4cm]{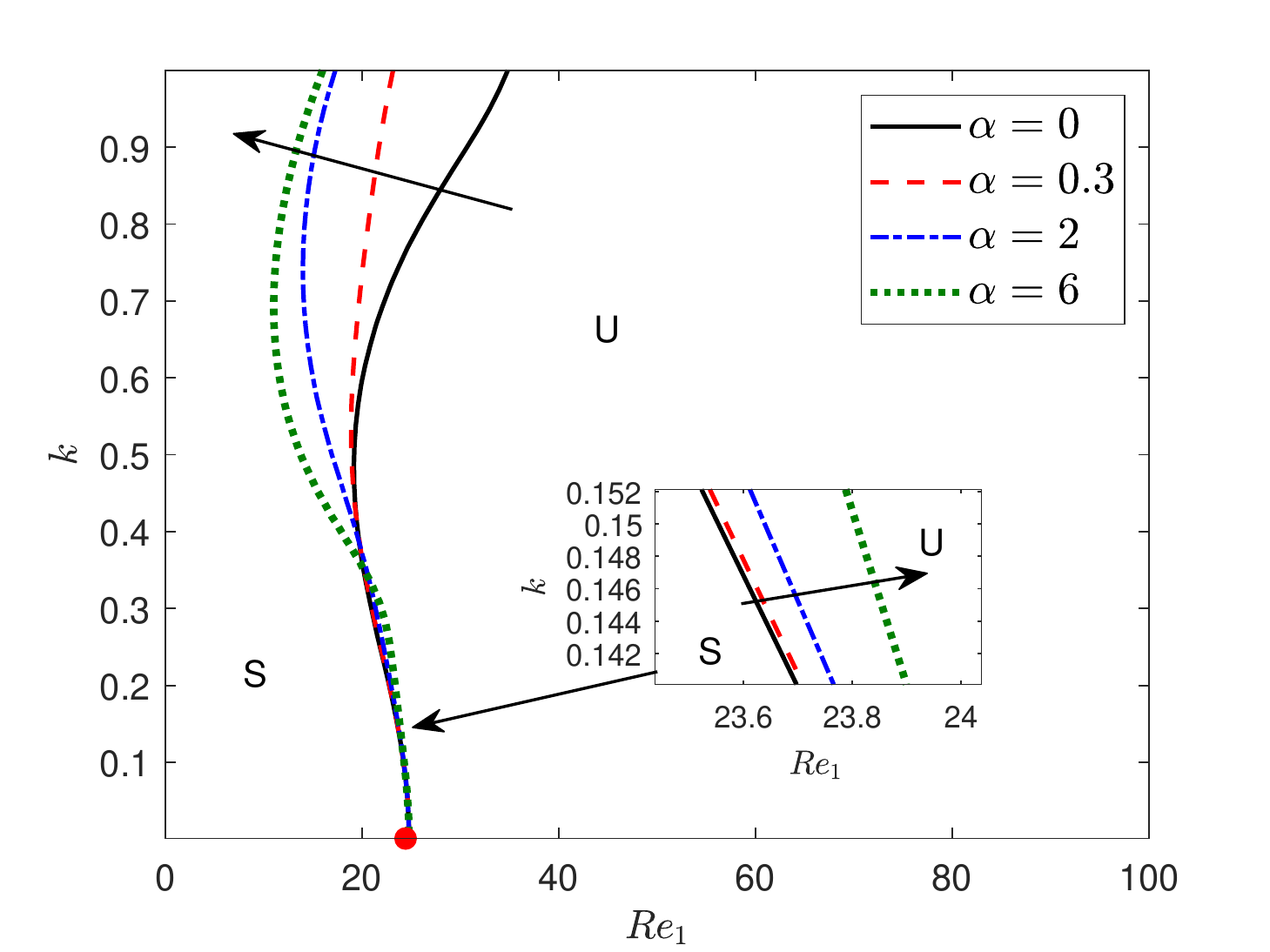}}
 	 \subfigure[$Ma=0.05$]{\includegraphics[width=5.4cm]{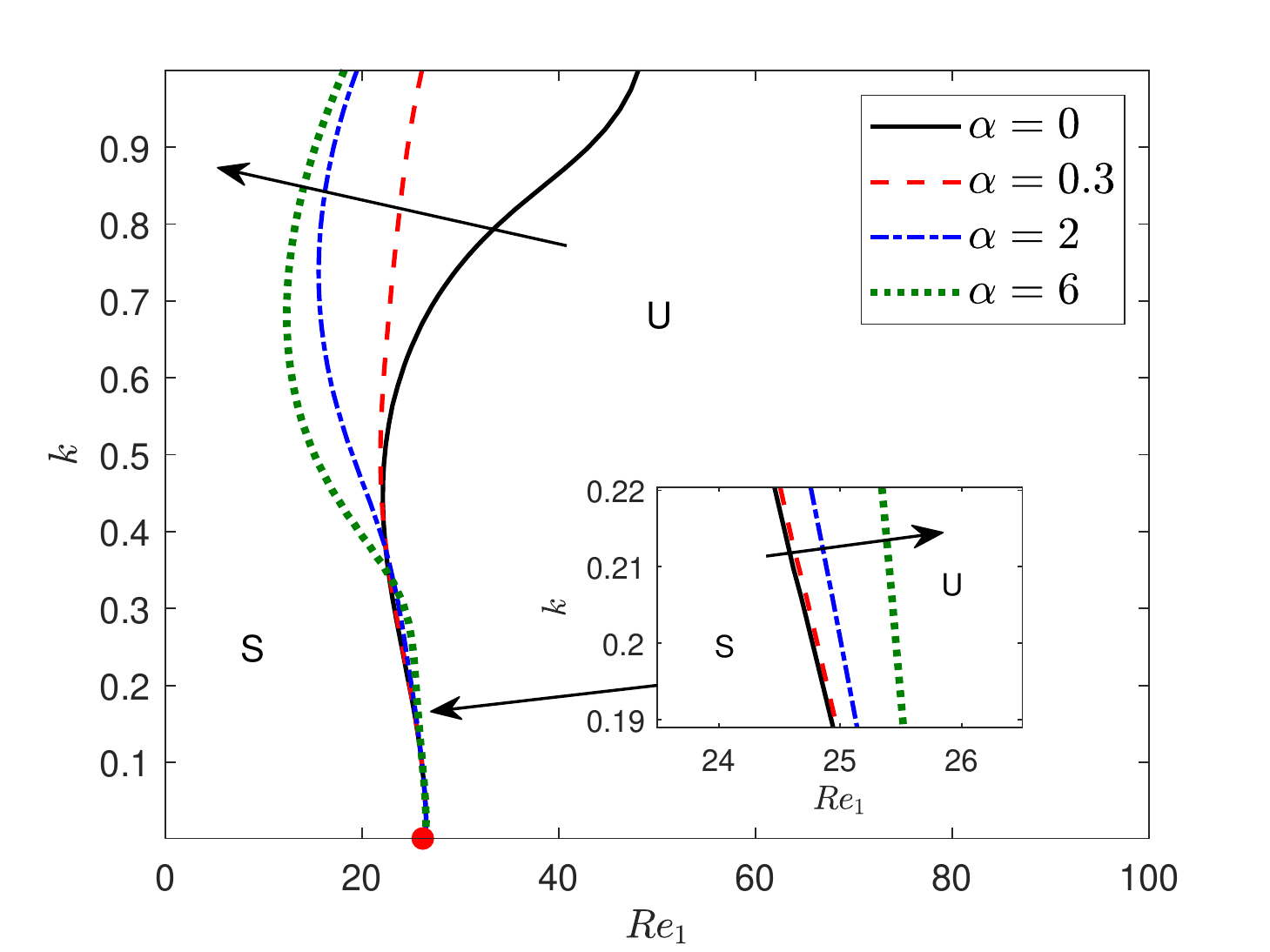}}
	\end{center}\vspace{-0.5cm}
	\caption{Effects of structural rigidity $\alpha$ on the neutral stability curves corresponding to the interface mode in ($Re_1-k$) plane ($mr>1$) for different Marangoni force ($Ma$) with different values of Capillary number ($Ca$). The Capillary numbers are $Ca=1$ and $Ca=4$ for the figures of top ((a)-(c)) and bottom ((d)-(f)), respectively. The other constant parameters are $\delta=1$, $r=1$, $m=1.5$, $Ca=1$, $\theta=0.2~rad$, $\beta=\frac{1}{Re_1}$, and $Pe=\infty$. The solid circular, rectangle, and star are the results of \citet{samanta2014effect} (Fig.6(a) of their paper). The right arrow in the inset plot and the left arrow represent the decreasing and increasing manner of the unstable zone. The solid lines are the result of a clean surface (i.e., $\alpha=0$ and $\gamma=0$) and the fixed value $\gamma=0.001$ when $\alpha\neq0$. The red circular points correspond to the critical Reynolds number.  
	 }\label{f6}
\end{figure}
To check the validity and accuracy of the co-existence IM result, the numerical observation is conducted in Fig.~\ref{f6} by fixing the values $\delta=1$, $r=1$, $m=1.5$, $Ca=1$, $\theta=0.2~rad$, $\beta=\frac{1}{Re_1}$, and $Pe=\infty$.
Here, the stability/instability boundary lines in the ($Re_1-k$) plane are illustrated for various rigidity parameters $\alpha$ as the $Ma$ value changes.

For $Ma=0$ (see, Fig.~\ref{f6}(a)), $Ma=0.05$ (see, Fig.~\ref{f6}(b)), and $Ma=0.1$ (see, Fig.~\ref{f6}(c)), the marginal curves totally recover the result obtained by \citet{samanta2014effect} in the case of free surface flow ($\alpha=0$ and $\gamma=0$). For this considered flow characteristic parameters, the uniform rigidity $\alpha$ of the flexible plate has a negligible influence on the IM in the longwave zone but has a significant non-monotonic impact in the smaller wavenumber range. Although, the results with a higher Capillary number ($Ca=4$) for the same set of other parameters show stabilization in the small wavenumber region for higher rigidity $\alpha$. Hence, Fig.~\ref{f6} displays the dual impact of $\alpha$ in the small wavenumber zone depending on the other flow characteristic parameters.
Furthermore, the Marangoni force stabilizes the IM instability by raising the critical Reynolds number.
Thus, for the double-layer flow down a rigid substrate, the Marangoni effect on the IM is almost similar to the surfactant contaminated single-layer fluid flow over a rigid substrate \cite{bhat2019linear}, where the Marangoni force also has stabilizing impact on the unstable SM. 
\begin{figure}[ht!]
	\begin{center}
\subfigure[$Ma=0$]{\includegraphics[width=5.4cm]{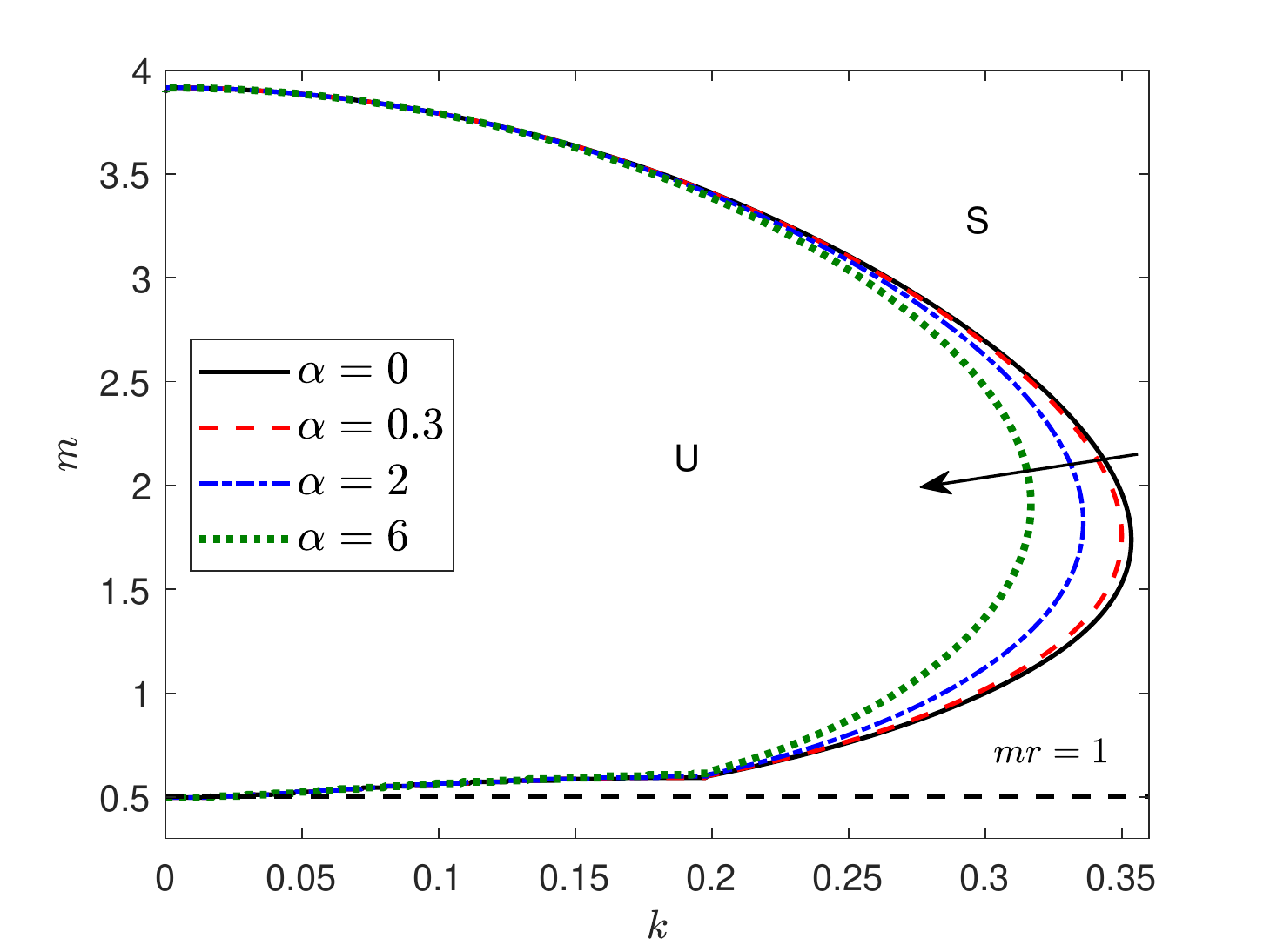}}
\subfigure[$Ma=0.05$]{\includegraphics[width=5.4cm]{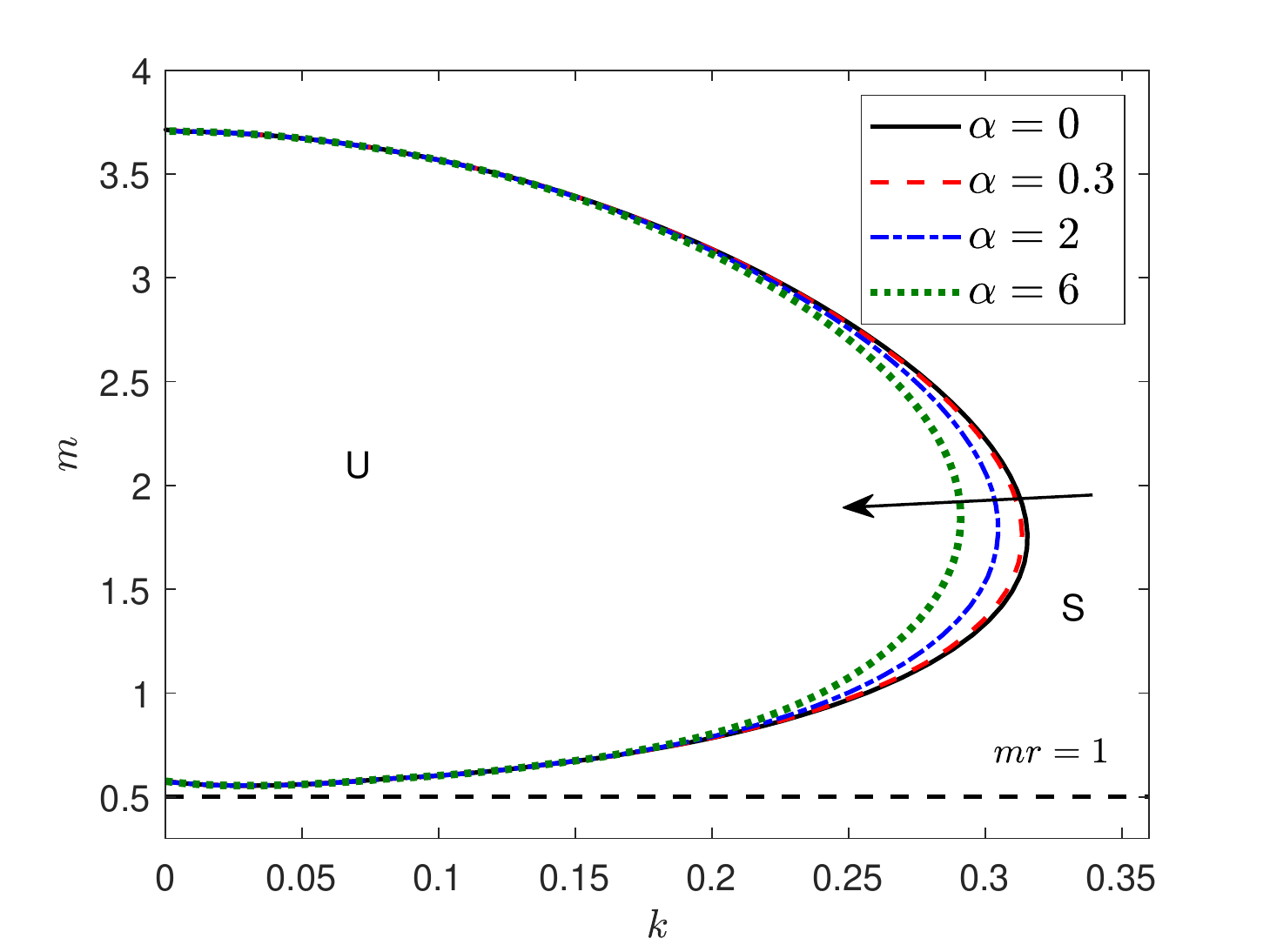}}
\subfigure[$Ma=0.1$]{\includegraphics[width=5.4cm]{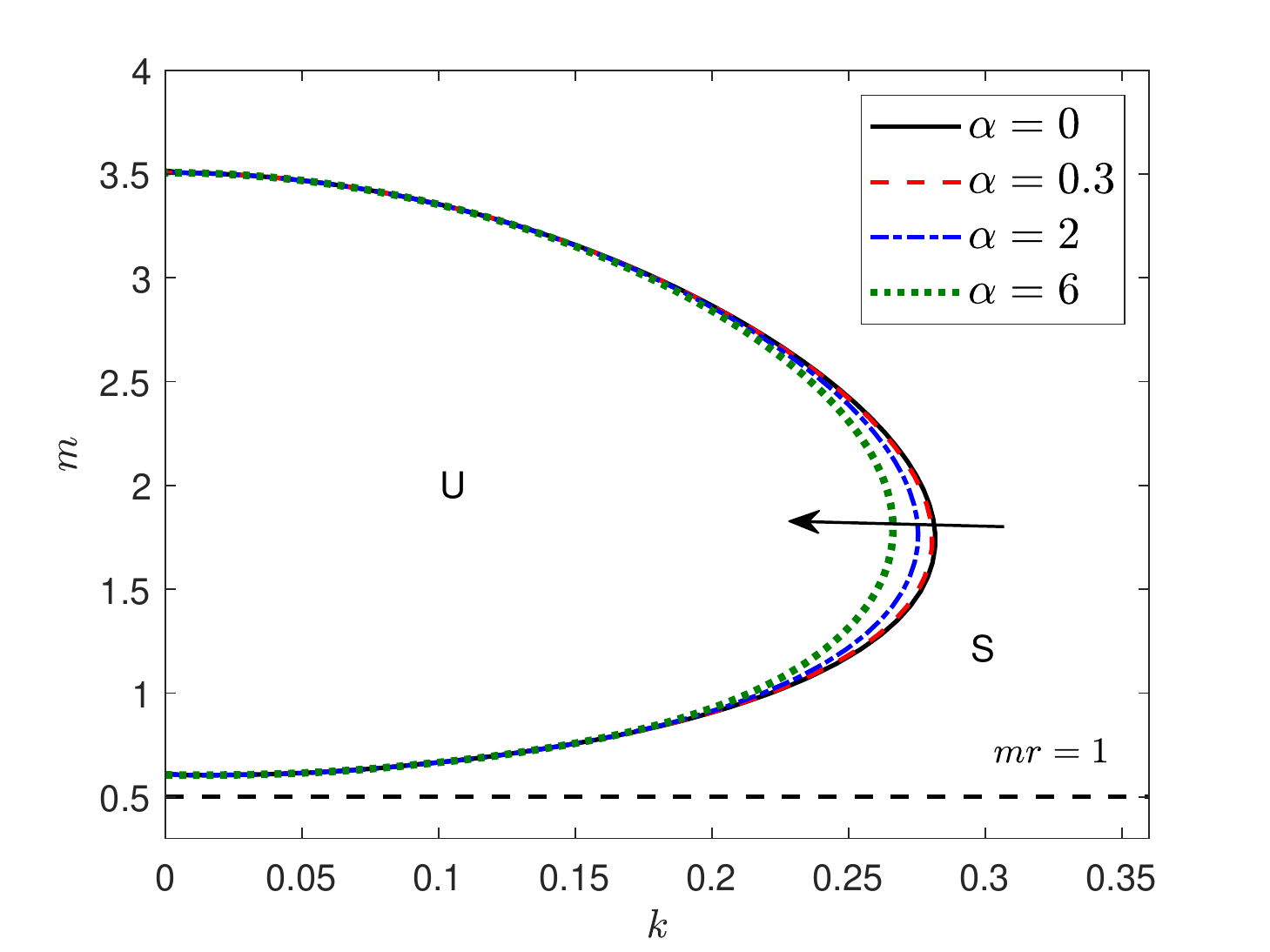}}
\subfigure[$Ma=0$]{\includegraphics[width=5.4cm]{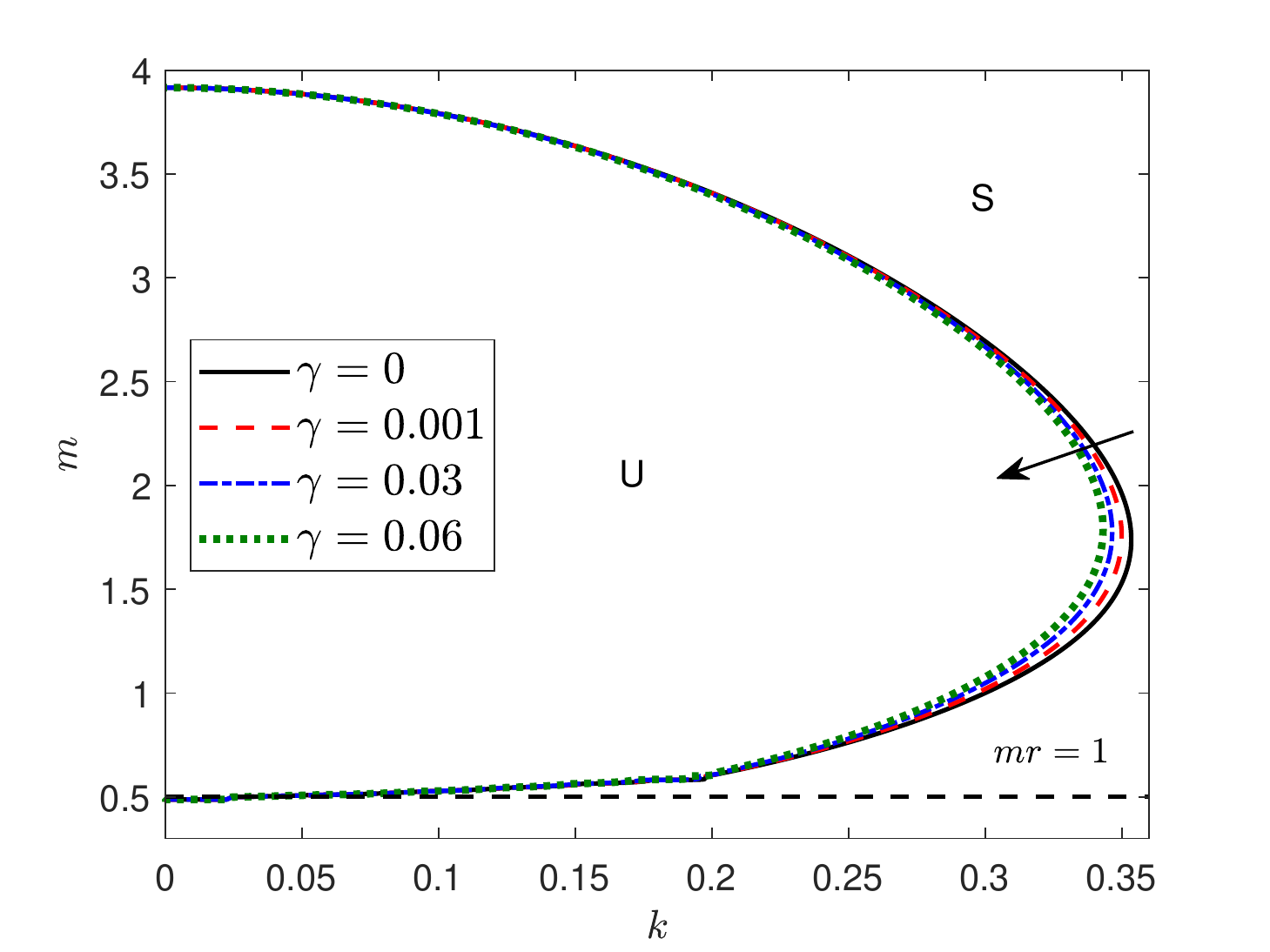}}
\subfigure[$Ma=0.05$]{\includegraphics[width=5.4cm]{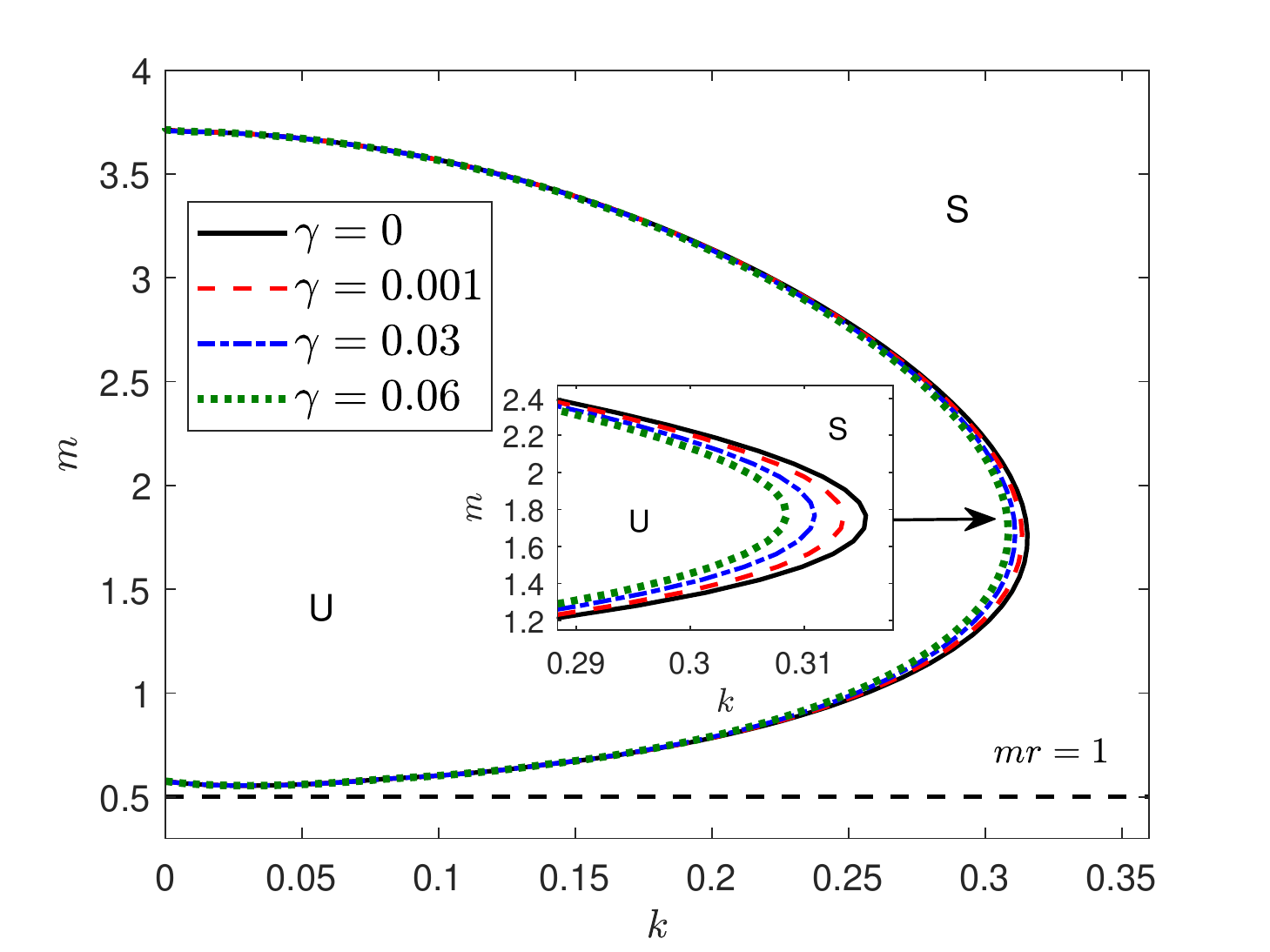}}
\subfigure[$Ma=0.1$]{\includegraphics[width=5.4cm]{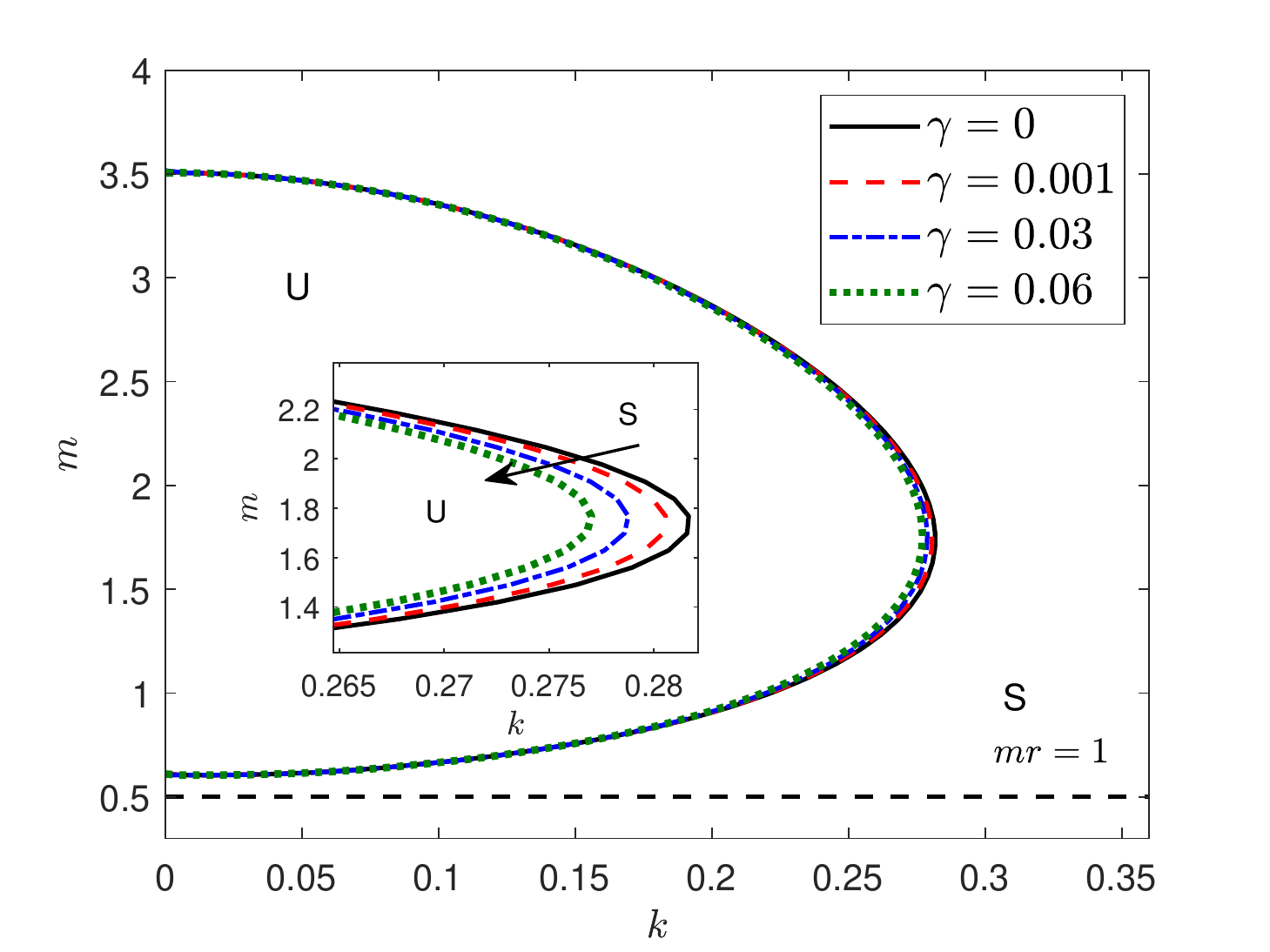}}
	\end{center}\vspace{-0.5cm}
	\caption{The neutral stability curves in ($m-k$) plane (for $mr>1$) corresponding to the interface mode for various values of ((a)-(c)) structural rigidity $\alpha$ and ((e)-(f)) uniform thickness $\gamma$. The other constant parameters are $Re_1=20$, $\delta=1$, $r=2$, $Ca=1$, $\theta=0.2~rad$, $\beta=\frac{1}{Re_1}$, and $Pe=\infty$. The arrow points out the decreasing manner of the unstable zone with respect to increasing $\alpha$ and $\gamma$. The solid line represents free surface flow with $\alpha=0$ and $\gamma=0$. The value of $\gamma=0.001$ when $\alpha\neq0$ and  $\alpha=0.3$ when $\gamma\neq0$.  }\label{f7}
\end{figure}
  
Fig.~\ref{f7} exhibits the influence of plate characteristic parameters $\alpha$ (Figs.~\ref{f7}(a)-(c)) and $\gamma$ (Figs.~\ref{f7}(d)-(f)) on the bandwidth of marginal stability curve in ($k-m$) plane corresponds to the unstable IM for various Marangoni numbers ($Ma$) with $mr>1$.
Notably, for each structural rigidity $\alpha$ (Figs.~\ref{f7}(a)-(c)) and thickness parameter $\gamma$ (Figs.~\ref{f7}(d)-(f)),  the neutral curves exhibit the existence of two critical points in the longwave region. This fact confirms the presence of sub-critical instability in the longwave region when the lower layer viscosity is chosen comparatively higher (i.e., $m>1$).
Also, the higher $\alpha$ value reduces the interface mode instability by decreasing the unstable zone, demonstrated in Figs.~\ref{f7}(a)-(c). A similar stabilizing effect of interface mode is possible by implementing the higher uniform plate thickness $\gamma$ (see, Figs.~\ref{f7}(d)-(f)). 

 Moreover, for both the Capillary values ($Ca=1$ and $Ca=4$), the unstable zone of the interface mode continuously decreases (see, Figs.~\ref{f7}(a) and (d) for clean interface ($Ma=0$), and for contaminated interface Figs.~\ref{f7}(b) and (e) when $Ma=0.05$ and Figs.~\ref{f7}(c) and (f) when $Ma=0.1$) with the increase in $Ma$ value. This implies that the Marangoni force stabilizes the unstable IM. The presence of the damped interface surfactant mode is the primary reason for this stabilizing nature of the interface mode. The Marangoni stress that results from the surfactant concentration gradient attenuates the energy transfer to the disturbed interface wave. Consequently, the Marangoni force is crucial in regulating the growth of interface wave instability. 

Next, the outcomes for the unstable IM are generated in the zone $m<1$. 
According to the study of \citet{bhat2020linear} and \citet{sani2020effect}, the characteristic of interface instability behaves differently in two different regions $mr<1$ and $mr>1$. 
\begin{figure}[ht!]
	\begin{center}
	  \subfigure[]{\includegraphics[width=7.2cm]{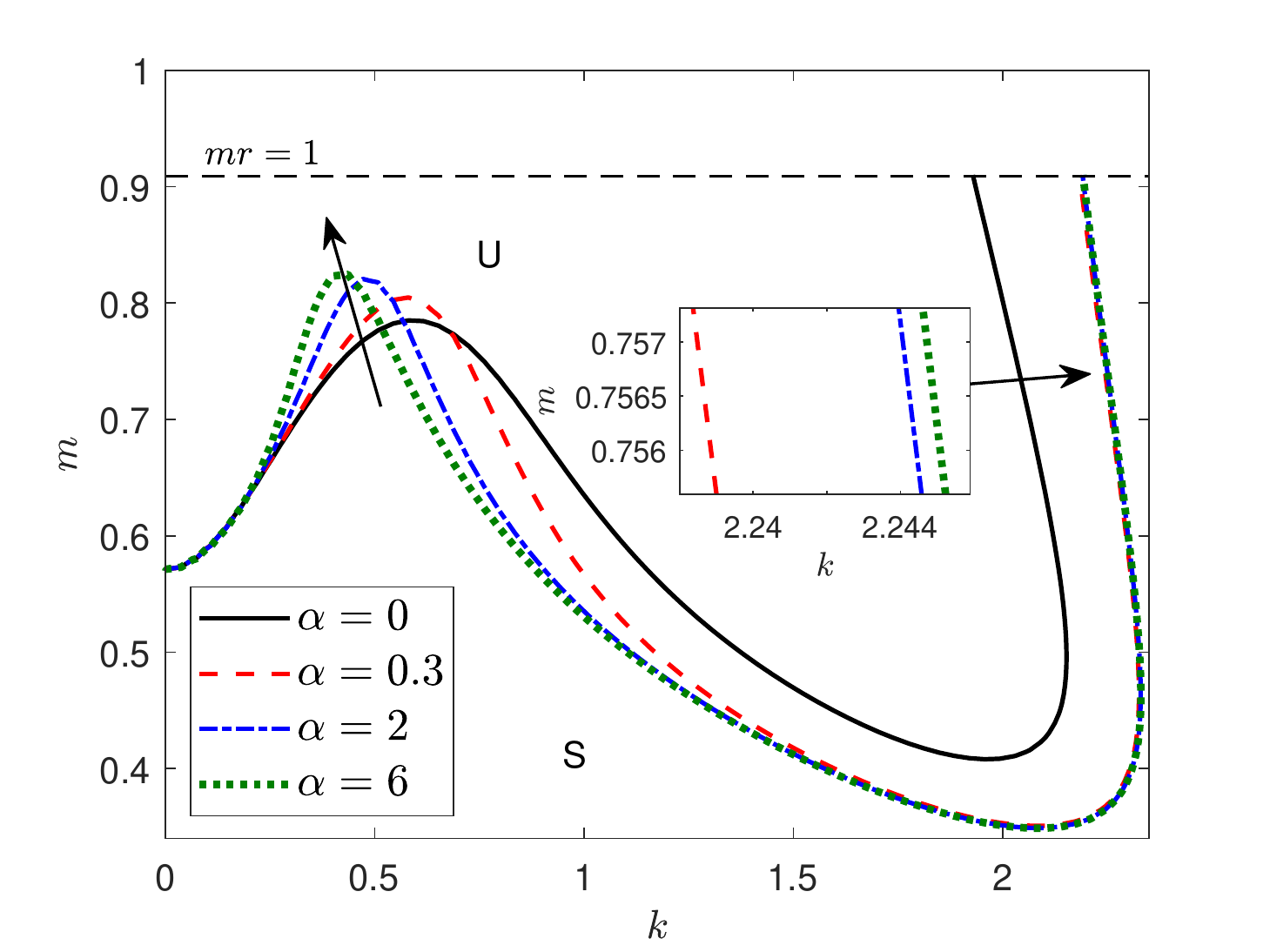}}
	  \subfigure[]{\includegraphics[width=7.2cm]{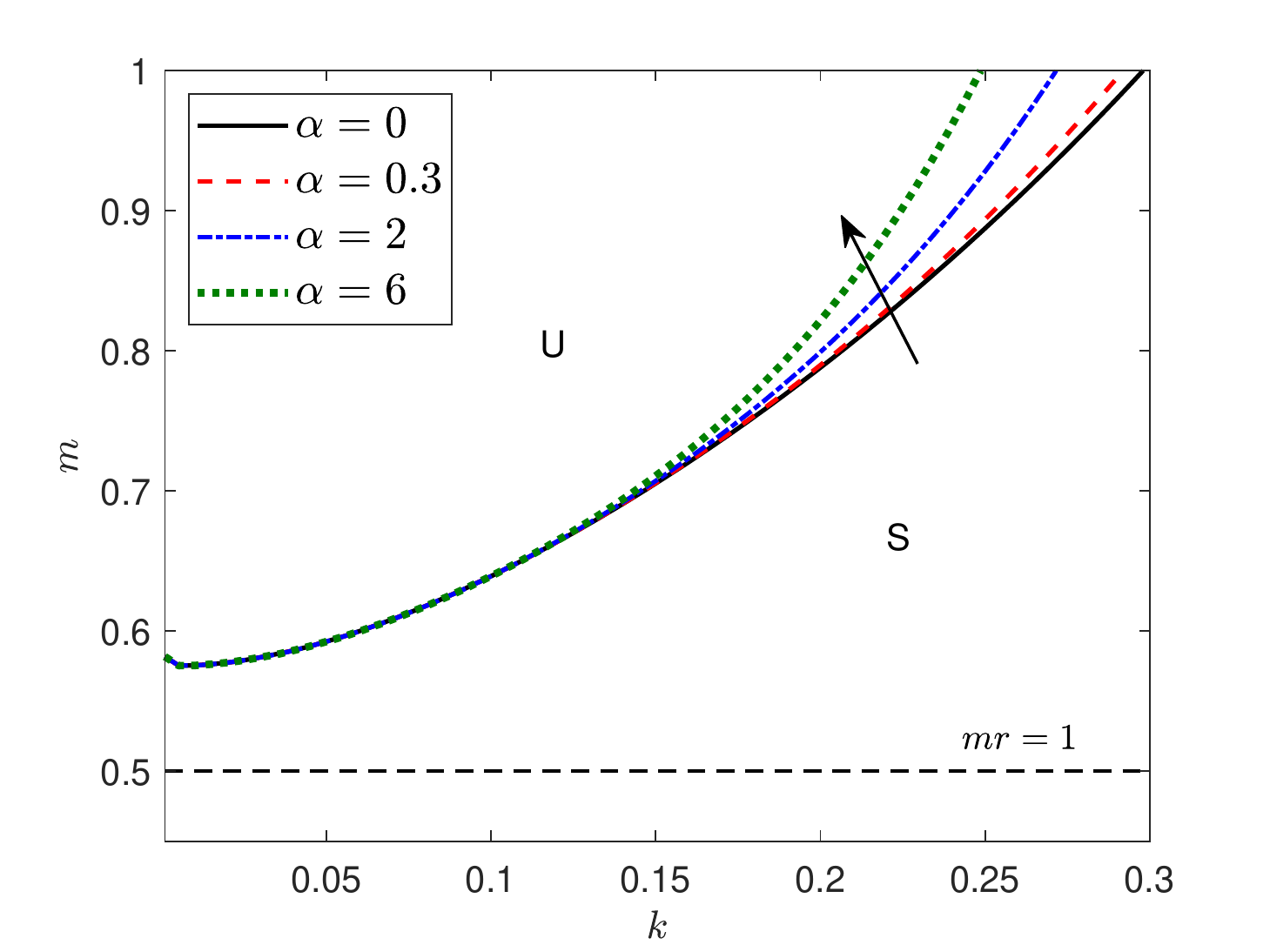}}
  \end{center}\vspace{-0.5cm}
	\caption{Effect of rigidity parameter $\alpha$ on the neutral stability curves corresponding to the interface mode in $(k-m)$ plane for (a) $mr<1$ with $r=1.1$ and $Re_1=20$ and (b) $mr>1$ with $Re=40$ and $r=2$. The other constant parameters are $\delta=1$, $Ca=1$, $Ma=0.1$, $\theta=0.2~rad$, $\beta=\frac{1}{Re_1}$, and $Pe=\infty$. The upward-directed arrow points out the decreasing manner of an unstable zone for a higher $\alpha$ value. The solid line refers to the free surface flow ($\alpha=0$ and $\gamma=0$) and the fixed value $\gamma=0.001$ when $\alpha\neq0$. 
	 }\label{f8}
\end{figure}
\begin{figure}[ht!]
	\begin{center}
	  \subfigure[]{\includegraphics[width=7.2cm]{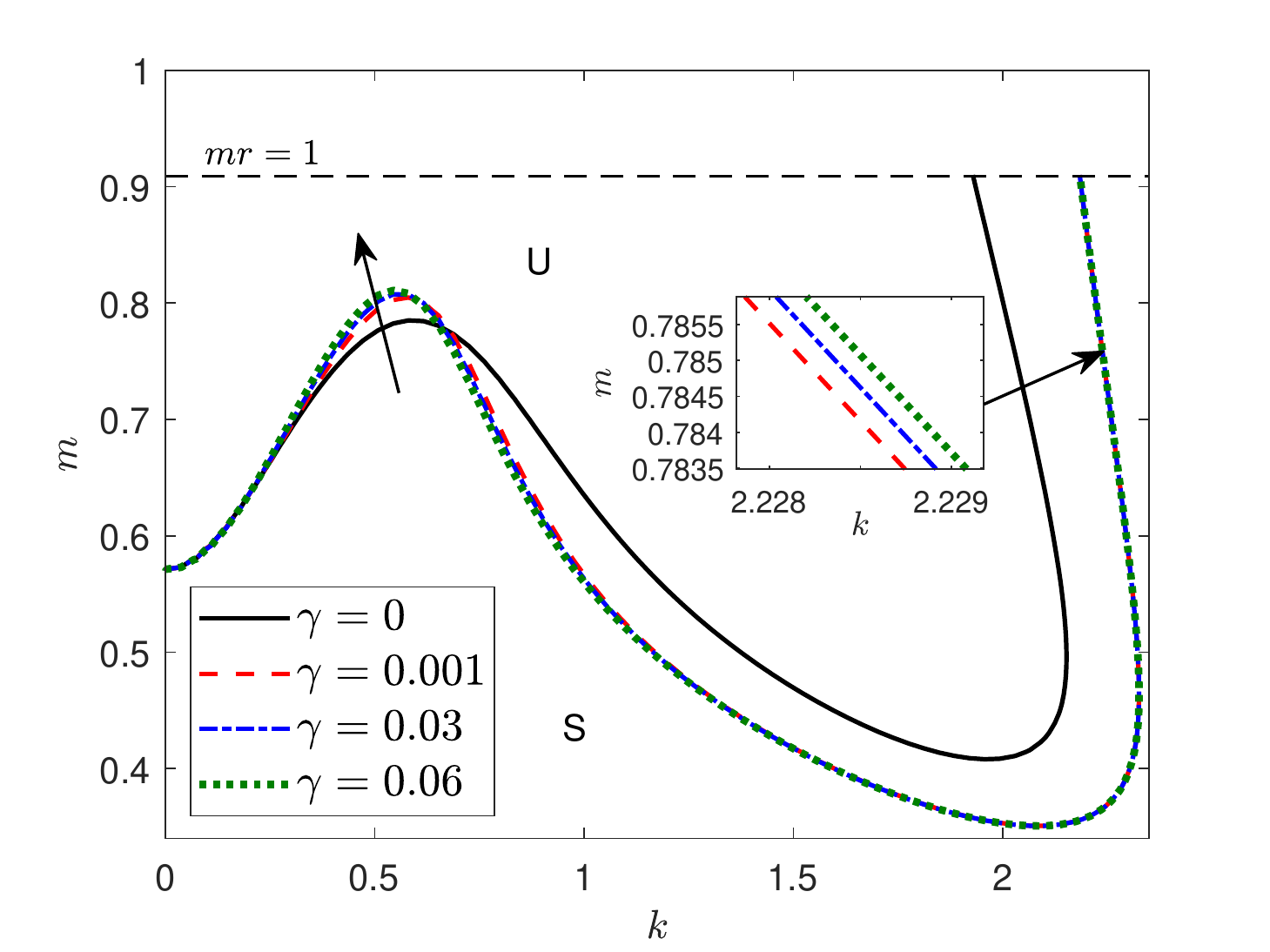}}
	  \subfigure[]{\includegraphics[width=7.2cm]{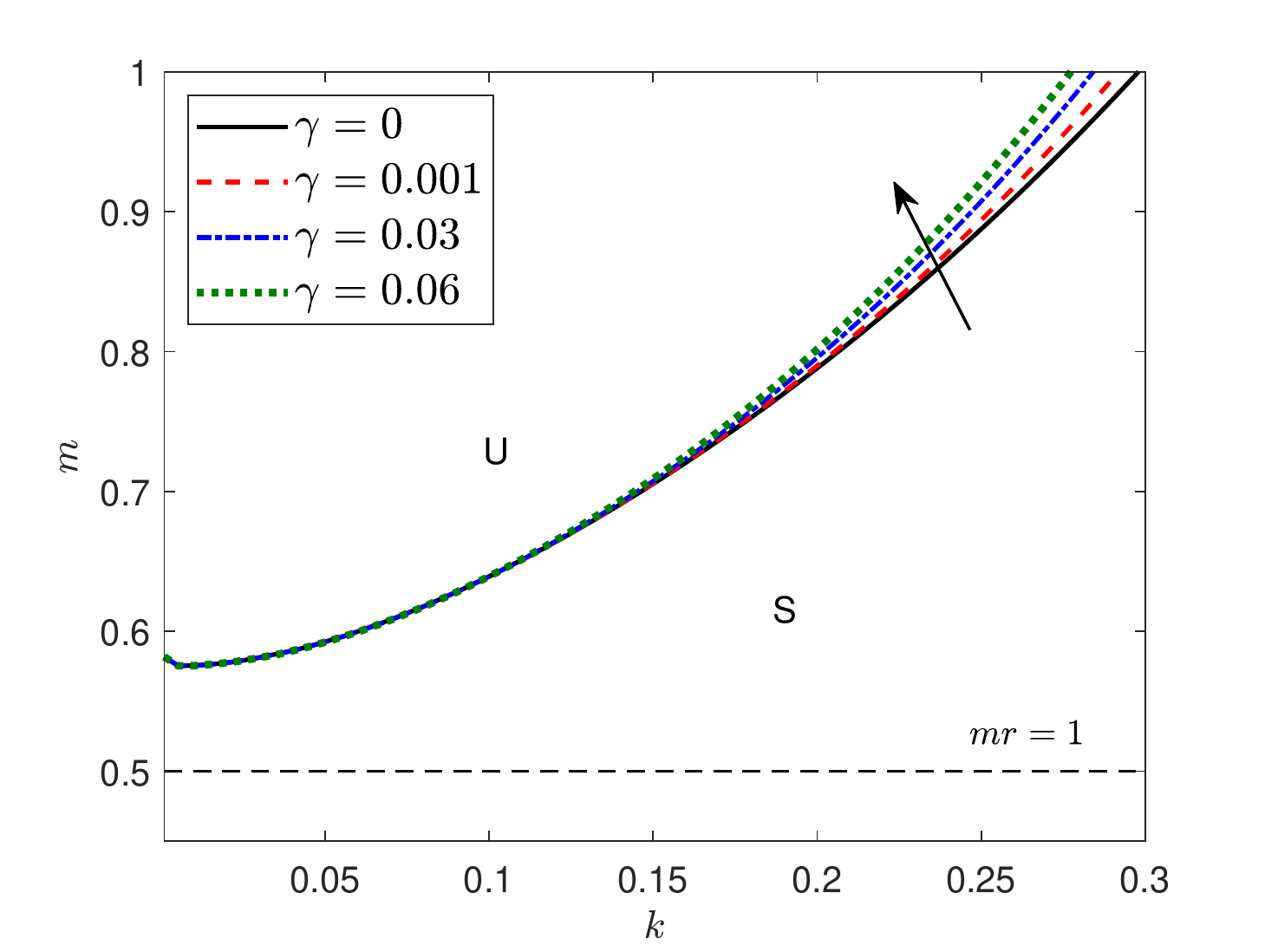}}
	\end{center}\vspace{-0.5cm}
	\caption{Effect of thickness parameter $\gamma$ on the neutral stability curves corresponding to the interface mode in $(k-m)$ plane for (a) $mr<1$ with $r=1.1$ and $Re_1=20$ and (b) $mr>1$ with $Re=40$ and $r=2$. The other constant parameters are $\delta=1$, $Ca=1$, $Ma=0.1$, $\theta=0.2~rad$, $\beta=\frac{1}{Re_1}$, and $Pe=\infty$. The upward-directed arrow points out the decreasing manner of an unstable zone for a higher $\gamma$ value. The solid line refers to the free surface flow ($\alpha=0$ and $\gamma=0$) and the fixed value $\alpha=0.3$ when $\gamma\neq0$. 
	 }\label{f9}
\end{figure}
 To check this fact, we have shown the neutral stability curves in the $k-m$ plane for various $\alpha$ values for the region $mr<1$, as in Fig.~\ref{f8}(a) and the region $mr>1$, as in Fig.~\ref{f8}(b), when top layer viscosity is much greater than the bottom layer ($m<1$). 
The dashed line $mr=1$ indicates the critical boundary line for the interfacial instability called the upper boundary (see, Fig.~\ref{f8}(a)) and the lower boundary (see, Fig.~\ref{f8}(b)) for the neutral stability domain. 
Fig.~\ref{f8}(a) shows that the shortwave instability is more dominant than the longwave instability in the region $mr<1$. Moreover, the stronger rigidity $\alpha$ shrinks the linear unstable zone up to a critical wavenumber, and then on destabilizing impact of $\alpha$ is observed. The dissipation of surface wave energy (see, Figs.~\ref{f4}(a) and (b)) by the strong rigidity of the plate is mainly responsible for the stabilizing nature of interface mode in the smaller wavenumber range (i.e., $k<k_c$). But, the stable SM in the shorter wave zone (i.e., larger $k$, $k>k_c$) gives the scope to the IM to become more unstable for higher rigidity of the plate due to the momentum conservation when the bottom layer viscosity becomes much lesser than the top layer ($m<1$) with $mr<1$.  
On the other side, the unstable interface mode bandwidth, depicted in Fig.~\ref{f8}(b), increases as the viscosity ratio $m$ increases ($m<1$) for the case of $mr>1$.
Moreover, the reduction of unstable mode bandwidth for higher rigidity parameter $\alpha$ is noticed.

Further, we observe the variation of marginal curves of the IM in two different regimes $mr<1$ and $mr>1$ for various $\gamma$ values from Fig.~\ref{f9}. In Fig.~\ref{f9}(a), it is clear that uniform thickness $\gamma$ also performs a double role in the interfacial wave dynamics just like uniform rigidity $\alpha$, when $mr<1$. However, the rapidly decreasing bandwidth of the unstable interface region with increasing $\gamma$ value in $mr>1$ ensures that thickness parameter $\gamma$ has the capacity to decrease the interface flow instability in the smaller wavenumber range (see, Fig.~\ref{f9}(b)). 

\subsection{\bf{Effect of the floating flexible plate on the interface surfactant mode}}
\begin{figure}[ht!]
\begin{center}
 \subfigure[]{\includegraphics[width=7.2cm]{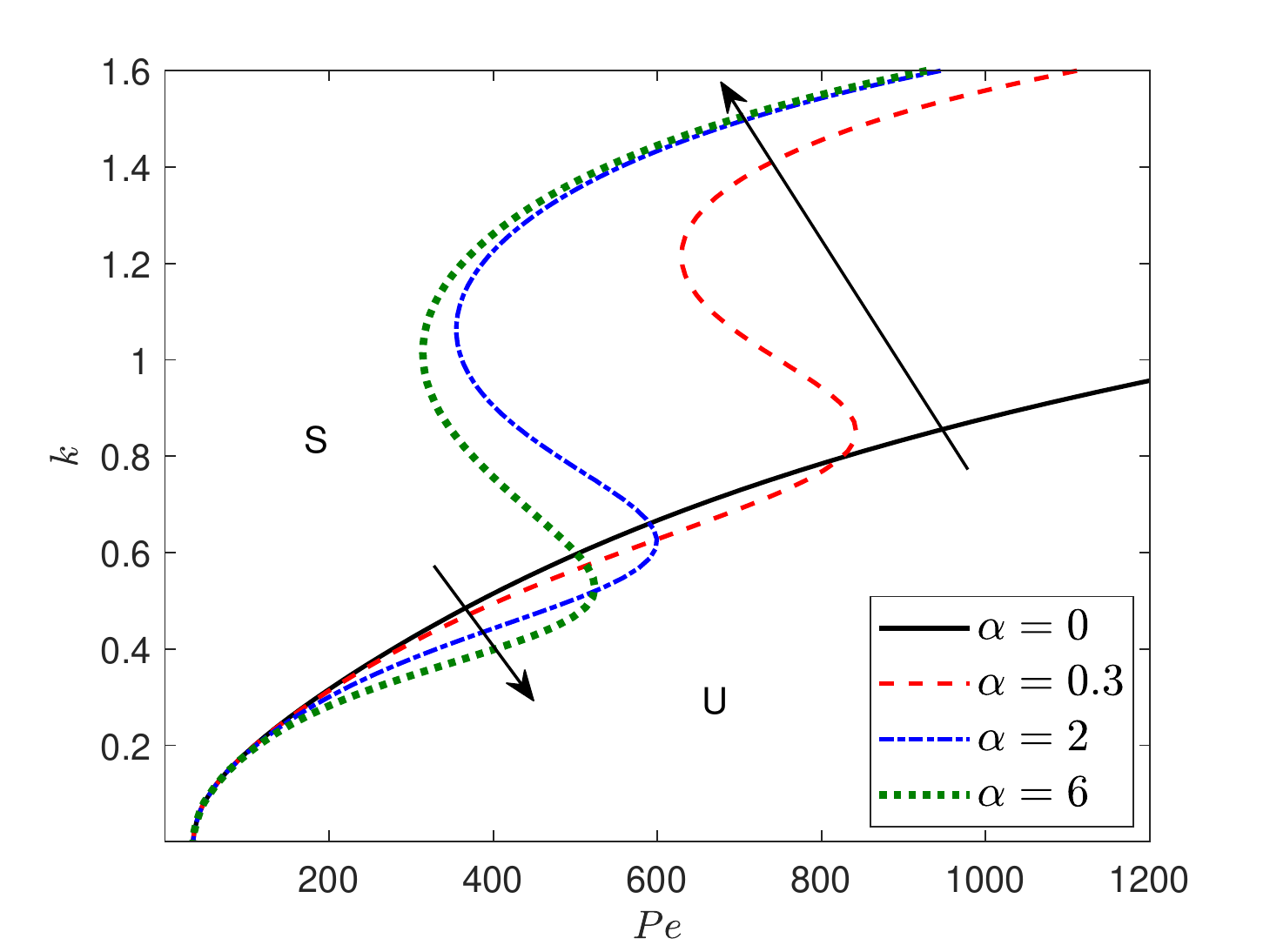}}
 \subfigure[]{\includegraphics[width=7.2cm]{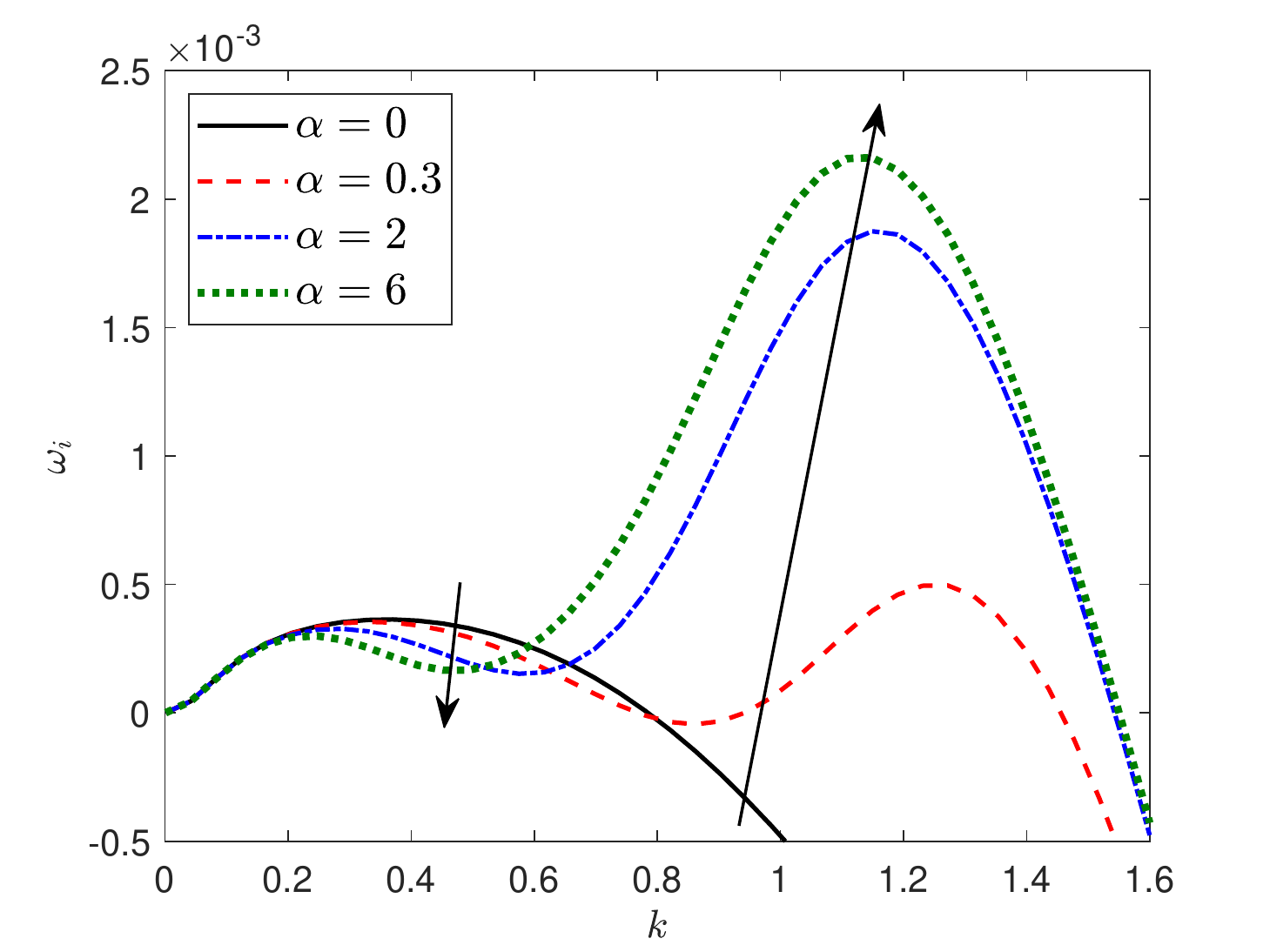}}
\subfigure[]{\includegraphics[width=7.2cm]{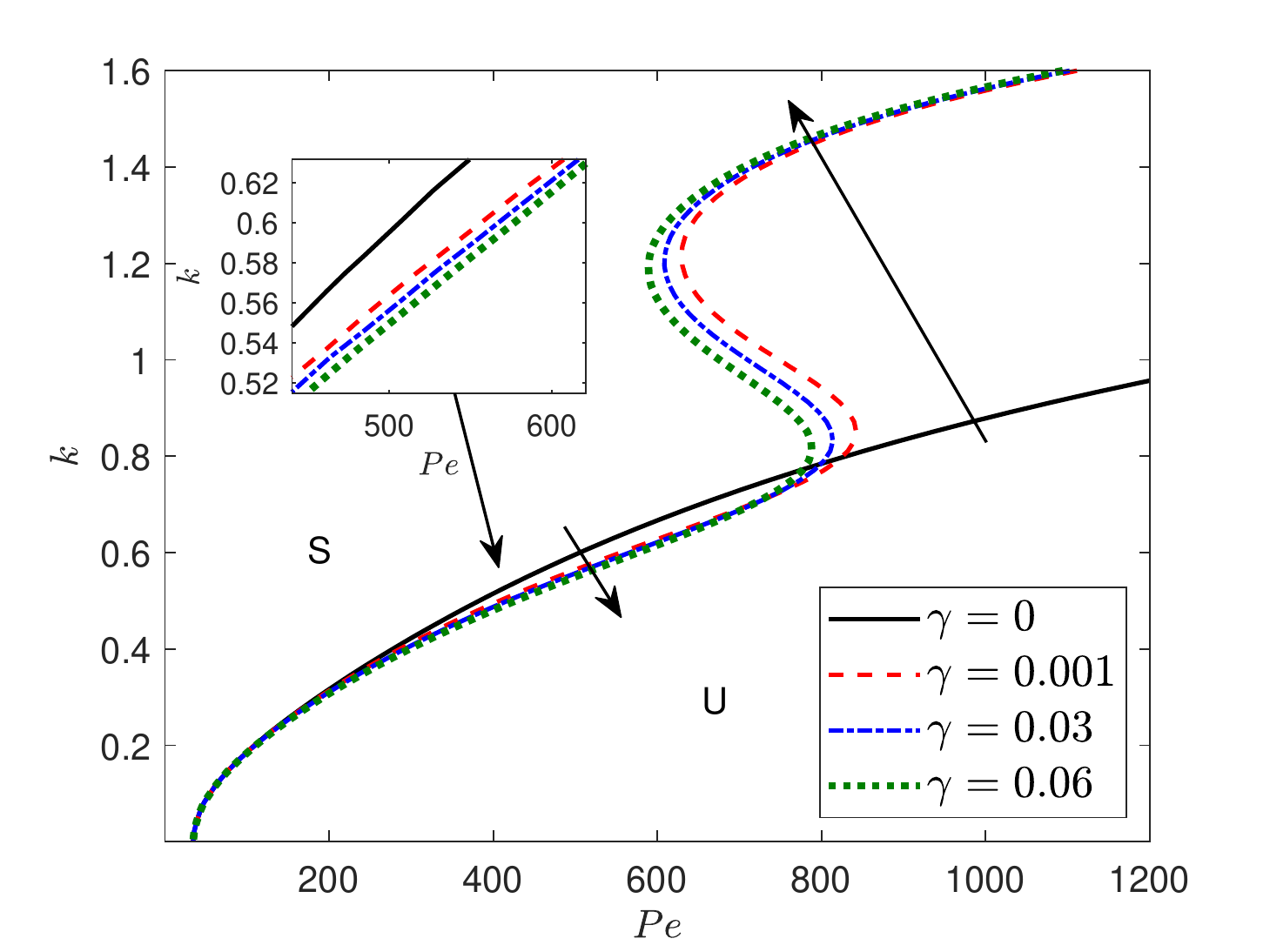}}
\subfigure[]{\includegraphics[width=7.2cm]{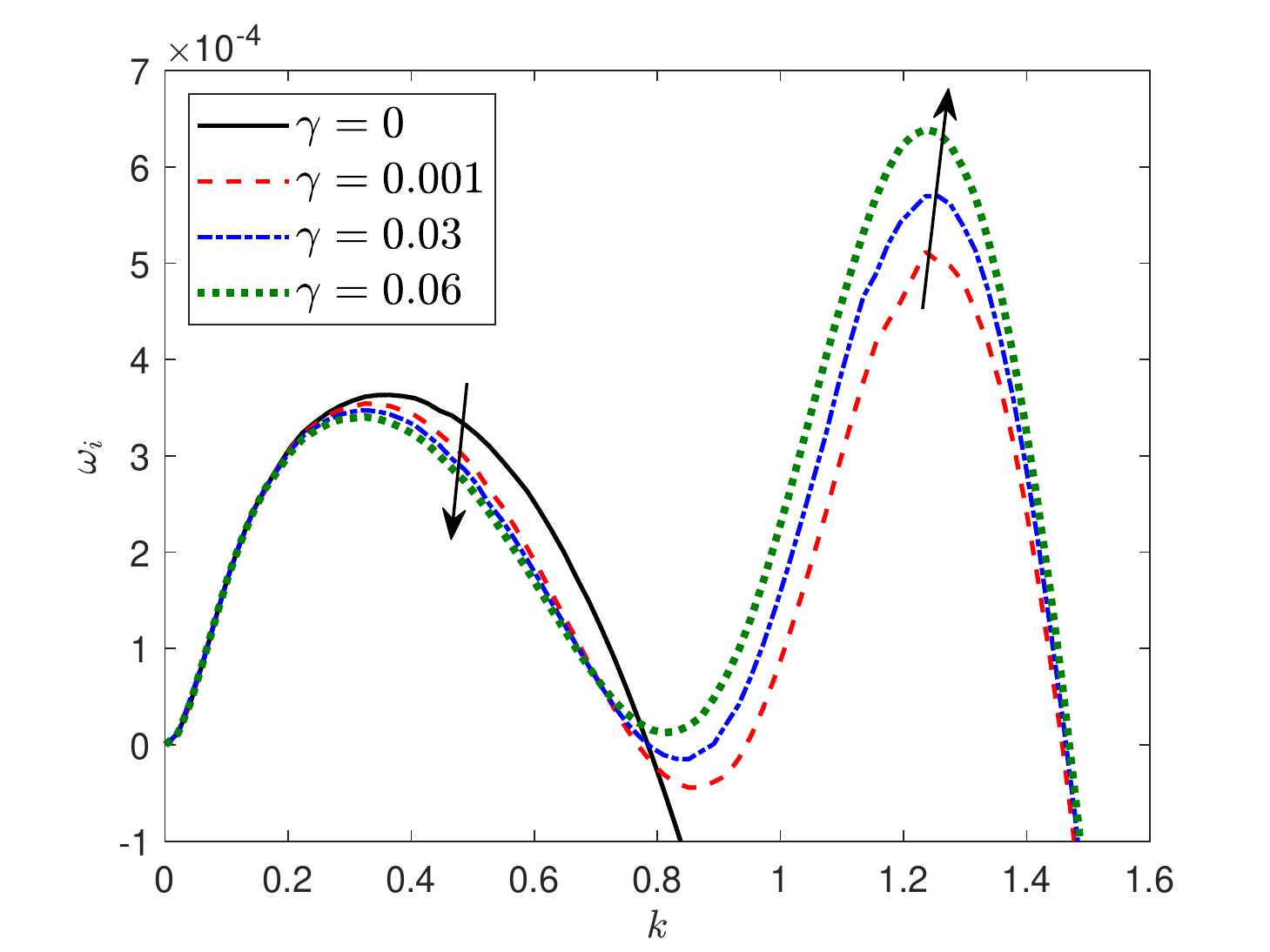}}
\end{center}\vspace{-0.5cm}
	\caption{((a) and (c)) The variation of marginal stability curves ($\omega_i=0$) of the interface surfactant mode in ($Pe-k$) plane for various values of structural rigidity  $\alpha$ (in (a)) and uniform thickness $\gamma$ (in (c)). The downward/upward directed arrow points out the decreasing/increasing manner of an unstable regime with the increasing of $\alpha$ and $\gamma$ values. ((b) and (d)) The  corresponding temporal growth rate curves ($\omega_i$) as a function of wavenumber ($k$) for various values of structural rigidity $\alpha$ (in (b)) and uniform thickness $\gamma$ (in (d)) with $Pe=800$. The downward/upward directed arrow points out the decreasing/increasing manner of the temporal growth rate. The other fixed parameters are $Re_1=20$, $\delta=1$, $r=1$, $m=1.5$, $Ca=1$, $Ma=0.01$, $\theta=0.2$ rad, and $\beta=\frac{1}{Re_1}$. The solid line marks the free surface flow (i.e., $\alpha=0$ and $\gamma=0$). The value of $\gamma=0.001$ when $\alpha\neq0$ and  $\alpha=0.3$ when $\gamma\neq0$.}\label{f10}
\end{figure}

This subsection addresses the instability behaviour of interface surfactant mode (ISM). 
For the plate-covered surface ($\alpha\neq0\  \textrm{and} \ \gamma\neq0$), two humps appear in the marginal stability curves in ($Pe-k$) plane  (see, Figs.~\ref{f10}(a) and (c)). However, the humps in the marginal curves disappear for free surface flow. Further, it is revealed that both $\alpha$ and $\gamma$ have negligible impact on the ISM in the longwave region, but a significant influence is observed in the finite wavenumber zone.
The rigidity parameter, $\alpha$ decreases the region of unstable ISM (Fig.~\ref{f10}(a)) up to a certain range of small wavenumber, whereas beyond this range, it increases. A similar kind of instability behaviour of ISM can be noticed for the higher thickness parameter $\gamma$ (Fig.~\ref{f10}(b)). These physical facts result in the dual nature of the ISM of the considered flow system. As in Figs.~\ref{f10}(a) and (c), the existence of two humps: one in the longwave zone and another in the finite wavenumber zone, is also confirmed by the corresponding growth rate curves illustrated in Fig.~\ref{f10}(b) (when $\alpha$ alters) and Fig.~\ref{f10}(d) (when $\gamma$ alters). It is evident from both Figs.~\ref{f10}(b) and (d) that floating elastic plate characteristic parameters $\alpha$ as well as $\gamma$ reduces the temporal growth of the ISM up to a certain range of small wavenumber and then on enhances the growth rate. 
\begin{figure}[ht!]
\begin{center}
 \subfigure[]{\includegraphics[width=7.2cm]{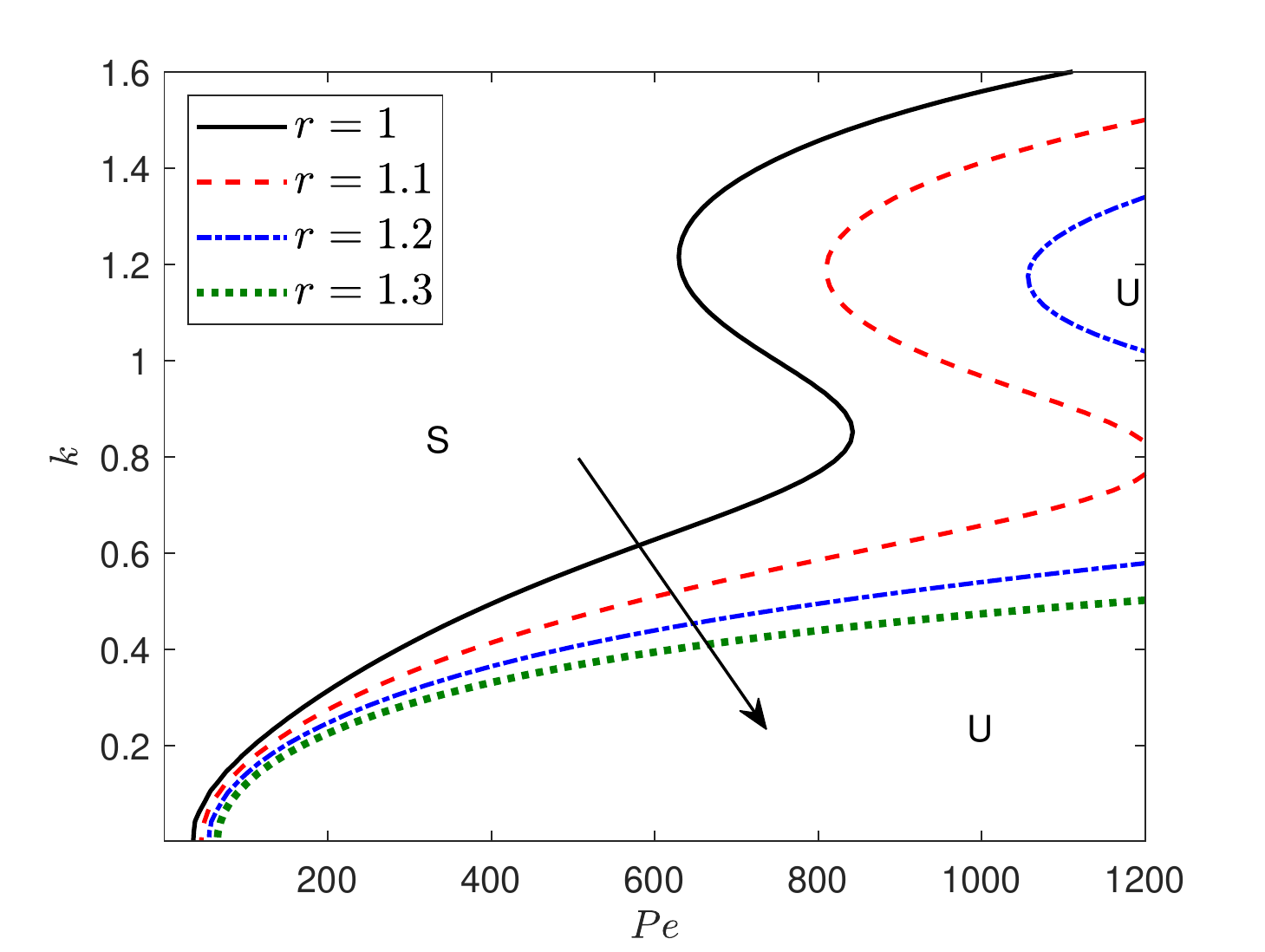}}
 \subfigure[]{\includegraphics[width=7.2cm]{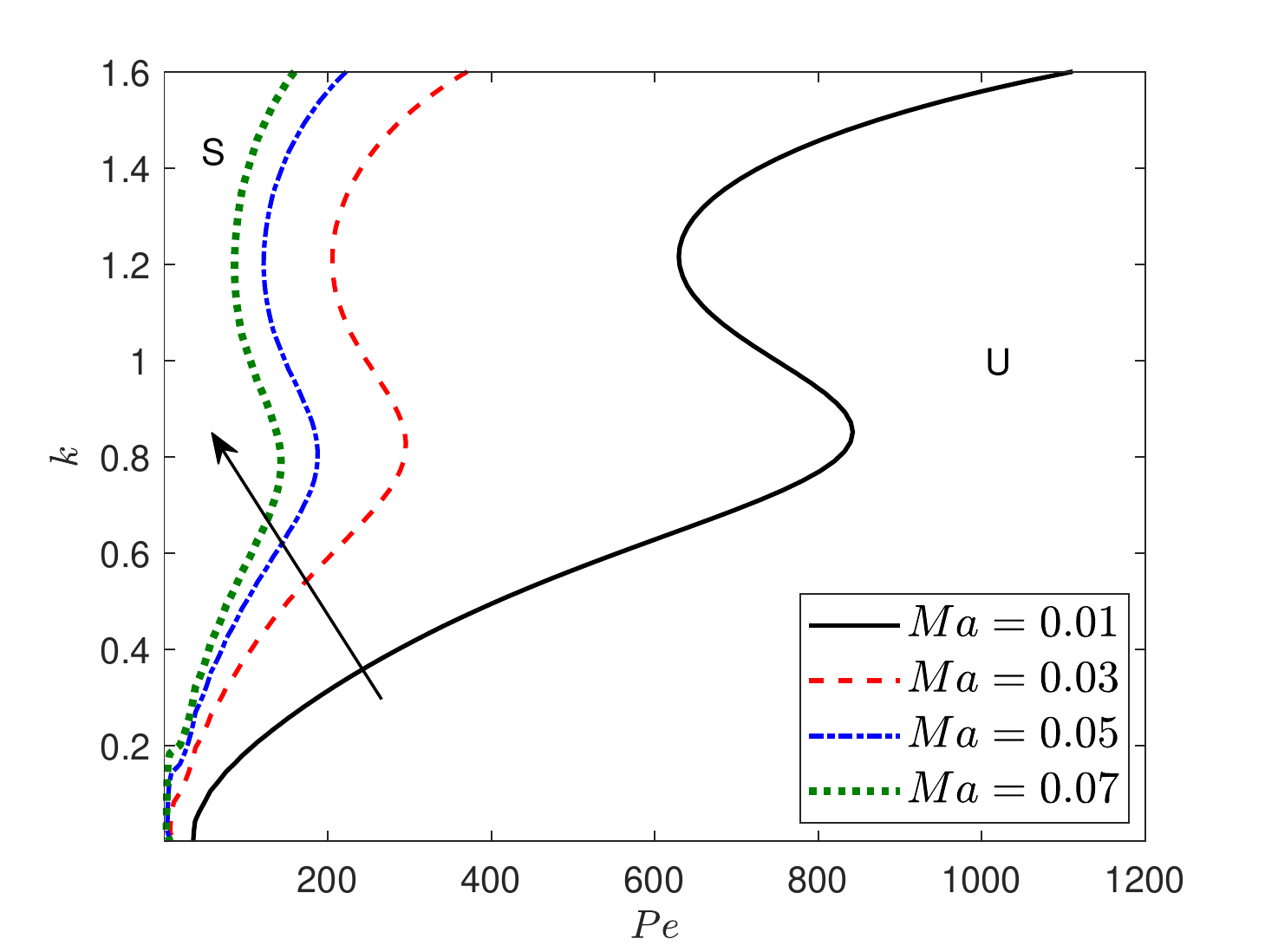}}
\end{center}\vspace{-0.5cm}
	\caption{The variation of the stability/instability boundaries of the interface surfactant mode in ($Pe- k$) plane for different (a) $r$ when $Ma=0.01$ and (b) $Ma$ when $r=1$. The remaining fixed values are $Re_1=20$, $\delta=1$, $m=1.5$, $Ca=2$, $\theta=0.2\,rad$, $\alpha=0.3$, $\beta=\frac{1}{Re_1}$, and $\gamma=0.001$. The downward and upward arrows indicate the decreasing and increasing manner, respectively, of the unstable zone.  
 	  }\label{f11}
\end{figure}

Fig.~\ref{f11}(a) exhibits the resulting neutral curves in the $Pe-k$ plane corresponding to the ISM when $r$ varies. The higher density ratio $r$ (i.e., lower layer density rapidly increases than the upper layer) attenuates the unstable regime by increasing the critical P\'eclet number. 
Hence, the ISM becomes more stable if the bottom layer is highly dense than the top layer.  
Next, the neutral curves of the ISM in the $Pe-k$ plane are demonstrated when the Marangoni number $Ma$ varies (Fig.~\ref{f11}(b)).
A stronger Marangoni force promotes ISM instability by increasing the unstable zone. Higher Marangoni force alters the surface tension gradient of lower-layer liquid, which increases the shear stress at the liquid-liquid interface. 
Consequently, the Marangoni force has a destabilizing impact owing to the advancement of disturbance energy of the ISM for the stronger Marangoni force.

\subsection{\bf{Effect of the floating flexible plate on the shear mode}}
\begin{figure}[ht!]
\begin{center}
\subfigure[]{\includegraphics[width=7.2cm]{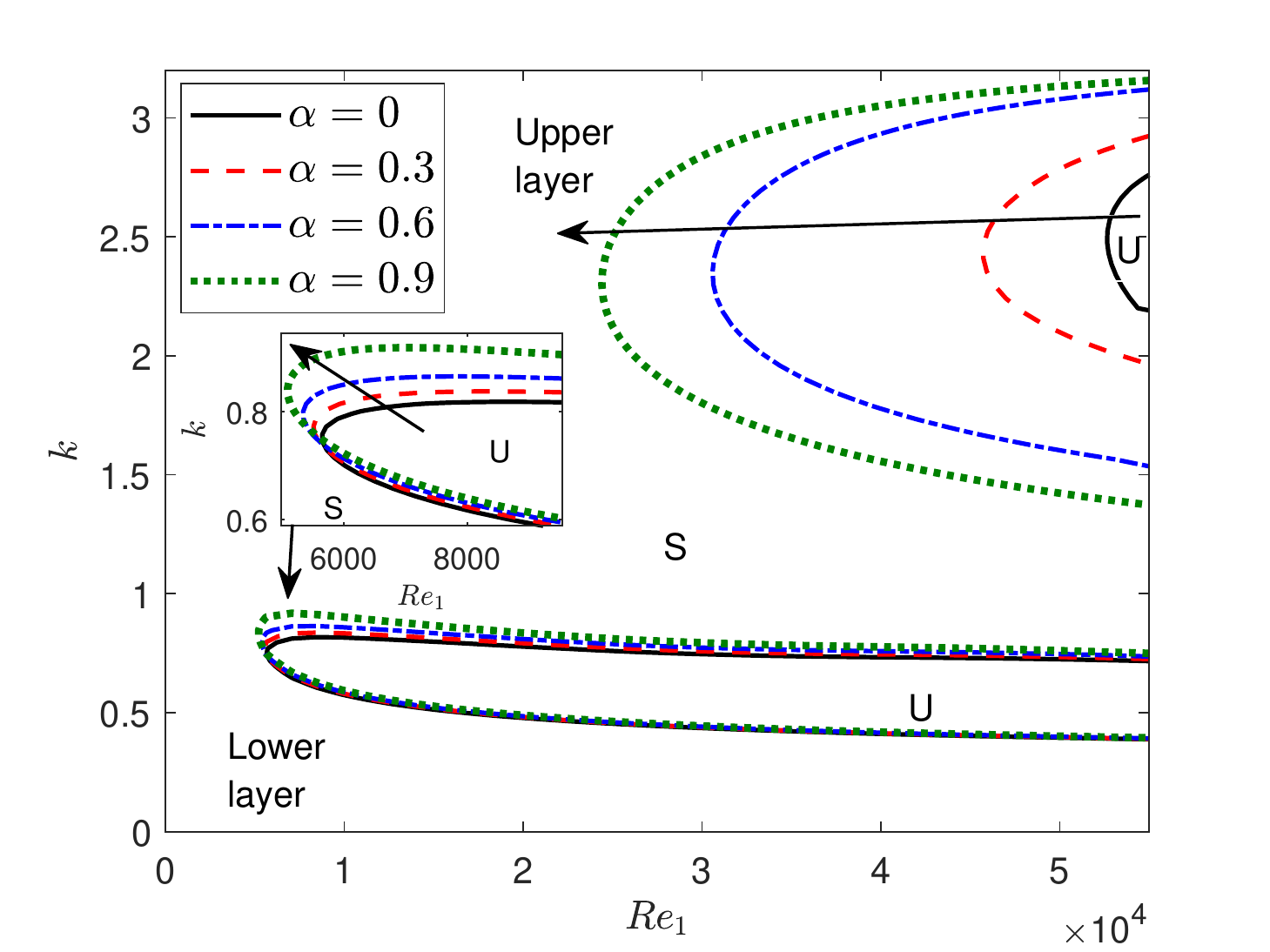}}
\subfigure[]{\includegraphics[width=7.2cm]{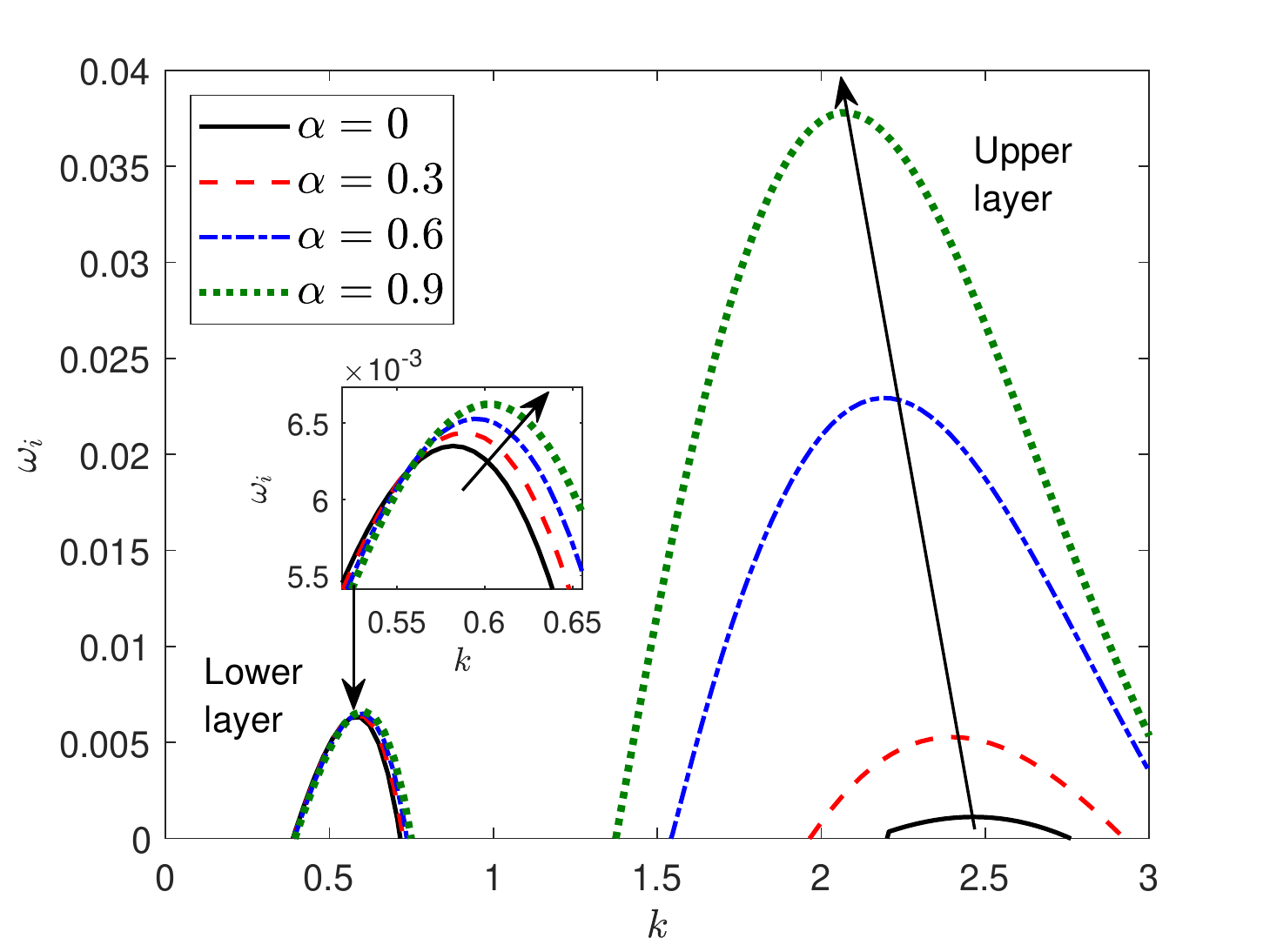}}
	\end{center}\vspace{-0.5cm}
	\caption{(a) The effect of structural rigidity ($\alpha$) on the marginal stability curve of shear modes. The left-hand arrow points out the increasing manner of the unstable zone. (b) The corresponding growth rate of the shear mode with $Re_1=55\times10^3$. The upward-directed arrow points out the increasing manner of temporal growth rate. The other constant parameters are $\delta=1$, $r=5$, $m=5$, $Ca=1$, $Ma=0.1$, $\theta=0.07~rad$, $\beta=\frac{1}{Re_1}$, and $Pe=\infty$. The solid line signifies the free surface flow ($\alpha=\gamma=0$) and the fixed value $\gamma=0.001$ when $\alpha\neq0$. }\label{f12}
\end{figure}

\begin{figure}[ht!]
\begin{center}
\subfigure[]{\includegraphics[width=7.2cm]{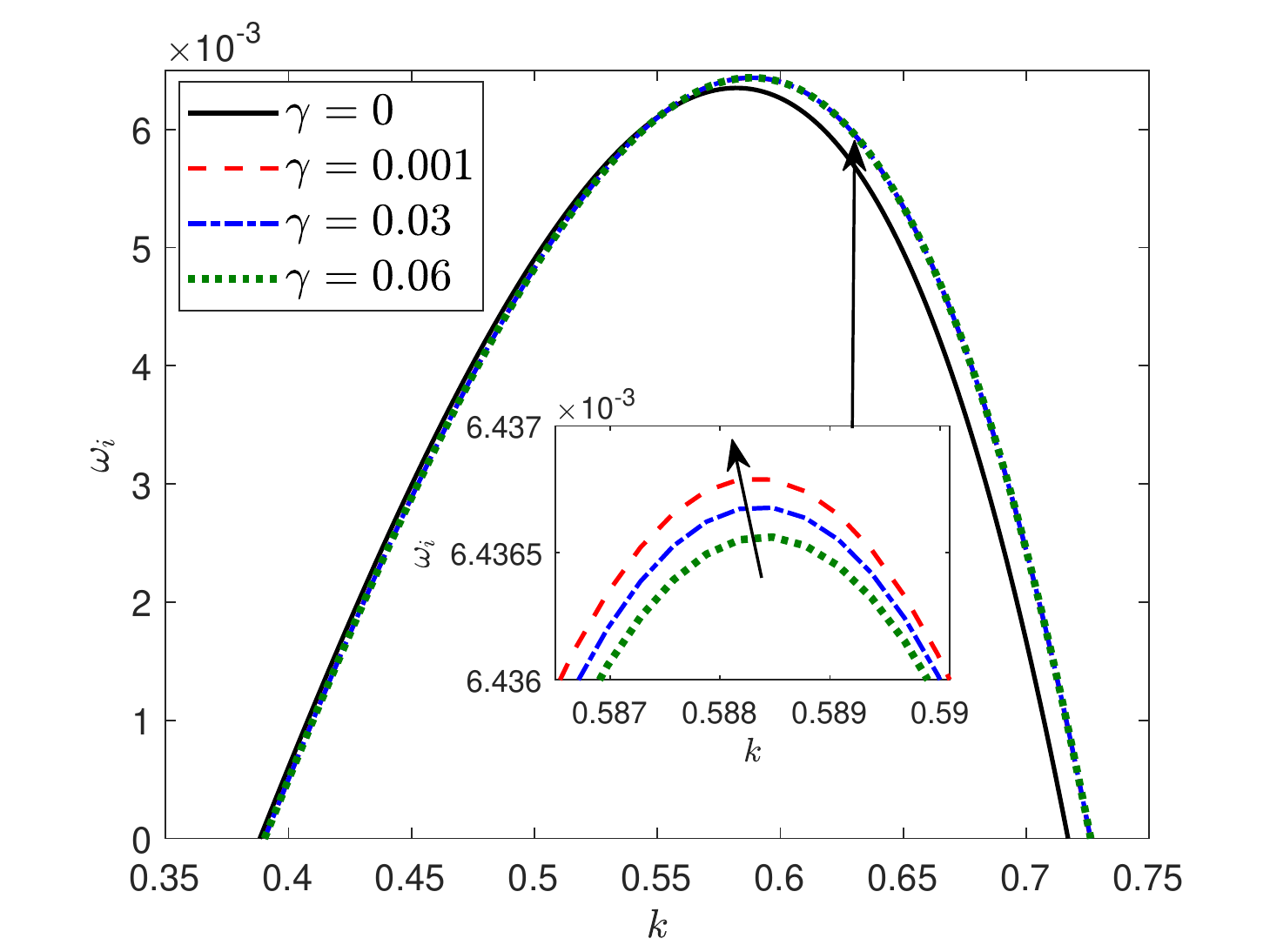}}
\subfigure[]{\includegraphics[width=7.2cm]{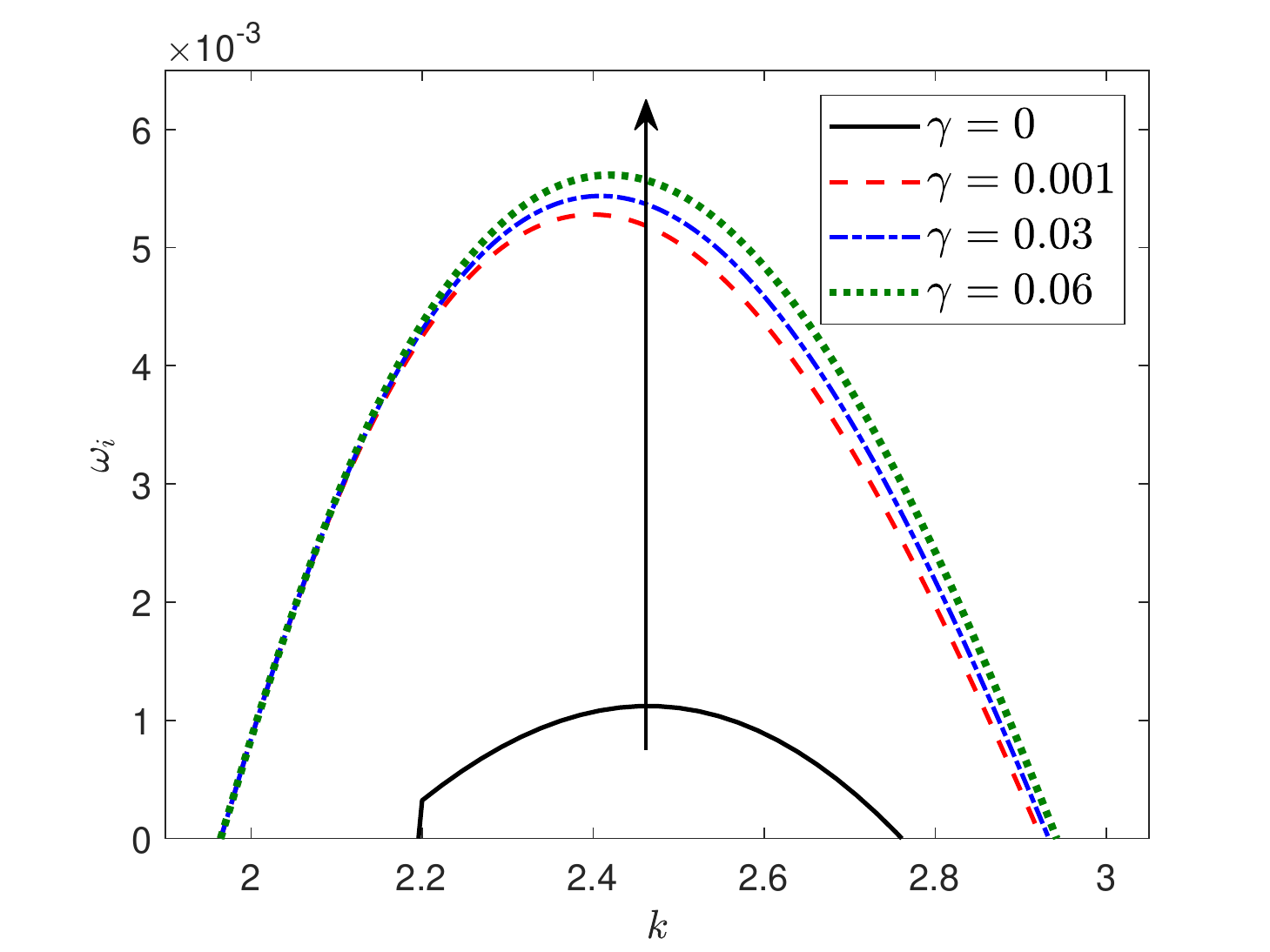}}
\end{center}\vspace{-0.5cm}
\caption{The effect of uniform thickness ($\gamma$) on the temporal growth rate of (a) the lower layer shear mode and (b) the upper layer shear mode as a function of $k$ when $Re_1=55\times10^3$. The upward-directed arrow in the inset plot points out the increasing manner of temporal growth rate. The other constant parameters are $\delta=1$, $r=5$, $m=5$, $Ca=1$, $Ma=0.1$, $\theta=0.07~rad$, $\beta=\frac{1}{Re_1}$, and $Pe=\infty$. The solid line signifies the free surface ($\alpha=\gamma=0$) and the fixed value $\alpha=0.3$ when $\gamma\neq0$.}\label{f13}
\end{figure}\vspace{-0.3cm}

Now, the numerical computation is repeated to study the shear wave dynamics of the two-layer fluid flow with the floating elastic plate at the top layer surface. Fig.~\ref{f2}(c) provides the assurance of the presence of the excited shear mode for low inclination angle when inertia force $Re_1$ of the upper layer is chosen very high \cite{bhat2019linear, bhat2020linear}. 
Fig.~\ref{f12} exhibits the influence of the plate on two shear modes sourcing from the upper and lower layer fluid when the inclination angle $\theta=0.07~rad$ and the upper layer of lesser density flows above the comparatively higher density liquid layer. 
The lower-layer instigated SHM occurs in the shortwave zone, and the upper layer's SHM exits in the higher wavenumber range (\citet{bhat2020linear}). 

In Fig.~\ref{f12}(a), the marginal curve of both the shear modes is illustrated in the $Re_1-k$ plane for different structural rigidity parameters $\alpha$. The unstable region of the upper-layer instigated SHM amplifies as rigidity parameter $\alpha$ increases.
However, the parameter $\alpha$ plays a double role in the SHM instability for the lower layer. The parameter $\alpha$ shrinks the SHM induced unstable zone of the lower layer up to a certain wavenumber $k$, and after that, it amplifies the unstable zone. So, the neutral curves corresponding to the lower layer SHM instability have a two-fold variation with respect to the rigidity parameter $\alpha$.
The corresponding temporal growth rate results further assure this stability/instability effect of $\alpha$ on the SHM, as shown in Fig.~\ref{f12}(b). 

We have demonstrated the effect of uniform thickness $\gamma$ on the growth rate incited by the lower layer shear wave, as shown in Fig.~\ref{f13}(a) and the upper layer shear wave, as in  Fig.~\ref{f13}(b). The growth rate for the plate-contaminated flow system (i.e., $\alpha\neq0$ and $\gamma\neq0$) becomes more than that for the clear surface (i.e., $\alpha=\gamma=0$) for both the shear modes. Further, on comparing Figs.~\ref{f13}(a) and \ref{f13}(b), it is found that the influence of plate thickness on the upper layer SHM is strong enough than the lower layer SHM. The physical reason is that the intensity of the normal force exerted by the plate thickness is superior on the upper layer surface, yielding a potentially destabilizing effect on the upper layer shear mode, whereas its influence is inferior at the interfacial wave and results in very weak destabilizing impact on the lower layer SHM. 

\begin{figure}[ht!]
	\begin{center}
		\subfigure[]{\includegraphics[width=7.2cm]{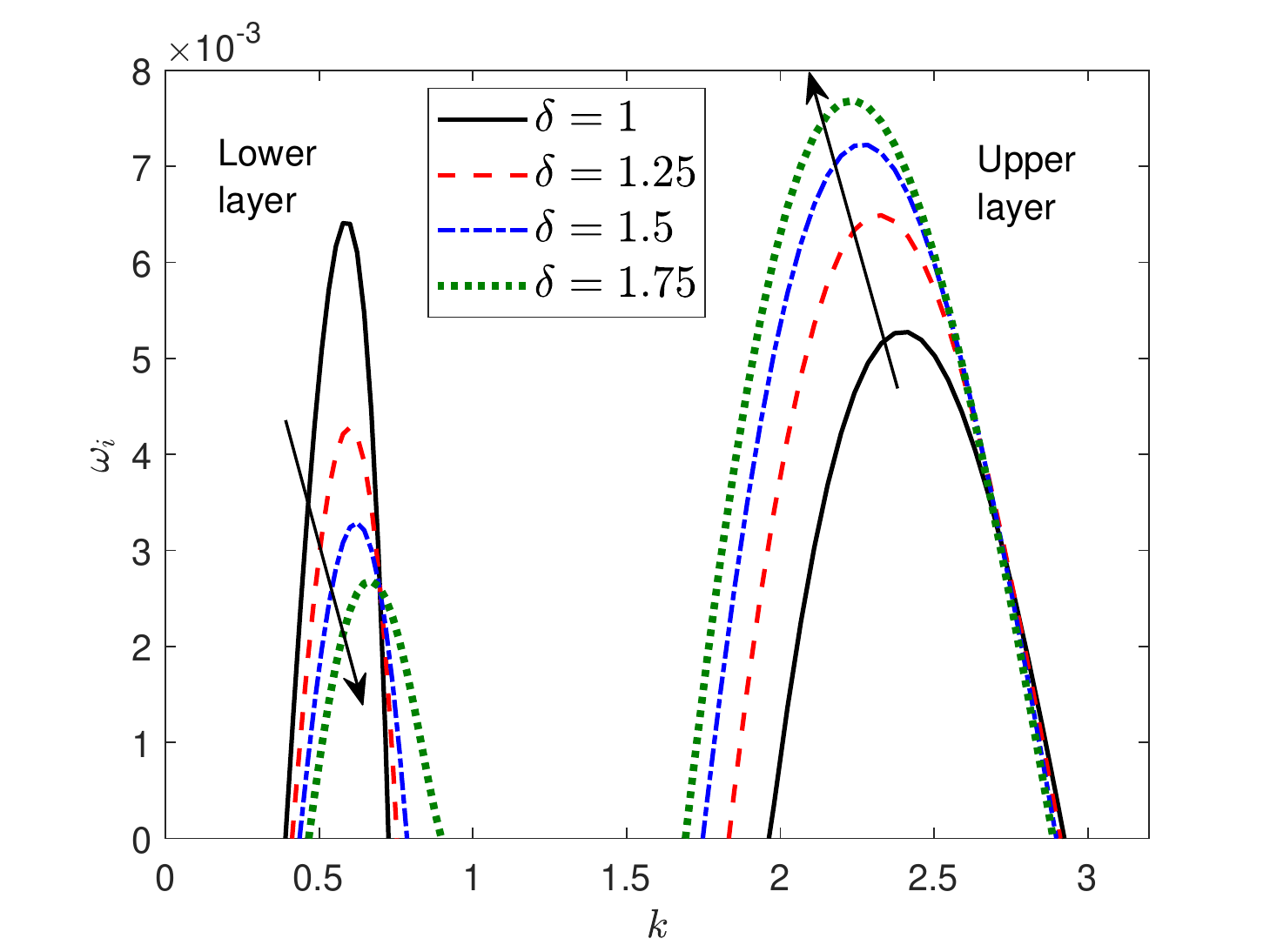}}
		\subfigure[]{\includegraphics[width=7.2cm]{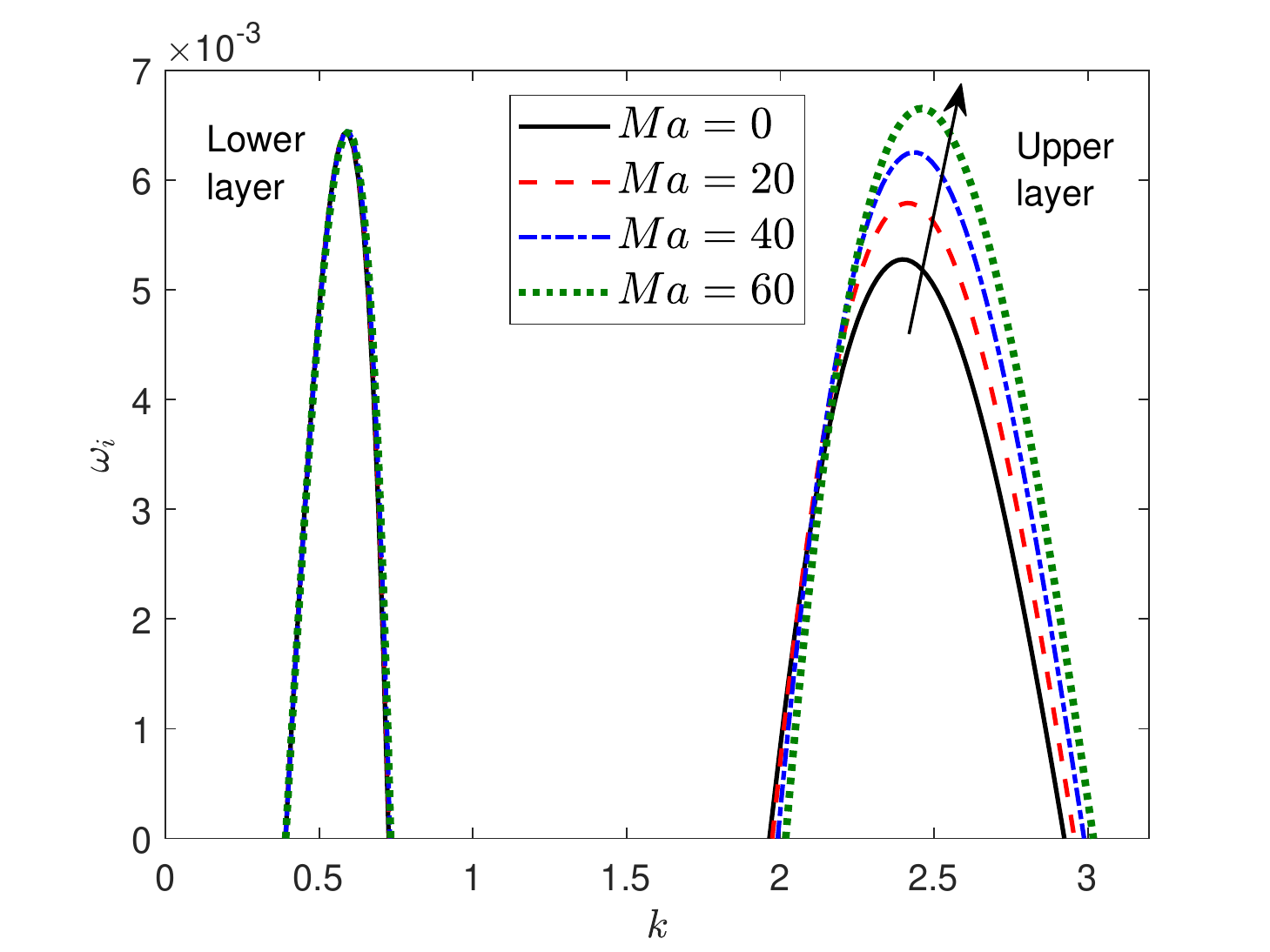}}
    \end{center}\vspace{-0.5cm}
	\caption{The effect of (a) thickness ratio $\delta$ with $Ma=0.1$ and (b) Marangoni number $Ma$ when $\delta=1$ on the temporal growth rate curves ($\omega_i$) of shear mode. The other constant parameters are  $Re_1=55\times10^3$, $r=5$, $m=5$, $\theta=0.07~rad$, $Ca=1$, $\beta=\frac{1}{Re_1}$, $Pe=\infty$, $\alpha=0.3$, and $\gamma=0.001$.  The upward/downward arrow sings the increasing/decreasing manner of the temporal growth rate. }\label{f14}
\end{figure}

The effect of the thickness ratio ($\delta$) of the lower to the upper layer and the Marangoni force ($Ma$) on the SHM instability triggered by the upper and lower layers are clearly observed in the temporal growth rate curves of Fig.~\ref{f14}(a) and Fig.~\ref{f14}(b), respectively. In Fig.~\ref{f14}(a), the thickness ratio $\delta$ shows the two-fold variations in the unstable SHM for the top and bottom layers. More specifically, in the case of the lower layer, the maximum growth rate induced by SHM reduces up to a certain wavenumber $k$, and then it enhances as the thickness ratio $\delta$ increases. But, for the upper layer induced SHM, the reverse trend is observed with respect to increasing $\delta$. Further, the Marangoni force $Ma$ has a stronger impact on the SHM related to the top layer instability as compared to the bottom layer (see, Fig.~\ref{f14}(b)). A stronger Marangoni force rapidly advances the SHM instability for the upper layer, whereas the Marangoni force has a very weak effect on the SHM instability of the lower layer.

\subsection{\bf{Competition between different co-existing modes}}
\begin{figure}[ht!]
	\begin{center}
		\subfigure[]{\includegraphics[width=7.2cm]{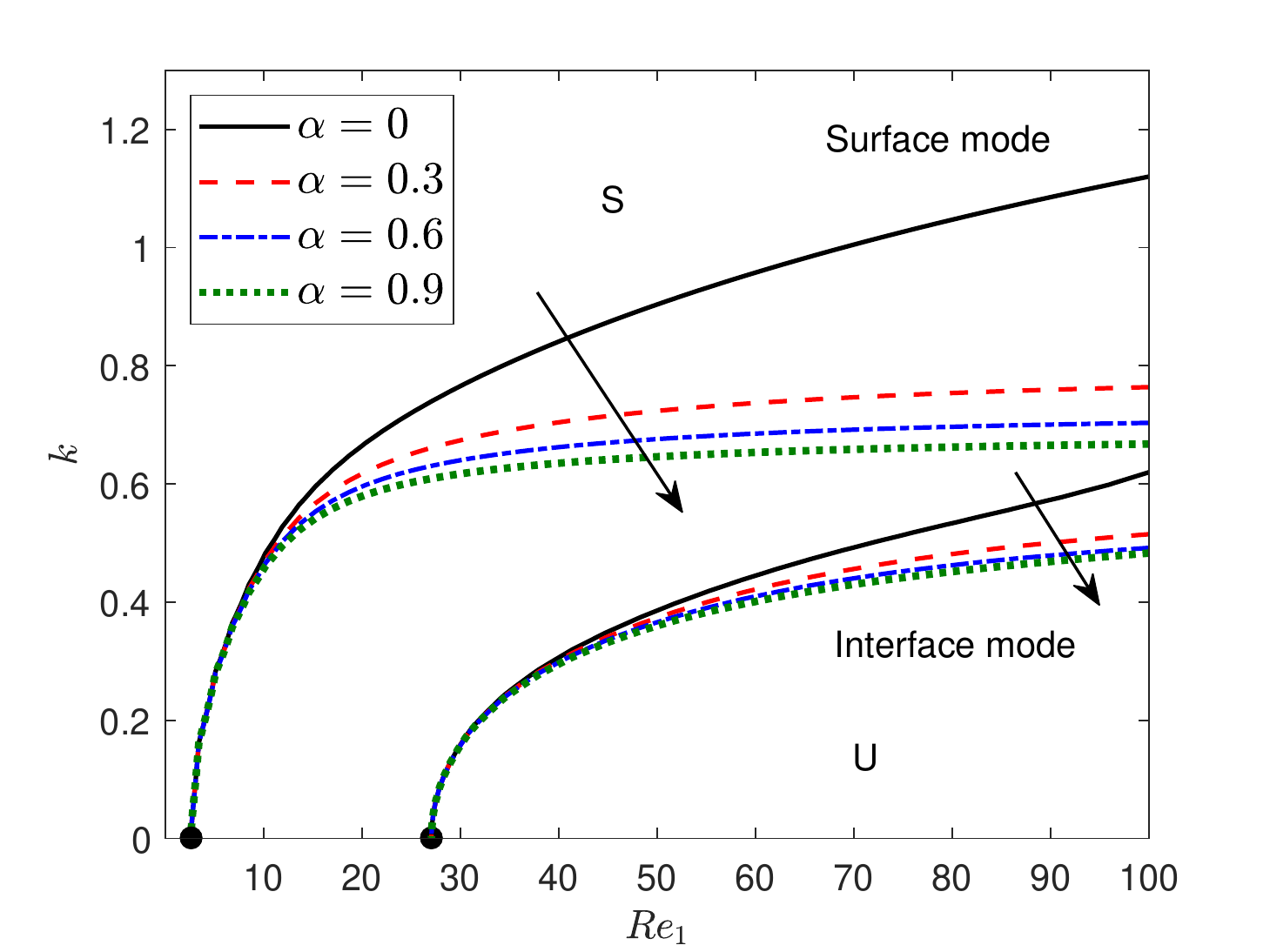}}
		\subfigure[]{\includegraphics[width=7.2cm]{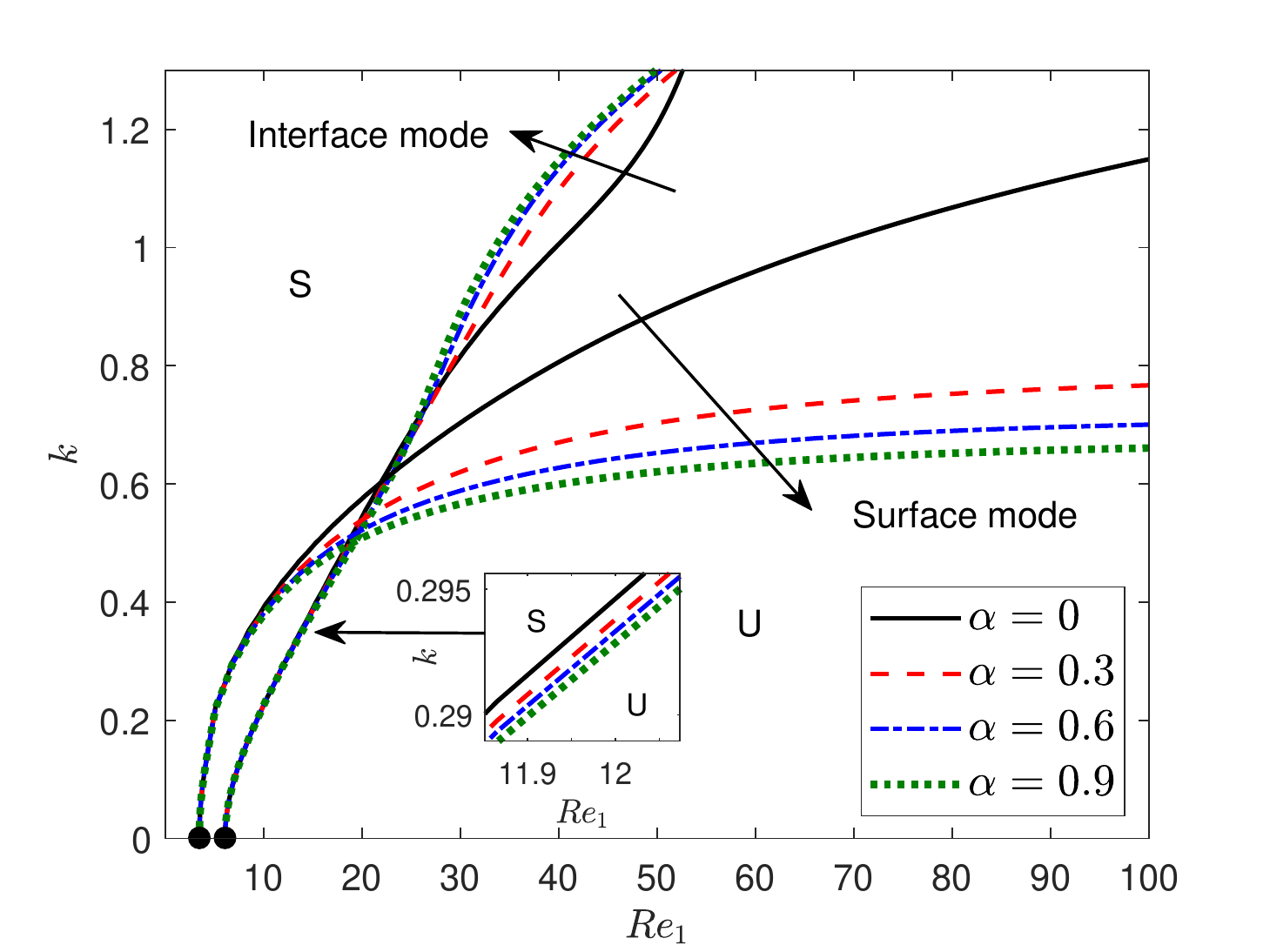}}
    \end{center}\vspace{-0.5cm}
	\caption{ Competition between marginal curves induced by the surface and interface modes for different $\alpha$ values when (a) $m=1.5$,  $r=1.1$ and (b) $m=3.5$,  $r=4.5$ with $\delta=1$, $Ca=1$, $Ma=0.05$, $\theta=0.2~rad$, $\beta=\frac{1}{Re_1}$, and $Pe=\infty$. The upward/downward arrow sings the increasing/decreasing unstable zone of the corresponding mode. The black rigid line represents the result of free surface flow ($\alpha=\gamma=0$) and $\gamma=0.001$ for $\alpha\neq0$.}\label{f15}
\end{figure}

The main goal here is to investigate the competition between all the unstable modes identified in different parameters range. In Fig.~\ref{f15}, the marginal stability curves of the SM and IM are plotted when $m=1.5$ and $r=1.1$, as in Fig.~\ref{f15}(a) and for $m=3.5$ and $r=4.5$, as in Fig.~\ref{f15}(b) when the remaining fixed values are $\delta=1$, $Ca=1$, $Ma=0.05$, $\theta=0.2~rad$, $\beta=\frac{1}{Re_1}$, and $Pe=\infty$. If the bottom layer's viscosity and density exceed those of the top layer, the unstable SM completely dominates the IM in the entire $k-Re_1$ domain, as illustrated in Fig.~\ref{f15}(a).
From the physical standpoint, SM instability becomes more than IM instability. Nevertheless, if the bottom layer becomes highly dense and viscous, then the SM loses its dominance over the IM except in the small wavenumber region (see, Fig.~\ref{f15}(b)). 
Further, as usual, the rigidity parameter $\alpha$ has a negligible effect on both unstable modes in the longwave range but has a significant influence in the finite wavenumber zone. 
\begin{figure}[ht!]
	\begin{center}
		\subfigure[]{\includegraphics[width=7.2cm]{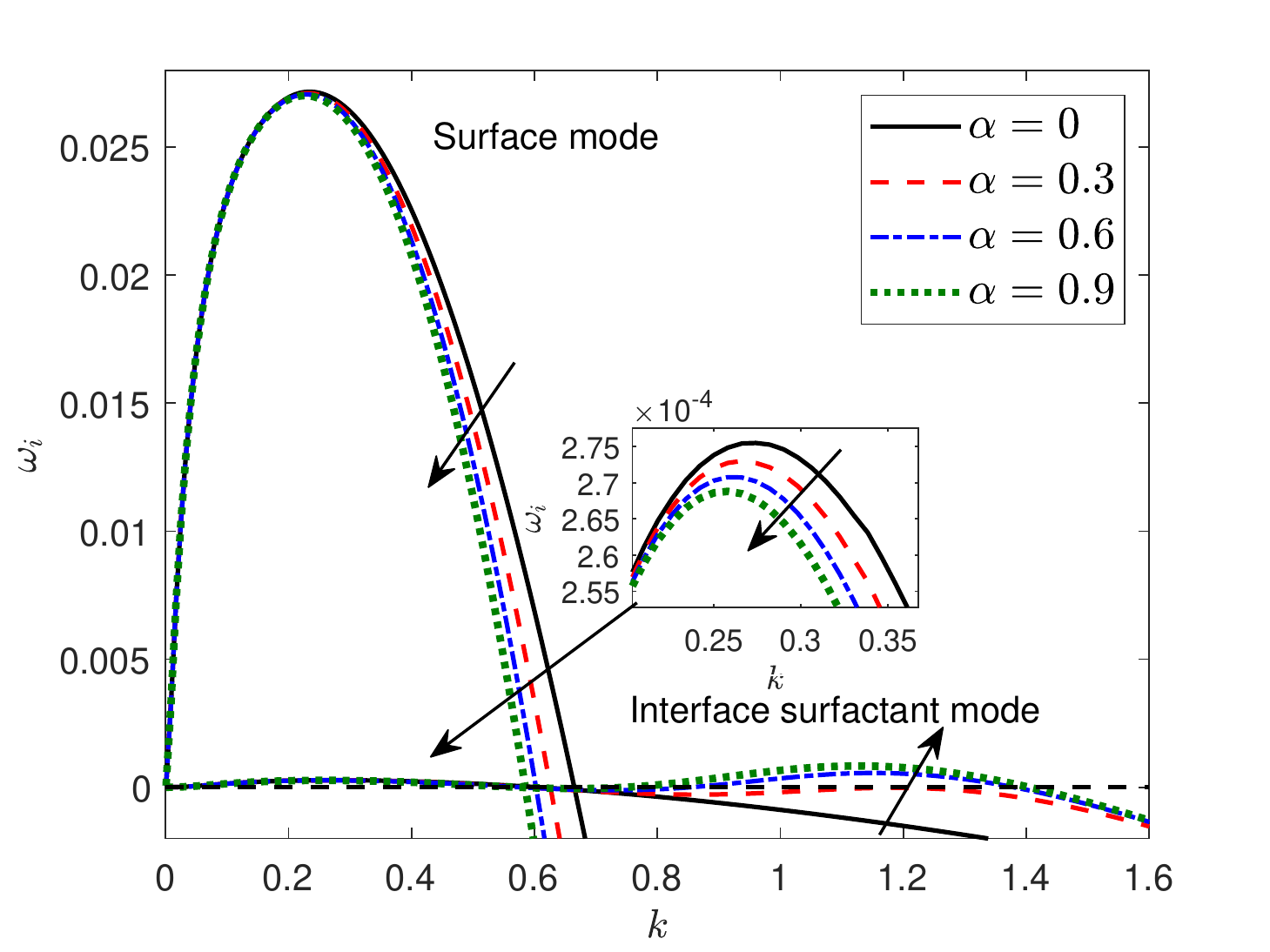}}
       \subfigure[]{\includegraphics[width=7.2cm]{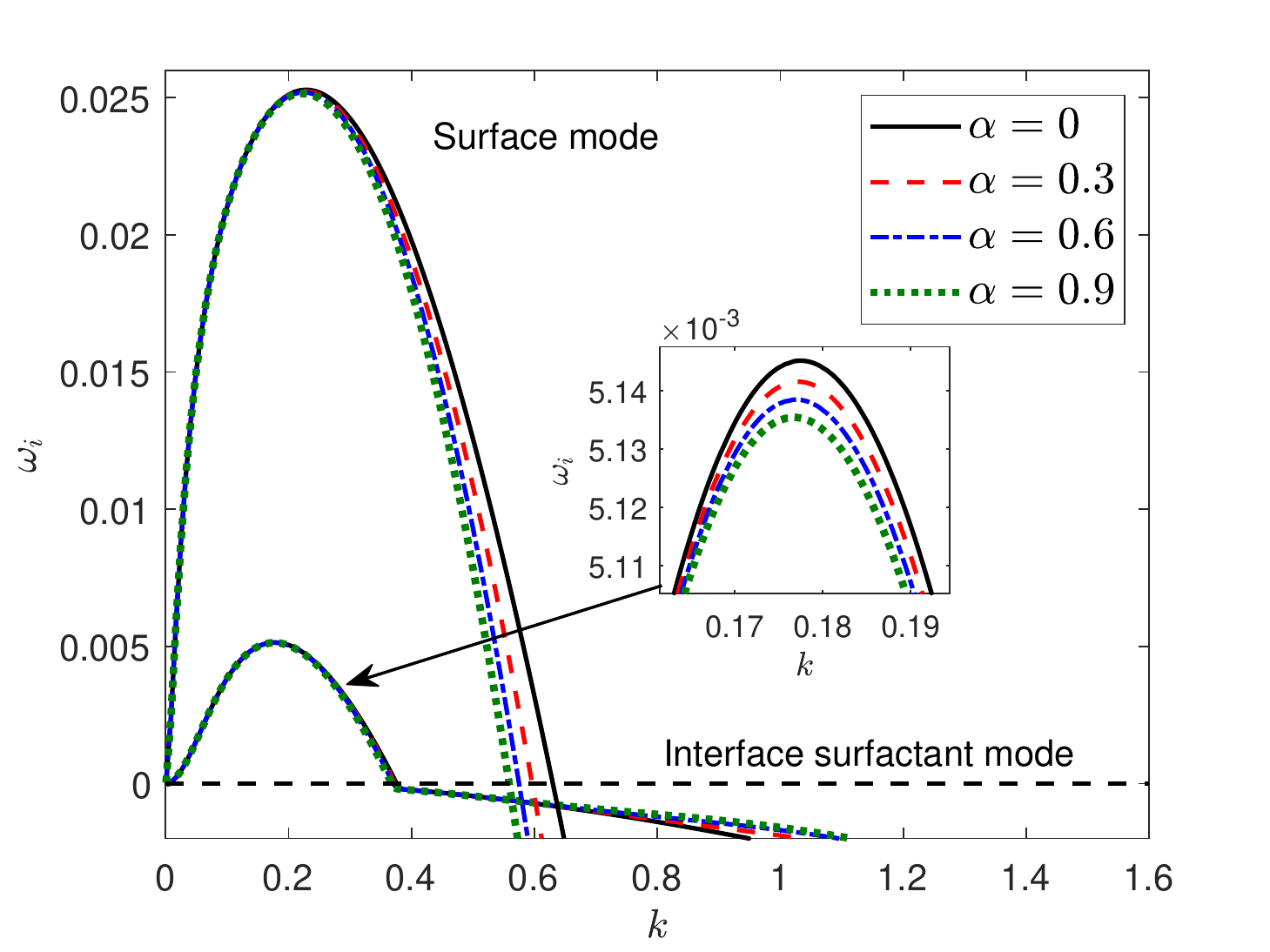}}
    \end{center}\vspace{-0.5cm}
	\caption{ Competition between temporal growth rate induced by the surface and interface surfactant modes for different $\alpha$ values when (a) $m=1.5$ and $r=1.1$, and (b) $m=2.5$ and $r=2.5$ with $Re_1=20$, $\delta=1$, $Ca=1$, $Ma=0.05$, $\theta=0.2~rad$, $\beta=\frac{1}{Re_1}$, and $Pe=\infty$. The upward/downward arrow sings the increasing/decreasing manner of the unstable zone of the corresponding mode. The black rigid line represents the free surface flow ($\alpha=\gamma=0$) and the value of $\gamma=0.001$ when $\alpha\neq0$.}\label{f16}
\end{figure}\vspace{-0.3cm}

Fig.~\ref{f16} exhibits the characteristic of temporal growth rate and assures the occurrence of both surface and interface surfactant mode when the fixed parameters are $Re_1=20$, $\delta=1$, $Ca=1$, $Ma=0.05$, $\theta=0.2~rad$, $\beta=\frac{1}{Re_1}$, and $Pe=\infty$. When the bottom layer viscosity and density become higher compared to the top layer ($m>1$ and $r>1$), the positive growth rate displayed in Fig.~\ref{f16}(a), of the SM becomes very high as compared to the ISM in the longwave regime, which confirms the SM's domination over the ISM. However, the temporal growth rate of SM in the shortwave zone takes a negative value for all values of $\alpha$, whereas the growth rate of the ISM becomes positive. This physical content assures that in the shortwave zone, the ISM fully dominates the SM.
However, as long as the density and viscosity of the lower layer increase, the positive temporal growth rate of the SM dominates the ISM (see, Fig.~\ref{f16}(b) for $m=2.5$ and $r=2.5$).
So, the competition between SM and ISM for dominancy is possible if one can significantly change the viscosity and density of the lower layer.       

\begin{figure}[ht!]
\begin{center}
\subfigure[]{\includegraphics[width=7.2cm]{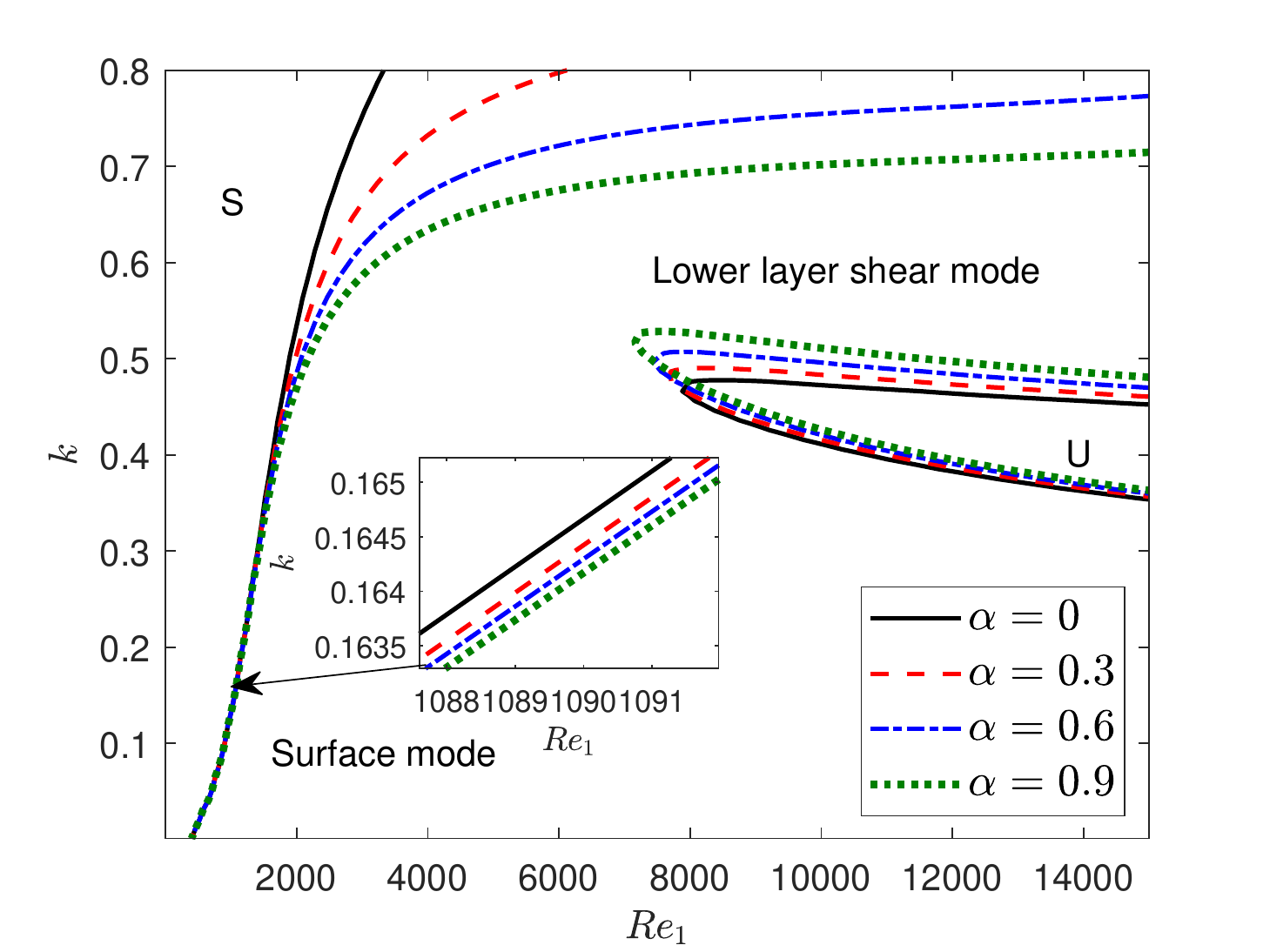}}
\subfigure[]{\includegraphics[width=7.2cm]{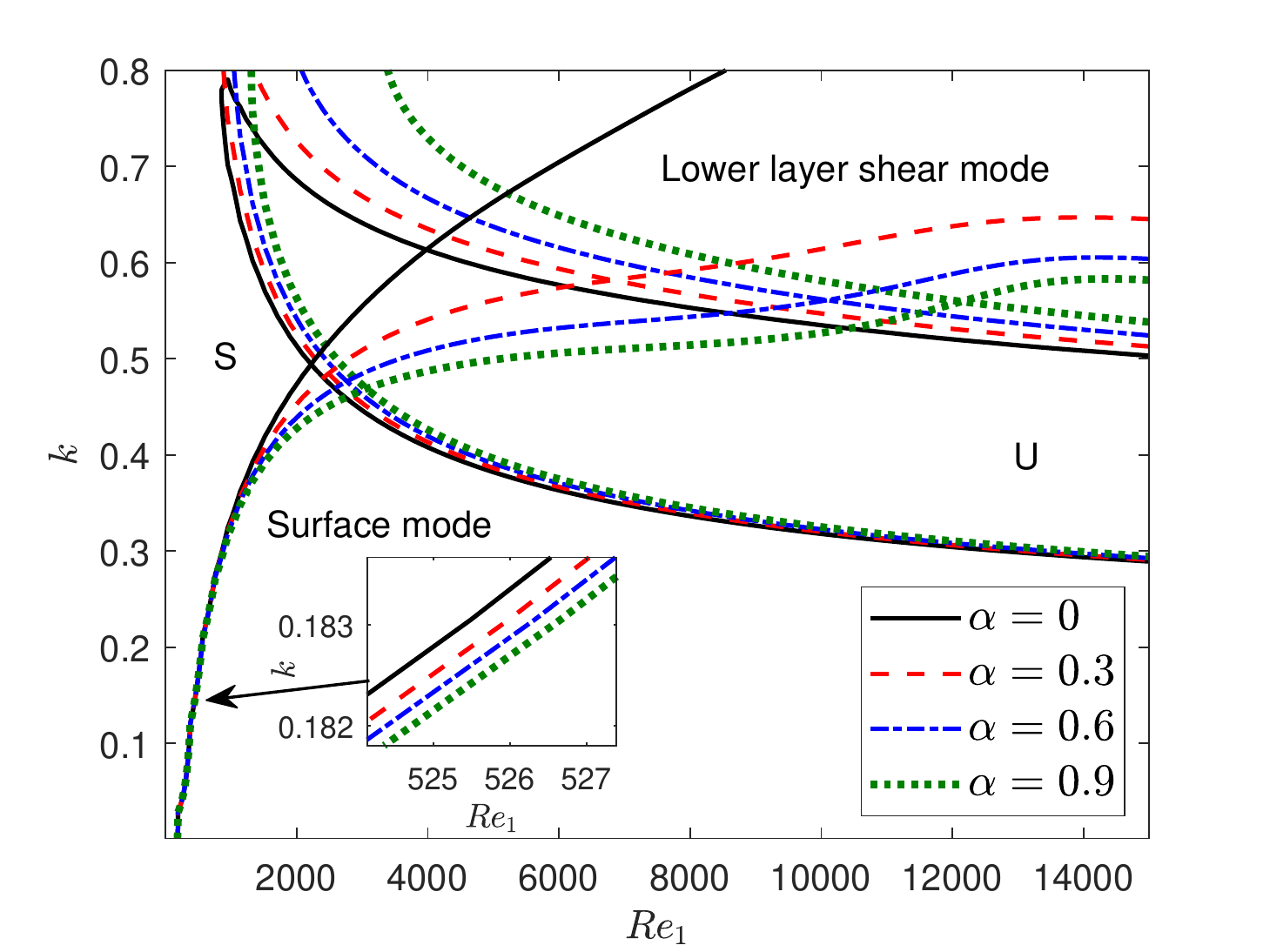}}
	\end{center}\vspace{-0.5cm}
	\caption{ Competition between marginal curves induced by the surface and lower layer shear mode for different $\alpha$ values when (a) $m=0.6$ and (b) $m=3$ with $\delta=1$, $r=1.1$, $Ca=1$, $Ma=0.01$, $\theta=0.0017~rad$, $\beta=\frac{1}{Re_1}$, and $Pe=\infty$. The black rigid line represents the free surface flow ($\alpha=\gamma=0$) and the value of $\gamma=0.001$ when $\alpha\neq0$.}\label{f17}
\end{figure}

The competition between SM and lower layer SHM is explained similarly to the earlier studies of \citet{bhat2018linear,hossain2022linear}. We have demonstrated the neutral curves in $Re_1-k$ plane related to SM and lower layer SHM for different rigidity $\alpha$ when the viscosity ratio (a) $m=3$ and (b) $m=0.6$ in Fig.~\ref{f17}. Here the fixed parameters $\delta=1$, $r=1.1$, $Ca=1$, $Ma=0.01$, $\theta=0.0017~rad$, $\beta=\frac{1}{Re_1}$, and $Pe=\infty$ are selected for this numerical mode competition test. Once the lower layer fluid becomes highly viscous ($m=3$), the neutral curves of the SM, as in Fig.~\ref{f17}(a), fully occupy the neutral curves corresponding to the lower layer SHM for all values of $Re_1$. This numerical result assures that the SM dominates the lower layer SHM. Besides, the higher structural rigidity of the floating flexible plate shrinks the unstable zone of the surface and shear modes and causes the stabilizing nature of both modes. However, if we choose a weaker viscous fluid in the lower layer ($m=0.6$), an opposite scenario occurs in Fig.~\ref{f17}(b). Here, the unstable SM dominates over the unstable lower layer SHM in the shorter wave range, but for higher wavenumber values, the SM loses its dominance over the lower layer SHM. \vspace{-0.1cm}
\section{Conclusions} \label{CON}
 \vspace{-0.2cm}
The current research focuses on the impact of the floating flexible plate in the double-layered film flow instability over an impermeable substrate. An insoluble surfactant is at the interface between the top and bottom layers. The flexible plate on the top layer surface modifies the normal stress balance between the upper layer and the surrounding air. Implementing the normal mode analysis corresponding to the infinitesimal disturbance of the fluid flow yields the Orr-Sommerfeld eigenvalue problem in terms of stream function. The Chebyshev collocation approach is employed to solve the eigenvalue problem numerically for disturbances with arbitrary wavenumbers.
The properties of various perturbation waves are exposed for different flow parameters. The three different unstable modes such as surface, interface, and interface surfactant modes are detected for moderate values of Reynolds number. The dominant surface mode induced by gravity waves becomes unstable for longwave to small wavenumber ranges. The surface mode travels with a higher phase speed ($c_r$) than the damped interface surfactant mode caused by the Marangoni effect. 
Note that the floating elastic plate has a negligible effect on all the above unstable modes in the longwave zone ($k\rightarrow0$) but has considerable influence in the finite wavenumber range. 
Based on the above findings, the effect of different fluid characteristic parameters on these modes is summarized as follows:

(i) The floating elastic plate has the potential to stabilize the unstable surface mode by reducing the corresponding unstable bandwidth. Also, each plate characteristic parameter (uniform rigidity and thickness) has an individual stabilizing influence on the surface wave instability in the finite wavenumber range. The unstable zone decreases as soon as the upper layer's viscosity increases. However, the opposite behaviour occurs when the upper layer density increases.    

(ii) When the weaker density fluid flows over the higher density fluid (i.e., $r>1$), the instability of the interface mode emerges up to a small wavenumber range for $mr<1$, whereas in the case of $mr<1$, interface mode instability is limited up to smaller wavenumber region, provided $m<1$. In the finite wavenumber domain, the rigidity and thickness of the floating plate play a dual role on the interfacial wave instability developed by the viscosity stratification when $mr<1$ with $m<1$. Further, no matter what $mr<1$ and $mr>1$, the interfacial wave instability in the smaller wavenumber region can be reduced individually by the rigidity and thickness parameters of the plate when $m<1$.   

(iii) The interface surfactant mode of the fluid flow system displays two humps: one in longwave and another one in finite wavenumbers zones. The interface surfactant mode exhibits dual behaviour with respect to the plate characteristic parameters. The flexible plate has the capability to reduce the interface surfactant mode instability for a significant range of small wavenumbers and then make it more unstable after that. 

Furthermore, when the inertia force is strong enough with a low inclination, the instability of the shear mode other than the above three modes appears when the density and viscosity of the lower layer become very high compared to the upper layer. The shear mode instability of the lower layer occurs in the small wavenumber range, but shear mode instability related to the upper layer occurs in the higher wavenumber zone \cite{bhat2020linear}. Also, it is noticed that the structural rigidity of the floating elastic plate amplifies the unstable zone corresponding to the induced by the shear mode of the upper layer, but manifests a double role on the lower layer shear mode. 

The idea of shear wave motion can be implicated in studying the flexural gravity waves since the shear wave instability emerges for a very small inclination angle (almost horizontal bottom bed). Thus, the findings of the shear wave instability will help to understand the stability mechanism of VLFS and the ice-breaking process in a stratified marine environment. 

Besides, the findings of the present investigation will assist the researchers to understand the linear instability mechanism of two-layer stratified flexural waves. This work will encourage the researchers for further experimental examination and verification of the elastic plate's effect on different complex fluid flow systems.   
\section*{Acknowledgment}\vspace{-0.25cm}
SG gratefully acknowledges the financial support from SERB, Department of Science and Technology, Government of India through Award No. MTR/2021/000442.

\section*{Data Availability}
 The data that supports the findings of this study are available
within the article, highlighted in the related figure captions and corresponding discussions.

\bibliographystyle{unsrtnat}
\bibliography{REF}
\end{document}